\documentclass[twocolumn, superscriptaddress, prb]{revtex4}
\usepackage{enumitem}

\usepackage[utf8]{inputenc} 
\usepackage[T1]{fontenc}    
\usepackage{hyperref}       
\usepackage{url}            
\usepackage{booktabs}       
\usepackage{amsfonts}       
\usepackage{nicefrac}       
\usepackage{microtype}      
\usepackage{lipsum}		
\usepackage{graphicx}
\usepackage{float}
\usepackage{natbib}
\usepackage{doi}
\usepackage{amsmath}
\usepackage{epstopdf}
\usepackage{bbm}
\usepackage{amsmath}
\usepackage{eqnarray}
\usepackage{siunitx}
\usepackage{soul}
\usepackage{xcolor}
\renewcommand{\figurename}{\textbf{Fig.}}

\usepackage{titlesec}

\titleformat{\chapter}[display]
  {\normalfont\fontsize{16pt}{18pt}\selectfont\mdseries}
  {\MakeUppercase{\chaptertitlename}\ \thechapter}{20pt}{\MakeUppercase}
\titleformat{\section}
  {\normalfont\fontsize{12pt}{14pt}\selectfont\bfseries}{\thesection}{1em}{}
  
\titleformat{\subsection}
  {\normalfont\fontsize{12pt}{14pt}\selectfont\itshape}{\thesubsection}{1em}{}

\titlespacing*{\section}
{0pt}{5.5ex plus 1ex minus .2ex}{4.3ex plus .2ex}
\titlespacing*{\subsection}
{0pt}{5.5ex plus 1ex minus .2ex}{4.3ex plus .2ex}

\newcommand{\eb}[2]{#2}

\begin{document}

\title{Electroluminescence and energy transfer mediated by hyperbolic polaritons}

\author{L. Abou-Hamdan}
\thanks{These two authors contributed equally to this work.}
\affiliation{Institut Langevin, ESPCI Paris, PSL University, CNRS, 1 rue Jussieu, F-75005 Paris, France}
\affiliation{DOTA, ONERA, Universit\'e Paris-Saclay, F-91123 Palaiseau, France}
\author{A. Schmitt}
\thanks{These two authors contributed equally to this work.}
\affiliation{Laboratoire de Physique de l'Ecole Normale Sup\'erieure, ENS, Universit\'e PSL, CNRS, Sorbonne Universit\'e, Universit\'e Paris-Cit\'e, 24 rue Lhomond, 75005 Paris, France}
\author{R. Bretel}
\affiliation{Laboratoire de Physique de l'Ecole Normale Sup\'erieure, ENS, Universit\'e PSL, CNRS, Sorbonne Universit\'e, Universit\'e Paris-Cit\'e, 24 rue Lhomond, 75005 Paris, France}
\author{S. Rossetti}
\affiliation{Institut Langevin, ESPCI Paris, PSL University, CNRS, 1 rue Jussieu, F-75005 Paris, France}
\affiliation{DOTA, ONERA, Universit\'e Paris-Saclay, F-91123 Palaiseau, France}
\author{M. Tharrault}
\affiliation{Laboratoire de Physique de l'Ecole Normale Sup\'erieure, ENS, Universit\'e PSL, CNRS, Sorbonne Universit\'e, Universit\'e Paris-Cit\'e, 24 rue Lhomond, 75005 Paris, France}
\author{D. Mele}
\affiliation{Laboratoire de Physique de l'Ecole Normale Sup\'erieure, ENS, Universit\'e PSL, CNRS, Sorbonne Universit\'e, Universit\'e Paris-Cit\'e, 24 rue Lhomond, 75005 Paris, France}
\affiliation{Univ. Lille, CNRS, Centrale Lille, Univ. Polytechnique Hauts-de-France, Junia-ISEN, UMR 8520-IEMN, F-59000 Lille, France.}
\author{A. Pierret}
\affiliation{Laboratoire de Physique de l'Ecole Normale Sup\'erieure, ENS, Universit\'e PSL, CNRS, Sorbonne Universit\'e, Universit\'e Paris-Cit\'e, 24 rue Lhomond, 75005 Paris, France}
\author{M. Rosticher}
\affiliation{Laboratoire de Physique de l'Ecole Normale Sup\'erieure, ENS, Universit\'e PSL, CNRS, Sorbonne Universit\'e, Universit\'e Paris-Cit\'e, 24 rue Lhomond, 75005 Paris, France}
\author{T. Taniguchi}
\affiliation{Advanced Materials Laboratory, National Institute for Materials Science, Tsukuba, Ibaraki 305-0047,  Japan}
\author{K. Watanabe}
\affiliation{Advanced Materials Laboratory, National Institute for Materials Science, Tsukuba, Ibaraki 305-0047,  Japan}
\author{C. Maestre}
\affiliation{Laboratoire des Multimat\'eriaux et Interfaces, UMR CNRS 5615, Univ Lyon, Universit\'e Claude Bernard Lyon 1, F-69622 Villeurbanne, France}
\author{C. Journet}
\affiliation{Laboratoire des Multimat\'eriaux et Interfaces, UMR CNRS 5615, Univ Lyon, Universit\'e Claude Bernard Lyon 1, F-69622 Villeurbanne, France}
\author{B. Toury}
\affiliation{Laboratoire des Multimat\'eriaux et Interfaces, UMR CNRS 5615, Univ Lyon, Universit\'e Claude Bernard Lyon 1, F-69622 Villeurbanne, France}
\author{V. Garnier}
\affiliation{Universit\'e de Lyon, MATEIS, UMR CNRS 5510, INSA-Lyon, F-69621 Villeurbanne cedex, France}
\author{P. Steyer}
\affiliation{Universit\'e de Lyon, MATEIS, UMR CNRS 5510, INSA-Lyon, F-69621 Villeurbanne cedex, France}
\author{J. H. Edgar}               
\affiliation{Tim Taylor Department of Chemical Engineering, Kansas State University, Durland Hall, Manhattan, KS 66506-5102, USA}
\author{E. Janzen}
\affiliation{Tim Taylor Department of Chemical Engineering, Kansas State University, Durland Hall, Manhattan, KS 66506-5102, USA}
\author{J-M. Berroir}
\affiliation{Laboratoire de Physique de l'Ecole Normale Sup\'erieure, ENS, Universit\'e PSL, CNRS, Sorbonne Universit\'e, Universit\'e Paris-Cit\'e, 24 rue Lhomond, 75005 Paris, France}
\author{G. Fève}
\affiliation{Laboratoire de Physique de l'Ecole Normale Sup\'erieure, ENS, Universit\'e PSL, CNRS, Sorbonne Universit\'e, Universit\'e Paris-Cit\'e, 24 rue Lhomond, 75005 Paris, France}
\author{G. Ménard}
\affiliation{Laboratoire de Physique de l'Ecole Normale Sup\'erieure, ENS, Universit\'e PSL, CNRS, Sorbonne Universit\'e, Universit\'e Paris-Cit\'e, 24 rue Lhomond, 75005 Paris, France}
\author{B. Plaçais}
\affiliation{Laboratoire de Physique de l'Ecole Normale Sup\'erieure, ENS, Universit\'e PSL, CNRS, Sorbonne Universit\'e, Universit\'e Paris-Cit\'e, 24 rue Lhomond, 75005 Paris, France}
\author{C. Voisin}
\affiliation{Laboratoire de Physique de l'Ecole Normale Sup\'erieure, ENS, Universit\'e PSL, CNRS, Sorbonne Universit\'e, Universit\'e Paris-Cit\'e, 24 rue Lhomond, 75005 Paris, France}
\author{J-P. Hugonin}
\affiliation{Laboratoire Charles Fabry, IOGS, Université Paris-Saclay, F-91123 Palaiseau, France}
\author{E. Bailly}
\affiliation{Laboratoire Charles Fabry, IOGS, Université Paris-Saclay, F-91123 Palaiseau, France}
\author{B. Vest}
\affiliation{Laboratoire Charles Fabry, IOGS, Université Paris-Saclay, F-91123 Palaiseau, France}
\author{J-J. Greffet}
\affiliation{Laboratoire Charles Fabry, IOGS, Université Paris-Saclay, F-91123 Palaiseau, France}
\author{P. Bouchon}
\affiliation{DOTA, ONERA, Universit\'e Paris-Saclay, F-91123 Palaiseau, France}
\author{Y. De Wilde}
\affiliation{Institut Langevin, ESPCI Paris, PSL University, CNRS, 1 rue Jussieu, F-75005 Paris, France}
\author{E. Baudin}
\thanks{Corresponding author: emmanuel.baudin@phys.ens.fr}
\affiliation{Laboratoire de Physique de l'Ecole Normale Sup\'erieure, ENS, Universit\'e PSL, CNRS, Sorbonne Universit\'e, Universit\'e Paris-Cit\'e, 24 rue Lhomond, 75005 Paris, France}
\affiliation{Institut Universitaire de France, France}

\begin{abstract}
\textbf{
Under high electrical current, some materials can emit electromagnetic radiation beyond incandescence. This phenomenon, referred to as electroluminescence, leads to the efficient emission of visible photons and is the basis of domestic lighting devices (\emph{e.g.}, light-emitting diodes) \cite{akasaki2015nobel,rosencher2002optoelectronics}. 
In principle, electroluminescence can lead to mid-infrared (mid-IR) emission of confined light-matter excitations called phonon-polaritons \cite{kalt2019semiconductor,low2017polaritons}, resulting from the coupling of photons with crystal lattice vibrations (optical phonons). In particular, phonon-polaritons arising in the van der Waals crystal hexagonal boron nitride (hBN) exhibit hyperbolic dispersion, which enhances light-matter coupling \cite{Dai2014science,Dai2015nnano}. For this reason, electroluminescence of hyperbolic phonon-polaritons (HPhPs) has been proposed as an explanation for the peculiar radiative energy transfer within hBN-encapsulated graphene transistors \cite{Wei2018Nnano,Baudin2020adfm}.
However, since HPhPs are locally confined, they are inaccessible in the far-field, and as such, any hint of electroluminescence has only been based on indirect electronic signatures and has yet to be confirmed by direct observation.
Here, we demonstrate far-field mid-IR ($\lambda=$~6.5~µm) electroluminescence of HPhPs excited by strongly biased high-mobility graphene within a van der Waals heterostructure, and we quantify the associated radiative energy transfer through the material. The presence of HPhPs is revealed via far-field mid-IR spectroscopy due to their elastic scattering at discontinuities in the heterostructure. 
The resulting radiative flux is quantified by mid-IR pyrometry of the substrate receiving the energy.
This radiative energy transfer is also shown to be reduced in hBN with nanoscale inhomogeneities, demonstrating the central role of the electromagnetic environment in this process.
}
\end{abstract}

\maketitle

Electroluminescence is the result of an electronic state driven strongly out-of-equilibrium which, due to light-matter coupling, radiates through the modes of its dielectric environment. In this regard, van der Waals materials are of particular interest as, owing to their extreme dielectric anisotropy, they naturally sustain low-loss HPhP modes \cite{Li2020jmcc}, which have long been sought after in metamaterial design \cite{Messina2016prb,Shi2015nlett}. These HPhP modes modify radiative energy transfer due to their substantial electromagnetic local density of states (LDOS) \cite{shen2009surface}. Nonetheless, this energy transfer remains inaccessible outside of the host material as the large wavevectors of HPhPs ($k\gg \omega/ c$) prevent them from coupling to far-field radiation due to momentum mismatch. The observation of HPhP electroluminescence requires three key ingredients. (i)~A~nearby strongly out-of-equilibrium electron gas acting as a radiating source, with this role played here by charge carriers in high-mobility graphene under large bias. (ii)~A~high-quality van der Waals material exhibiting ultra-low-loss HPhP modes such as hBN. (iii) The presence of scatterers in the heterostructure to allow HPhP elastic scattering to the far-field.

\begin{figure*}[t!]
\centering
\includegraphics[width=\linewidth]{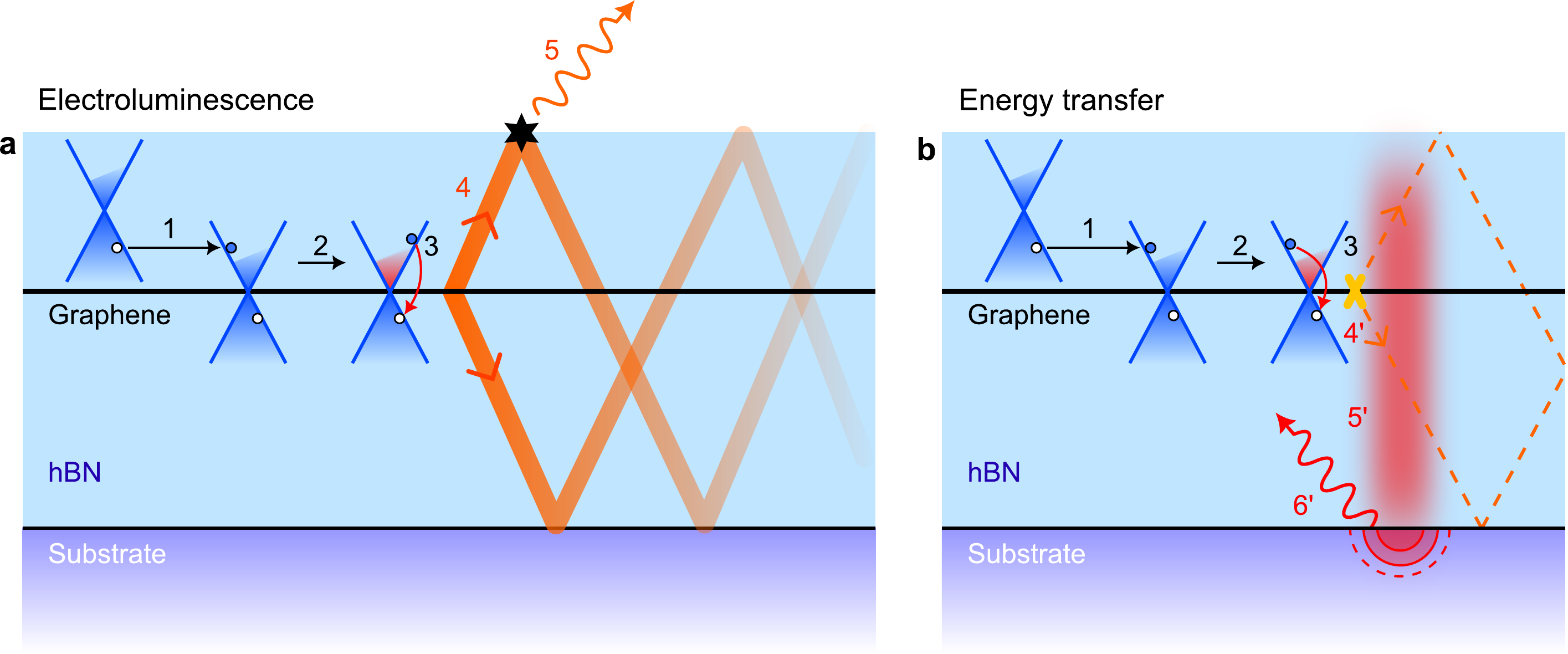}
\caption{\textbf{Electroluminescence and radiative energy transfer from graphene in an hBN encapsulate.}
\textbf{a},~Electroluminescence from graphene to the far-field results from the emission of small wavevector polaritons of hBN and follows a 5-step process:
(Step~1) Electrical Zener-Klein interband pumping under large applied bias. 
(Step~2) Intraband thermalization. 
(Step~3) Emission of a hyperbolic phonon-polariton (HPhP) of hBN by interband electron-hole recombination.
(Step~4) HPhPs propagate through the encapsulating medium. 
(Step~5) Propagating small-wavevector HPhPs are elastically scattered, leading to far-field photon emission.
\textbf{b}, Energy transfer occurs contemporaneously:
(Step~4') Dissipation of large wavevector HPhPs over nanometric length scales and heating of hBN.
(Step~5') Heat conduction through hBN results in a local temperature increase of the SiO$_{2}$ substrate.
(Step~6') Incandescent emission from the hot SiO$_{2}$ substrate.
}
\label{fig1}
\end{figure*}

The study of graphene as a hot thermal emitter has already revealed that heat transfer in hBN/graphene/hBN heterostructures is dominated by HPhP radiative heat transfer \cite{Tielrooij2018Nnano,Principi2017prl,Pogna2021ACS}.
However, in high-mobility graphene ($\mu\geq 2\ \rm{m^2.V^{-1}.s^{-1}}$) under large bias, electron-phonon decoupling enables graphene's electrons to reach a non-thermal state and radiate into hBN with an efficiency that surpasses that of Planckian radiative heat transfer. Two processes have been proposed to give rise to HPhP electroluminescence, which are illustrated below.

The first HPhP emission process is due to intraband electronic transitions: Under large bias, the large electron drift velocity gives rise to Cherenkov HPhP emission \cite{maciel2020probing,guo2023hyperbolic}, which corresponds to inelastic electron-HPhP scattering. The second HPhP emission process is due to interband electronic transitions: Under large bias, graphene's electrons are in a strongly out-of-equilibrium distribution due to electron-hole pair injection \cite{Wei2018Nnano, Schmitt2023NPhys,Vandecasteele2010prb} as a result of interband Zener-Klein (ZK) tunneling. Recombination of these electron-hole pairs then results in HPhP emission.

HPhP emission in hBN may be observed in its two \emph{Reststrahlen} bands:
the lower-energy type~I HPhPs of \emph{Reststrahlen} band~I (energy range $\hbar \omega_{I} \simeq 90-100$~meV) occurs as a result of both intraband and interband emission processes. 
In contrast, the higher energy type~II HPhPs of \emph{Reststrahlen} band~II ($\hbar \omega_{II} \simeq 169- 200$~meV) can only occur via interband emission since electronic drift velocities in graphene are limited by type~I HPhPs, as revealed by current saturation \cite{Wei2018Nnano}. The interband electron-HPhP energy transfer results in a saturation of the heating of the electron gas (see Extended Data Fig. 11) as revealed by monitoring the electronic noise \cite{Wei2018Nnano}, a process reminiscent of electroluminescent cooling \cite{Baudin2020adfm,sadi2020thermophotonic}.

In this study, we present a direct observation of hyperbolic radiative energy transfer by investigating hBN-encapsulated graphene devices in ambient conditions under large bias. We use mid-IR spectroscopy \cite{Li2018prl,Abou2021ol,Abou2022acs} to reveal the spectral signatures of the type~II HPhP energy carriers scattered to the far-field by discontinuities in the heterostructure, such as folds and metal contacts (Fig.~\ref{fig1}a). Then, we monitor the out-of-plane power transferred to the device's SiO$_{2}$ substrate by pyrometry, unequivocally demonstrating radiative energy transfer mediated by HPhPs (Fig.~\ref{fig1}b). Finally, by measuring graphene's electronic temperature via near-IR thermometry, we demonstrate that the crystalline quality of the hBN encapsulant has a major impact on the out-of-plane energy transfer efficiency with limited effect on electronic transport.

\section*{Electroluminescence of hyperbolic phonon-polaritons}

\begin{figure*}[t!]
\centering
\includegraphics[width=\linewidth]{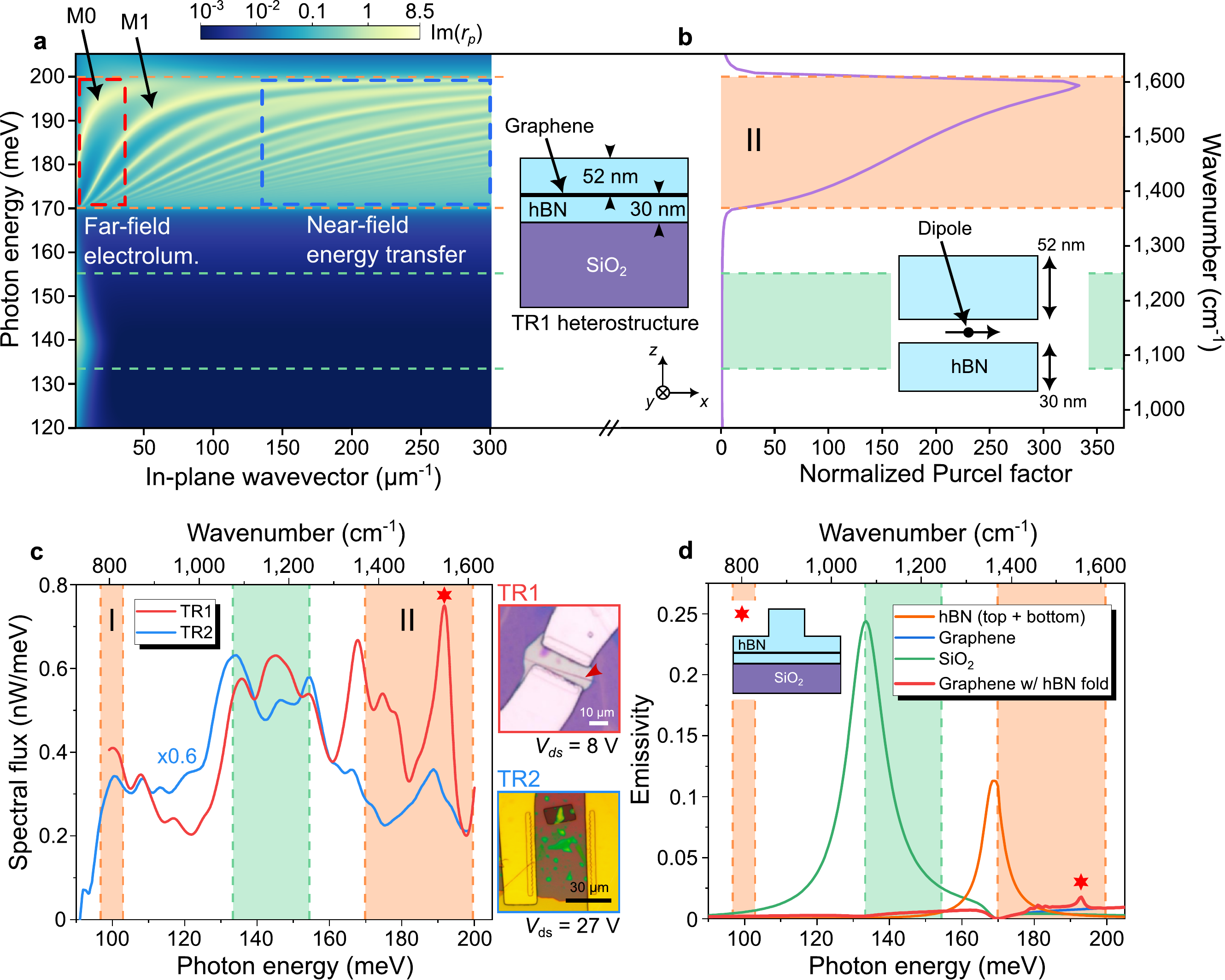}
\caption{\textbf{Spectral emission from hBN-encapsulated graphene devices under large bias: identification of the electroluminescent signal.}
\textbf{a},~Hyperbolic phonon-polariton dispersion branches within hBN's second \emph{Reststrahlen} band calculated for the device sketched in the central inset (TR1 in panel \textbf{c}) and a Fermi energy $E_{F}=30\ \mathrm{meV}$ (see supplementary information, section~\ref{polariton dispersion} for full details). Vertical confinement results in the emergence of HPhP modes (labelled M0 and M1, and indicated by arrows). The small wavevector region (dashed red box) is responsible for the far-field electroluminescence, while the large wavevector region (dashed blue box) is responsible for near-field energy transfer from graphene's electrons to hBN. 
\textbf{b},~Energy-dependent Purcell factor computed for the geometry described in the inset for an oscillating dipole at a distance of 3.55~\AA~ from each layer. The Purcell factor is normalized by its value at $1,000\ \mathrm{cm^{-1}}$ to highlight its enhancement within hBN's \emph{Reststrahlen} band (orange-shaded region). The green-shaded region indicates SiO$_{2}$'s \emph{Reststrahlen} band. \textbf{c},~Mid-infrared (mid-IR) emission spectra of two electrically-biased hBN-encapsulated graphene devices near charge neutrality (Fermi energy, $E_{F} \simeq 30$~meV): the small ($L\times W = 9.2 \times 15.5$~$\mu$m$^{2}$) TR1 device ($V_{ds}/L= 0.87~\mathrm{V/\mu m}$) and the large ($L\times W = 35 \times 35$~$\mu$m$^{2}$) TR2 device ($V_{ds}/L=0.77~\mathrm{V.\mu m^{-1}}$). A fold in the top hBN layer in TR1 (indicated by a red arrowhead in the top micrograph to the right of panel \textbf{c}) acts as an efficient scatterer of the HPhP modes of the heterostructure (the electroluminescent peak is marked by a red star).  Furthermore, in this smaller device, the edge contacts are visible within the field of view of the mid-IR microscope, contributing to scattering as well.  
\textbf{d},~Simulated emissivity of the individual layers of the hBN/graphene/hBN/SiO$_{2}$/Si heterostructure for the pristine heterostructure (blue, orange, and green curves) and for the case of a 0.25~$\mu$m-wide and 0.1~$\mu$m-thick hBN fold in the top layer (red curve, see supplementary information, section~\ref{AFM sec} for an atomic-force microscopy characterisation of the hBN fold) displaying a scattering peak at 192~meV (red star) corresponding to the coupling of confined HPhPs to the far-field.
}
\label{fig2}
\end{figure*}

When pushed out of equilibrium, graphene's electrons (in hBN/graphene/hBN heterostructures) emit HPhPs over a broad range of wavevectors. Small- and large-wavevector HPhPs (red and blue boxes in Fig.~\ref{fig2}a, respectively) can be schematically distinguished:
The small-wavevector HPhPs are thin film modes of the hBN cavity, which propagate over micrometric distances (see Fig. SI-~\ref{6modespropagation}). In particular, they convey the coupling between graphene's electrons and the far-field via defects in hBN and/or scatterers on top of the hBN encapsulant. In contrast, the large-wavevector HPhPs have a vanishing group velocity which leads to their dissipation into acoustic phonons and subsequent conversion into heat over extremely short distances (on the order of a fraction of the Fermi wavelength: $\lambda_{F}/4\pi \sim 11$~nm at $E_F=30~\mathrm{meV}$) \cite{Tielrooij2018Nnano}.
The quasi-flat dispersion of the large-wavevector HPhPs (see Fig.~\ref{fig2}a) results in a significant boost to the LDOS and consequently to the Purcell factor (Fig.~\ref{fig2}b), which is defined as the ratio between the total power emitted by a dipole in the structure and the total power emitted by a dipole in vacuum. The energy transfer associated with such large-wavevector HPhP modes leads to a heating of the hBN layer and subsequently to the heating of the underlying substrate through thermal conduction, resulting in incandescent emission from the substrate.

We begin by experimentally revealing the far-field signature of type~II HPhP electroluminescence of high-mobility graphene devices under large bias. To this end, we investigate the emission from two hBN-encapsulated graphene devices, both fabricated on an $\mathrm{SiO_2/Si}$ substrate, but with different characteristics. The first device (referred to as TR1) has a rather small graphene channel ($15 \times 9$~µm$^2$) and a fold in the top hBN layer (see the top micrograph to the right of Fig.~\ref{fig2}c).
The second device (hereafter referred to as TR2) has a much larger graphene channel ($35 \times 35$~µm$^2$) and presents a uniform hBN encapsulant. The mid-IR spectra of the two devices under electrical bias are shown in Fig.~\ref{fig2}c. 

The large and uniform device TR2 presents a predominant emission from $\mathrm{SiO_2}$'s \emph{Reststrahlen} band (see Fig.~\ref{fig2}c, blue curve) which arises due to the heating of the heterostructure under large bias. This incandescent emission results from the finite emissivities of $\mathrm{SiO_2}$ and hBN which are computed in Fig.~\ref{fig2}d, and is exploited in the next section to perform mid-IR pyrometry (the reader is referred to supplementary information, section VII for an advanced discussion of the spectral features). However, emission in hBN's type II \emph{Reststrahlen} band is limited as expected from numerical simulations. Contrary to TR2's emission, TR1's measured spectrum (Fig.~\ref{fig2}c, red curve) is characterised by an increased spectral flux in hBN's second \emph{Reststrahlen} band (orange-shaded region II). Emission in this spectral band can be decomposed into two regions: At the lower bound of the \emph{Reststrahlen} band ($167-180$~meV), emission originates both from hBN's bulk incandescence at the transverse optical phonon frequency (Fig.~\ref{fig2}d, orange curve) and from the scattered type II HPhPs as confirmed by the calculated emissivity of hBN and graphene within the heterostructure (see supplementary information, section~\ref{spec features}). The upper bound of hBN's \emph{Reststrahlen} band exhibits a solitary peak emerging at 192~meV (marked by a red star in Fig.~\ref{fig2}c), which is unambiguously attributed to type II HPhPs scattered to the far-field by the fold in the top hBN layer \cite{gilburd2017}, as revealed by the emissivity simulation plotted in Fig.~\ref{fig2}d (red curve). 
Finally, a small peak at $\sim 100$~meV is also observed in the measured spectra, which is due to type~I HPhPs, and is routinely observed in other devices as well (see Extended Data Fig.~\ref{ED_fig8} and \ref{ED_fig9}) \cite{guo2023hyperbolic}.

The large spectral flux observed at 192~meV in comparison to the relatively small computed emissivity peak amplitude at that energy qualitatively indicates the electroluminescent nature of this emission. To prove that the type~II HPhP emission at $192$~meV in TR1 is due to electroluminescence, we experimentally rule out its incandescent origin using the framework of the generalized Kirchhoff law \cite{Greffet2018prx} (see Methods).
We first compared the device's emission under bias to its thermal emission when heated to the temperature of hBN's optical phonons, which was found to be $\sim 75\mathrm{^\circ C}$ by Stokes anti-Stokes Raman thermometry (see Methods and Extended Data Fig.~\ref{ED_fig5}). The polariton peak around 190~meV was markedly absent from the measured thermal radiation spectrum (see Extended Data Fig.~\ref{ED_fig6}), indicating that the device's measured emission under bias is uncorrelated with hBN's incandescence. Then, we estimated the mid-IR incandescence from graphene's hot electrons using a Planck law fit, which was determined from near-IR measurements of graphene's electron temperature as a function of applied bias (see Extended Data Fig.~\ref{ED_fig7}). 
The device's measured signal as a function of increasing bias was found to increase about 2 times faster than that of the Planck law fit in the bias range between $6-8$~V, effectively ruling out graphene's incandescence as the source of emission.

We finally mention here that while a fold in hBN is sufficient to reveal HPhP electroluminescence in the far-field, we have found that for a small device, the device's edge contacts naturally act as scatterers (see Extended Data Fig.~\ref{ED_fig9}). It is also possible to use intentionally deposited scatterers, such as gold pillars that are either randomly distributed (see Extended Data Fig.~\ref{ED_fig8}) or organized in a periodic lattice \cite{guo2023hyperbolic}. 
However, systematic numerical simulations (see Fig. SI-~\ref{scatterer_types}) indicate that water/hydrocarbon bubbles, which are frequent in van der Waals heterostructures and can be observed in TR2 (see green bubbles in the bottom micrograph in Fig.~\ref{fig2}c), are inefficient scatterers, which explains why TR2 only displays a faint emission peak at 190~meV.

\section*{Pyrometry of the electroluminescent energy transfer} \label{hotSiO2}

\begin{figure*}[t!]
\centering
\includegraphics[width=\linewidth]{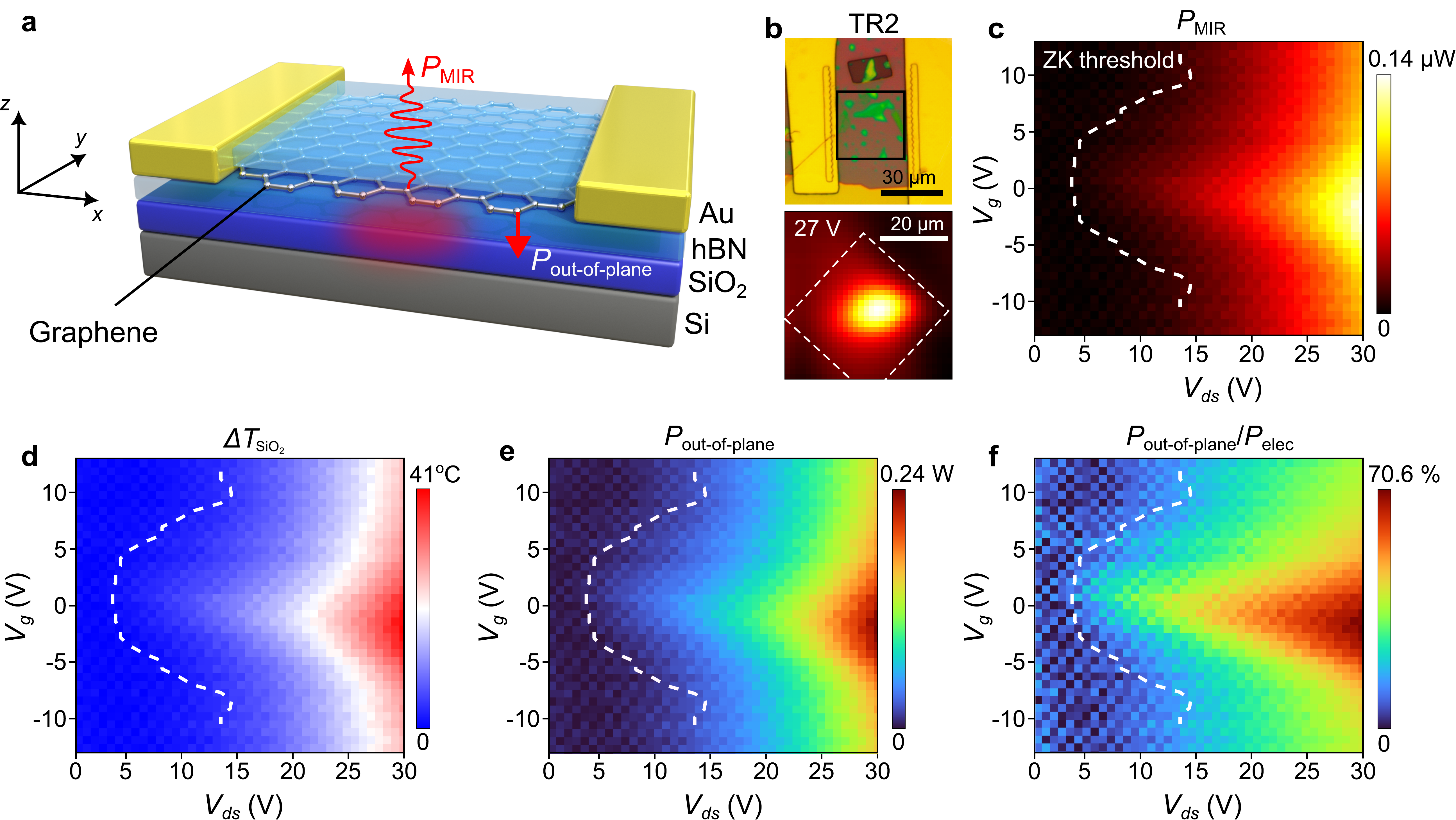}
\caption{\textbf{Monitoring the out-of-plane energy transfer in a graphene device through its mid-infrared incandescent emission.} \textbf{a}, Schematic depiction of out-of-plane energy transfer in a high-mobility graphene field-effect device with an SiO$_{2}$ back-gate (substrate). The back-gate warms up slightly due to this out-of-plane energy transfer and subsequently emits mid-infrared (mid-IR) thermal radiation. One can directly observe the device's out-of-plane energy transfer by monitoring the substrate's mid-IR emission. \textbf{b}, Microscope image (top) and spatial scan (bottom) of the spectrally integrated mid-IR signal ($P_{\mathrm{MIR}}$, integrated over the detector's spectral bandwidth, $\lambda = 6-14$~µm) emitted by the TR2 device under bias. The dashed white box indicates the orientation of the device's graphene channel (black box in the top micrograph) in the image. The bright emission spot, between the source and drain electrodes (left and right gold pads in the top micrograph), indicates that the signal is most intense in that region, revealing that the emission originates from beneath the device channel. By contrast, the device's gold electrodes appear dark due to their low emissivity. \textbf{c}, Scan of $P_{\mathrm{MIR}}$ as a function of electrical bias. The white dashed curve indicates the device's Zener-Klein (ZK) tunneling threshold.  \textbf{d}, Measured temperature increase in the SiO$_{2}$ substrate as a function of bias. The temperature increase is determined by calibrating the detected mid-IR signal as a function of sample temperature (see Methods and Extended Data Fig.~\ref{ED_fig4} for full details). \textbf{e}, Measured out-of-plane power $P_{\mathrm{out\mbox{-}of\mbox{-}plane}}$ as a function of applied bias. \textbf{f}, Ratio between $P_{\mathrm{out\mbox{-}of\mbox{-}plane}}$ and the total electrical power ($P_{\mathrm{elec}}= V_{ds}I_{ds}$, where $V_{ds}$ and $I_{ds}$ are the drain-source bias voltage and current, respectively) injected into the TR2 device.}
\label{fig3}
\end{figure*}

A direct indication of out-of-plane radiative energy transfer (Fig. 1b) in high-mobility graphene devices with an SiO$_{2}$/Si back-gate can be readily obtained by monitoring the device's substrate far-field mid-IR emission as a function of electrical bias. 
Indeed, a fraction $P_{\mathrm{out-of-plane}}$ of the total electrical power injected into the device reaches the SiO$_{2}$ back-gate via electroluminescent energy transfer from graphene to the first few nanometers of hBN and subsequently, via thermal conduction through the heterostructure.
This results in a slight temperature increase of the heterostructure, which is then signaled by incandescent mid-IR emission from the SiO$_{2}$ substrate as depicted schematically in Fig.~\ref{fig3}a. 

We use a HgCdTe photo-detector with maximal detectivity centred at SiO$_{2}$'s \emph{Reststrahlen} band (green shaded region in Fig.~\ref{fig2}) to quantify the incandescent emission. Within this spectral range, emission from hBN and graphene is overshadowed by SiO$_{2}$'s incandescence. Concordantly, the measurements presented in this section primarily serve as a pyrometric analysis of the SiO$_{2}$ substrate \cite{footnote1}, which is obtained from the spectrally-integrated mid-IR signal emitted by the large TR2 device under bias. Although the electroluminescence signal from scattered HPhPs also contributes to the integrated signal, it lies at the periphery of our photo-detector's spectral range of maximal detectivity, thus minimally impacting the current measurements, amounting to about 11\% of the integrated signal of the TR2 device.

We first verify through a spatial scan of the detected mid-IR signal ($P_{\mathrm{MIR}}$), that the signal is most intense at the channel location (see Fig.~\ref{fig3}b, bottom map), indicating that the detected signal indeed originates from the heated SiO$_{2}$ substrate beneath the device channel. We then quantify the contribution of out-of-plane energy transfer to the device's total power budget. The total power balance is given by $P_{\mathrm{elec}} = P_{\mathrm{in\mbox{-}plane}} + P_{\mathrm{out\mbox{-}of\mbox{-}plane}}$, where $P_{\mathrm{elec}} = V_{ds}\,I_{ds}$ ($V_{ds}$, drain-source bias voltage; $I_{ds}$, drain-source current) is the electrical power and $P_{\mathrm{out\mbox{-}of\mbox{-}plane}}$ ($P_{\mathrm{in\mbox{-}plane}}$) is the  power transferred in the out-of-plane (in-plane) direction. The power transferred in the out-of-plane direction is computed from $P_{\mathrm{out\mbox{-}of\mbox{-}plane}} = \Delta T_{\mathrm{SiO}_{2}} /R_{\mathrm{th}}$, where $R_{\mathrm{th}}$ is a thermal resistance obtained by solving the heat equation of the heterostructure (see supplementary information, section~\ref{model_therm}).
The temperature increase of the SiO$_{2}$ substrate $\Delta T_{\mathrm{SiO}_{2}}$ is determined by comparing the device's spectrally-integrated mid-IR signal under bias (Fig.~\ref{fig3}c) with the device's signal at null bias and known substrate temperature (see Methods and Extended Data Fig.~\ref{ED_fig4}), where we equate the apparent radiation temperature to the temperature of the SiO$_{2}$ substrate (see supplementary information, section~\ref{uni heat}).
As a result, we obtain the distribution of the temperature increase as a function of electrical bias, which is on the order of $\Delta T_{\mathrm{SiO}_{2}} \simeq 20-40$~$^{\circ}$C beyond the Zener-Klein threshold (see Fig.~\ref{fig3}d).  \\
\indent Following this model, $P_{\mathrm{out\mbox{-}of\mbox{-}plane}}$ and the ratio of the out-of-plane transferred power to the total electrical power injected into the device ($P_{\mathrm{out\mbox{-}of\mbox{-}plane}}/P_{\mathrm{elec}}$) can be computed as a function of electrical bias and doping, and the result is presented in Fig.~\ref{fig3}e and f. The maps reveal that the investigated out-of-plane energy transfer mechanism amounts to a significant portion of the device's total power budget, with a maximal contribution of about $71\, \%$. We underscore here that this corresponds to an out-of-plane power per surface area of $\sim 19.5~\mathrm{kW.cm^{-2}}$. This value is three orders of magnitude larger than the highest cooling power experimentally reported for a parent mechanism, namely the electroluminescent cooling in conventional semi-conductors \cite{Olsson2016ieee,Radevici2018ssc,Sadi2020NPhot}. \\
\indent Finally, we emphasize here that the behaviour of $P_{\mathrm{out\mbox{-}of\mbox{-}plane}}$ as a function of bias and gate voltage $(V_{ds}\, \text{and}\,  V_{g})$ in the large TR2 device is exemplary of electroluminescent energy transfer. Firstly, the onset of out-of-plane energy transfer matches the threshold of Zener-Klein interband tunneling extracted from DC transport measurements (see supplementary information, section~\ref{elec trans sec}), as indicated by the dashed line in Fig.~\ref{fig3}e. In fact, $P_{\mathrm{out\mbox{-}of\mbox{-}plane}}$ follows the same trend as that of the radiated power ($P_{\mathrm{MIR}}$, Fig.~\ref{fig3}c), whose onset is also given by the Zener-Klein tunneling threshold. In contrast, the electrical power shows the expected decrease near the charge-neutral point at large bias (see supplementary information, Fig. SI-~\ref{all powers}a).
Secondly, the out-of-plane power is largest at large bias near the charge-neutral point ($V_{g} =0$~V)  and falls off away from it due to Pauli blocking in the interband tunneling process. This is in stark contrast to the intraband mechanism, for which emission is more efficient at finite doping. This demonstrates the interband origin of type II HPhP electroluminescence in strongly biased hBN-encapsulated graphene devices.

\section*{Engineering the hyperbolic energy transfer}

Having demonstrated hyperbolic radiative energy transfer from graphene to hBN, we show in this section how to engineer this mechanism. To do so we consider two different sources of hBN crystals with contrasting properties: The first hBN source consists of crystals synthesized at high pressure and high temperature \cite{Taniguchi2007jcg} (denoted by HPHT-hBN in the following). The second hBN source consists of hBN crystals synthesized through a specific method at lower pressure, the polymer-derived ceramics route \cite{Maestre20222D} (denoted by PDC-hBN). HPHT and PDC crystals show comparable crystalline quality with yet a higher density of crystal defects in the latter, presumably involving carbon or oxygen interstitial atoms \cite{roux2021}. The presence of such defects is expected to affect the radiative energy transfer. 
Notably, devices made with PDC-hBN on a gold-coated substrate present similar transport properties to those made with HPHT-hBN, and feature, in particular, a large electronic mobility (up to $ 8~\mathrm{m^2.V^{-1}.s^{-1}}$  in ambient conditions) which is necessary to reach the electronic interband pumping regime. 

\begin{figure}[t!]
\centering
\includegraphics[width=\linewidth]{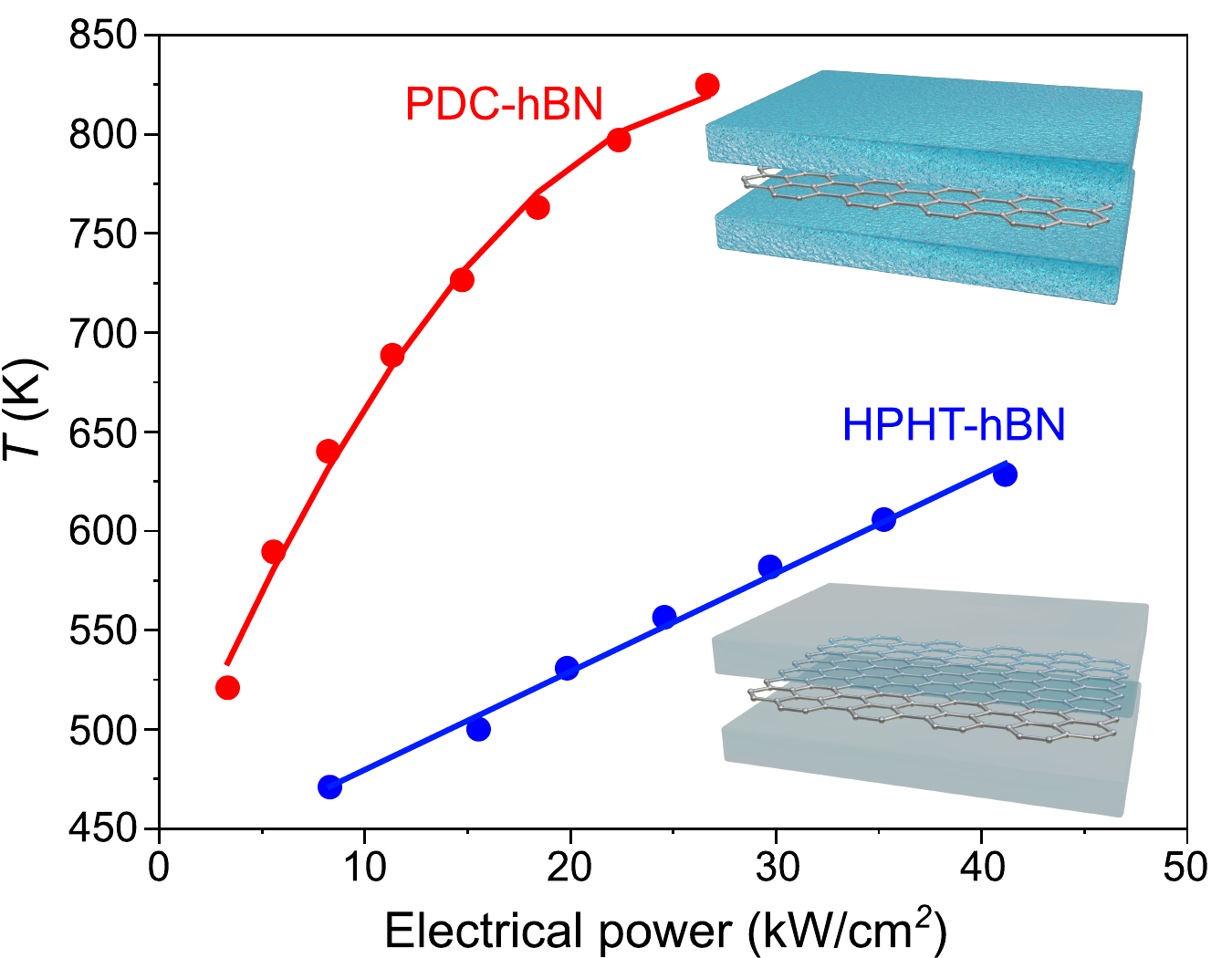}
\caption{\textbf{Engineering the energy transfer efficiency by tuning hBN crystalline quality.} Comparison of graphene's electron temperature, obtained from its near-IR emission versus \eb{bias}{surface Joule power} (details are outlined in Ref.~\cite{schmitt2023electroluminescence}), for a device made with PDC-hBN (Lyon1) and another made with HPHT-hBN (F161) [see Extended Data Table~\ref{liste_devices}].
}
\label{fig4}
\end{figure}

The influence of defects on the investigated energy transfer is confirmed by a near-infrared (near-IR) determination of graphene's electron temperature (Fig.~\ref{fig4}, experimental details are outlined elsewhere in Ref.~\cite{schmitt2023electroluminescence}), in which the electronic temperature is noticeably higher in the PDC-hBN device at a given electrical power dissipation per unit area. From this measurement, we observe that reaching a temperature of $T\sim 600$~K requires $34.3\ \mathrm{kW.cm^{-2}}$ with the HPHT-hBN device, and only $6.5\ \mathrm{kW.cm^{-2}}$ in the PDC-hBN device, the two devices having similar dimensions. This corresponds to a whopping 5.3-fold decrease in the out-of-plane energy transfer. This behaviour is also corroborated by $1-10$~GHz noise thermometry measurements (see Methods and Extended Data Fig.~\ref{ED_fig11}), in which we compare the electronic noise temperature of a device made with PDC-hBN and another made with HPHT-hBN.

Interestingly, far-field measurements performed on both hBN crystals revealed similar far-field optical properties in the investigated mid-infrared spectral range, which suggests that scattering sites in PDC-hBN are present on a subwavelength scale. Beyond mean-field, it has been shown that, in strongly disordered media, multiple scattering can result in a strong modification of the electromagnetic LDOS \cite{sapienza2011} leading to drastic changes in energy transfer.  In isotropic materials, this effect leads to an increase in the Purcell effect. In contrast, we observe a reduced out-of-plane energy transfer in hyperbolic hBN, highlighting the need for a re-evaluation of this phenomenon in the context of hyperbolic materials. By enabling the suppression of out-of-plane energy transfer, such materials with nanoscale inhomogeneities open the door to new regimes with extremely out-of-equilibrium electronic distribution.

\section*{Conclusion}

In summary, we have directly observed the type~II HPhP electroluminescence of hBN-encapsulated graphene devices under large bias as a result of scattering by discontinuities. We have also shown that this electroluminescence is responsible for the large out-of-plane energy transfer to the substrate, as revealed by its key features (electron-hole injection threshold and gate voltage dependence). Through pyrometric measurements, we quantified the efficiency of the electroluminescence-driven energy transfer for a $35\times35\ \rm{\mu m^2}$ transistor. Remarkably, our findings indicate that this process contributes to as much as $\sim 71\,\%$ of the device's total power budget.

These two phenomena illustrate the dual nature of a single fundamental mechanism in high-mobility graphene, as they both arise from the non-equilibrium electronic pumping of the electron gas and its subsequent radiative recombination through the emission of phonon-polaritons. Notably, this mechanism holds promise for the development of novel bright sources of phonon-polaritons and is not necessarily limited to the encapsulant's \emph{Reststrahlen} bands: Through nano-photonic engineering, it should be possible to evoke electroluminescence -- and potentially optical gain and lasing -- in the far-field at other wavelengths where damping is absent.

Finally, we demonstrated the engineering of the radiative energy transfer by tuning the local disorder in the hyperbolic hBN encapsulant and as such the LDOS of HPhPs. This effect was confirmed by comparing the electronic noise thermometry and near-IR thermometry profiles of two hBN sources of different crystalline quality. 
The control of energy transfer by engineering material disorder is appealing from a technological point of view, \emph{e.g.}, it has been shown that the acoustic turbidity of a material enables the control of its thermal conduction without altering its macroscopic properties \cite{Kim2006prl,poudel2008high}. 
Our experiments demonstrate the same concept for radiative energy transfer and could find interesting applications in nano-thermal management.

\newpage


\section*{Methods}

\noindent \textbf{Sample fabrication and electrical characterisation} 

\vspace{0.25cm}
\noindent
The hexagonal-boron-nitride (hBN)-encapsulated graphene heterostructures are fabricated with the standard pick-up and stamping technique, using a polydimethylsiloxane (PDMS) / polypropylene carbonate (PPC) stamp.
When needed, the gate electrode is first fabricated on a high-resistivity Si substrate covered by a 285~nm-thick $\text{SiO}_{\text{2}}$ layer. It consists of a pre-patterned gold pad (80~nm thick) designed by laser lithography and Cr/Au metalization. The hBN/Graphene/hBN heterostructure is then deposited on top of the back-gate.
This is followed by acetone cleaning of the stamp residues, Raman spatial mapping, and an atomic-force microscope characterisation of the stack.
Graphene edge contacts are then created by means of laser lithography and reactive ion etching, securing low contact resistance $\lesssim\, 1 \, \mathrm{k\Omega}.\mu \text{m}$.
Finally, metallic contacts to the graphene channel are designed via a Cr/Au Joule evaporation.

The transistors' dimensions were optimized to obtain the highest optical signal, while their high mobility at room temperature ensures a moderate channel electric field $E= V/L \gtrsim 10^{5}\ \rm{V.m^{-1}}$ for the threshold of electroluminescent near-field radiative energy transfer. The DC electrical measurements were performed using a Keithley 2612 voltage source to apply gate and bias voltages and measure the bias current. 
Gate polarization $V_g$ controls the electron charge density via $n=\epsilon_0 \left[ (\varepsilon_{\rm{hBN}}/t_{\rm{hBN}})^{-1}+(\varepsilon_{\rm{SiO_2}}/t_{\rm{SiO_2}})^{-1}\right]^{-1} (V_g - V_{\rm{CNP}})$, where $V_{\rm{CNP}}$ is the charge-neutral point of the channel, $t_{\rm{hBN}}$ and $t_{\rm{SiO_2}}$ are the thicknesses of the bottom hBN layer and the $\rm{SiO_2}$ layer, and $\varepsilon_{\rm{hBN}}=3.4$ and $\varepsilon_{\rm{SiO_2}}= 3.7$ are the static out-of-plane dielectric permittivities of hBN \cite{pierret2022dielectric} and $\rm{SiO_2}$, respectively. The properties of the various devices utilised in this study are summarized in Extended Data Table~\ref{liste_devices}. A comparison of the mid-IR spectral response of the various hBN crystals used here is shown in Extended Data Fig.~\ref{ED_fig10}.

\vspace{0.5cm}

\noindent \textbf{Pyrometric and spectroscopic signatures of electroluminescent energy transfer: mid-infrared experimental techniques} 

\vspace{0.25cm}
\noindent
The far-field mid-IR emission of the devices investigated here is collected via an infrared microscope fitted with a Cassegrain objective ($\text{NA} = 0.78$, angular collection interval $\in [33^{\circ}, 52^{\circ}]$). The collected signal is modulated before reaching a liquid nitrogen-cooled HgCdTe detector and the detected signal is demodulated by a lock-in amplifier at the frequency of modulation.
Two signal modulation techniques are used. The first technique is an infrared spatial modulation technique, in which the sample is laterally modulated by a piezoelectric translation stage. The second technique employs an optical chopper as sketched in Extended Data Fig.~\ref{ED_fig2}. Spectroscopy or integrated signal mapping is then performed on the detected signal, as outlined in Ref.~\cite{schmitt2023electroluminescence}.

The spectral response function of the setup is obtained by measuring the blackbody thermal emission of a reference gold substrate patterned with emissive carbon black ink squares of known emissivity ($\epsilon_{\mathrm{CB}}$) at various sample temperatures. The response function is then given by $ r (\nu) = \frac{S (\nu, T_{2}) - S (\nu, T_{1})}{  \epsilon_\text{CB} [ M (\nu, T_{2}) - M (\nu, T_{1}) ]}$, where $\nu =1/\lambda$ is the wavenumber, $S (\nu, T_{1})$ and $S (\nu, T_{2})$ are the measured spectral fluxes at the sample temperatures  $T_{1}$ and $T_{2}$ ($T_{2}> T_{1}$), and $M (\nu, T_{1})$ and  $M (\nu, T_{2})$ are the corresponding Planck blackbody spectral radiance integrated over the field of view of the microscope and the collection angles of the Cassegrain objective (see supplementary information of Ref.~\cite{schmitt2023electroluminescence}).
The spectral response functions of the two detection techniques used here are shown in Extended Data Fig.~\ref{ED_fig3}.

The mid-IR spectrum of the TR1 device was measured via infrared spatial modulation spectroscopy (IR-SMS) \cite{Li2018prl,Abou2021ol,Abou2022acs}, as described in Ref.~\cite{schmitt2023electroluminescence}, while that of TR2 was measured with the optical chopping detection technique. The latter was subtracted by a baseline spectrum to remove the contribution of the optical chopper to the detected signal.
The imaging resolution of the spectrally integrated signal spatial map (Fig.~\ref{fig3}b) measured via the setup with the optical chopping scheme is estimated to be $\simeq 15.7$~µm (see supplementary information, section~\ref{opt res sec}). 

The out-of-plane radiative power of the device $P_{\mathrm{out\mbox{-}of\mbox{-}plane}}$ is determined from the device's detected mid-IR signal as a function of electrical bias (Extended Data Fig.~\ref{ED_fig4}c).
This is done by calibrating the amplitude of the detected mid-IR signal with respect to the sample temperature by placing the sample on a hot sample holder and measuring its thermal emission at various temperatures (see Extended Data Fig.~\ref{ED_fig4}a and b).
As a result of this calibration procedure, we find that the temperature increase of the SiO$_{2}$ substrate as a function of bias in HPHT-hBN devices ranges between $\Delta T_{\mathrm{SiO}_{2}} \simeq 20-40^{\circ}$C  beyond the Zener-Klein threshold (see Extended Data Fig.~\ref{ED_fig4}d).

\vspace{0.5cm}

\noindent \textbf{Demonstration of the electroluminescent origin of the mid-IR signal} 

\vspace{0.25cm}
\noindent
We demonstrate the electroluminescent nature of the mid-IR signal emitted by the devices, following the approach adopted in Ref.~\cite{schmitt2023electroluminescence}. Namely, we experimentally rule out the possibility that the detected signal originates from the incandescence of hBN's optical phonons or graphene's hot electrons. To do so, we measure the temperature of the two aforementioned quasi-particles, which is then used to estimate their blackbody emission. The temperature of hBN's optical phonons is determined via Raman Stokes-anti-Stokes thermometry \cite{schmitt2023electroluminescence} (see Extended Data Fig.~\ref{ED_fig5}). We then compare the mid-IR spectrum of the device under electrical bias with that of its mid-IR thermal emission when the whole device is heated via a hot plate to the temperature of hBN's optical phonons (see Extended Data Fig.~\ref{ED_fig6}). As mentioned in the main text, the polariton peak around 190~meV is absent from the measured thermal radiation spectrum (Extended Data Fig.~\ref{ED_fig6}, green curve), effectively ruling out hBN's incandescence as the source of the measured emission. As such, the measured signal originates from the graphene layer. 

Having identified that emission results from graphene, we now use the theoretical expression for electroluminescent emission as given by the generalized Kirchhoff law (outlined in Ref.~\cite{Greffet2018prx}): $\frac{dP_{e}}{d\omega} = \sigma_{\text{abs}}(-u_{r},\omega)\frac{I_{b}(T,\omega)}{2}d\Omega$ where
$\frac{dP_{e}}{d\omega}$ is the spectral radiance per unit volume, $\sigma_{\text{abs}}(-u_{r},\omega)$ is the absorption cross-section of the graphene layer and $I_{b}(T,\omega)=\frac{\omega^{2}}{4\pi^{3}c^{2}}\frac{\hbar \omega}{\exp\big(\frac{\hbar \omega - \mu_{c}}{k_{B}T}\big)-1}$ is a modified Bose-Einstein distribution that accounts for the electrochemical potential imbalance between the conduction and the valence band $\mu_{c}$, which represents the out-of-equilibrium character of the electronic distribution with respect to the thermal state. To demonstrate electroluminescence at 192 meV, we can simply compare the experimental and thermal emission (obtained by setting $\mu_{c} = 0$ meV and $\hbar \omega = 192$ meV in the above formula): if an excess emission is detected, it proves the role of out-of-equilibrium electrons in the emission, and therefore of the electroluminescent nature of the detected signal. 

The spectrum of the excess mid-IR emission with respect to the null bias condition is measured via IR-SMS as a function of increasing bias and the resulting spectral flux at the polariton peak, marked by an asterisk in Extended Data Fig.~\ref{ED_fig7}a, is compared with a Planck blackbody distribution (see Extended Data Fig.~\ref{ED_fig7}b) using the temperature of graphene's hot electrons, which was measured by characterising the device's near-infrared incandescence as outlined in Ref.~\cite{schmitt2023electroluminescence}. This comparison shows that the measured spectral flux with increasing bias is about two times larger than that expected from incandescence in the bias range $6-8$~V (see Extended Data Fig.~\ref{ED_fig7}b, inset), thus precluding graphene's incandescence as a possible source of the measured signal. We emphasize here the fact that this fitting procedure represents a worst-case scenario, in which it is assumed that the measured mid-IR flux at a low value of bias (4.3~V) is equal to that of the incandescence flux and the evolution of the two fluxes is compared as a function of increasing bias.

We conclude from the analysis above that the mid-IR signal emitted by the device under electrical bias is electroluminescent in nature.

In addition to the spectra of TR1 and TR2 shown in the main text, the mid-IR spectra of two additional electroluminescent devices made with HPHT-hBN are presented in Extended Data Fig.~\ref{ED_fig8} and \ref{ED_fig9}. In these devices, the scattering of HPhPs to the far-field occurs due to gold discontinuities.

\vspace{0.5cm}

\noindent \textbf{Description of TR1 and TR2}

\vspace{0.25cm}
\noindent
The TR1 and TR2 devices serve two different purposes. TR1 is smaller with contacts proximal to the channel, and a fold in the top hBN that can be observed in Extended Data Fig.~\ref{ED_fig1}, such features contribute to a more pronounced outward coupling of HPhPs compared to TR2.  
The design of TR2, with its channel size matching the field of view of the infrared microscope, was specifically chosen to simplify the analysis of out-of-plane energy transfer, minimizing scattering and thus the electroluminescence signal from HPhPs.
Additionally, TR2 contains bubbles, which appear as green spots on the micrograph. These bubbles are common in nanofabrication, they can be due to hydrocarbons or water and do not perturb electronic transport at room temperature. However, these bubbles can be responsible for the emission peak due to the scattering of the $\mathrm{SiO_2}$ SPhPs that occur at the bubble-$\mathrm{SiO_2}$ interface.
Finally, the two devices have different hBN layer thicknesses which are summarized in Extended Data Table~\ref{liste_devices}.

\vspace{0.5cm}

\noindent \textbf{Noise temperature measurements} 

\vspace{0.25cm}
\noindent
Following the approach described in Refs.~\cite{Wei2018Nnano,Schmitt2023prb} we performed a noise thermometry characterisation of hBN-encapsulated graphene transistors under large bias to quantify the efficiency of near-field radiative energy transfer. Noise temperature measurements were performed at room temperature in a cryogenic probe station adapted to radio-frequency measurements up to 67~GHz, using a Tektronix DPO71604C ultra-fast oscilloscope for measurements up to 16~GHz. The high-frequency signal coming from the device is amplified by a CITCRYO1-12D Caltech low noise amplifier in the $1-10$~GHz band, whose noise and gain have been calibrated against the thermal noise of a 50~$\Omega$ calibration resistance measured at various temperatures between 10~K and 300~K in the probe station. Electronic noise thermometry measurements of the investigated devices are presented in Extended Data Fig.~\ref{ED_fig11}.

\vspace{0.5cm}

\noindent \textbf{Raman Stokes-anti-Stokes thermometry} 

\vspace{0.25cm}
\noindent
In order to precisely and unequivocally determine the temperature of hBN's optical phonons, we performed Stokes-anti-Stokes thermometry. We used a \emph{Renishaw Raman} spectrometer with a notch filter that allows simultaneous measurement of Stokes~(S) and anti-Stokes~(AS) peaks. The graphene device is excited with a $\lambda =$~532~nm laser ($P=$~5~mW). 
We relied on the intensity ratio between the S and AS peaks at a temperature $T$ given by $\frac{I_{AS}}{I_{S}}=e^{-\frac{\hbar \Omega}{k_{B}T}}$, where $\hbar \Omega$ is the energy of the optical phonon mode (for the hBN optical phonon mode $E_{2g}$, the energy is 1,365~cm$^{-1} =$~169~meV). The measurement at zero bias voltage allows for an absolute calibration. By comparing the AS/S ratios at zero bias and non-zero bias, we deduced the temperature of the optical phonons of hBN as a function of doping and bias voltage (see Extended Data Fig.~\ref{ED_fig5}). 

A Stokes-anti-Stokes characterisation of a graphene device made with PDC-hBN is presented in supplementary information, section~\ref{PDC SAS sec}, in which hBN optical phonon temperature was found to be similar to that of HPHT-hBN.

\vspace{0.5cm}

\noindent \textbf{Numerical methods} 

\vspace{0.25cm}
\noindent
Figures \ref{fig2}b and d of the main text were generated using the open-source numerical electromagnetic solver Reticolo, developed by Jean-Paul Hugonin and Philippe Lalanne \cite{hugonin2021reticolo} and using the hBN and $\mathrm{SiO_2}$ dielectric permittivity models described in the supplementary information (section~\ref{polariton dispersion}). In Fig.~\ref{fig2}d, we presented the layer-by-layer emissivity of the TR1 layer geometry, integrated over the mid-IR microscope objective aperture. The red curve gives the emissivity of the graphene layer in the heterostructure with a fold in the top hBN layer. Figure~\ref{fig2}b shows the normalized Purcell factor computed for a horizontal dipole source situated 3.5~\AA~above a 30 nm thick hBN slab and 3.5~\AA~below a 52 nm thick hBN slab, chosen to mimic the cutoff wavevector emission of the interband electronic transitions involved in type II HPhP electroluminescence.

The reflection coefficient of the TR1 heterostructure, whose imaginary part is depicted in Fig.~\ref{fig2}a, was analytically derived in the supplementary information (section~\ref{polariton dispersion}) and numerically evaluated using MATLAB. All corresponding codes are available in the Zenodo repository linked below.

\section*{Data availability}

\noindent Data are publicly available on Zenodo at the doi 10.5281/zenodo.10625437‌

\section*{Acknowledgments}

\noindent The research leading to these results has received partial funding from the European Union Horizon 2020 research and innovation program under grant agreement No.881603 "Graphene Core 3", and from the French  ANR-21-CE24-0025-01 "ELuSeM". This work has received support under the program “Investissements d’Avenir” launched by the French Government. JHE and EJ appreciate support for monoisotopic hBN crystal growth from the U. S. Office of Naval Research under award number N00014-22-1-2582. We thank Romain Pierrat and Rémi Carminati for their scientific insights, as well as Laure Tailpied and Baptiste Fix for supplemental scattering-scanning near-field optical microscopy characterisations.

\section*{Authors contributions}

LAH, AS, BP, YDW, and EmB conceived the experiments. AS conducted device fabrication and electrical measurements under the guidance of DM, AP, and MR in the early developments. TT, KW, CM, CJ, BT, VG, PS, JHE, and EJ provided the hBN crystals. 
LAH and SR performed the mid-infrared emission measurements with the help of AS and RB.
AS performed the Raman thermometry measurements. 
MT performed the near-infrared measurements. JPH, JJG, ElB, BV, LAH, and EmB performed the numerical simulations.
LAH, AS, RB, SR, JJG, ElB, BV, YDW, CV, BP, JMB, GF, PB, GM, and EmB analysed the data and developed the theoretical interpretation. 
JPH wrote the numerical codes. 
LAH, AS, RB, YDW, and EmB wrote the manuscript with contributions from the co-authors.

\section*{Competing interests}

\noindent The authors declare no competing interests.

\newpage

\renewcommand{\figurename}{\textbf{Extended Data Fig.}}
\renewcommand{\tablename}{\textbf{Extended Data Table}}
\setcounter{figure}{0}

\begin{figure*}[!htb]
\centering
\includegraphics[width=0.8\linewidth]{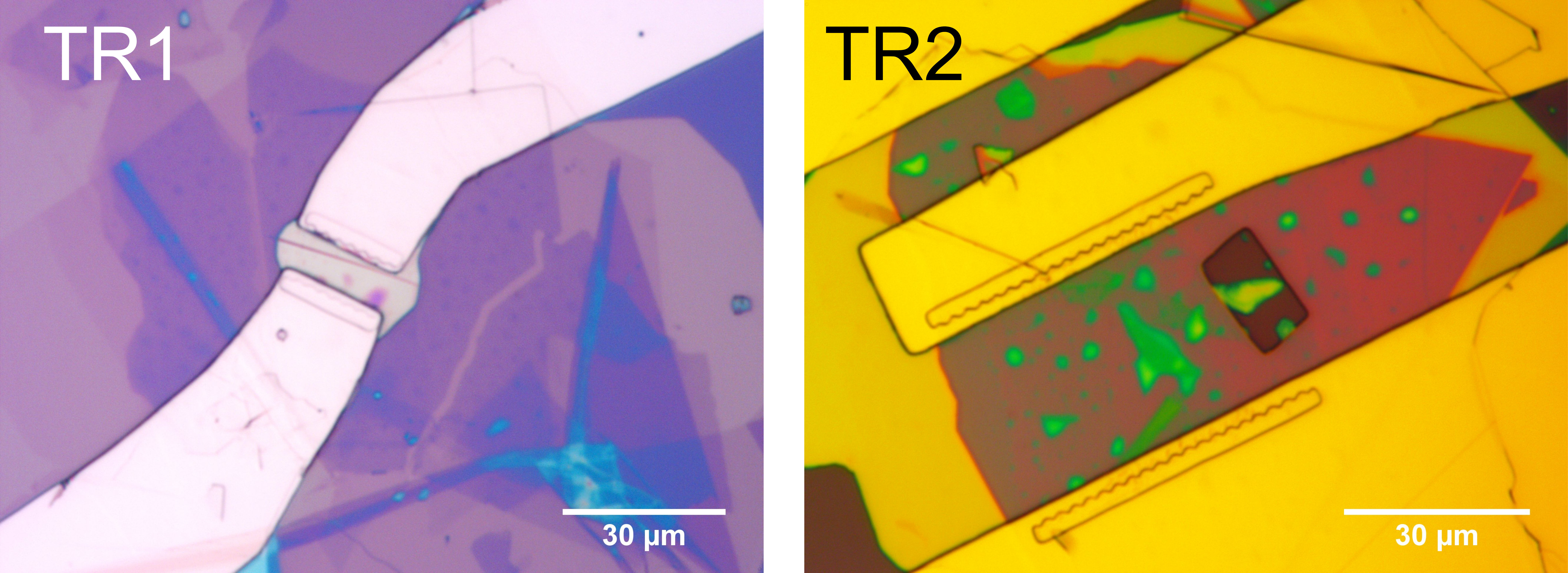}
\caption{\textbf{Optical microscope images of transistors TR1 (left) and TR2 (right)} resolved with a $\times 50$ objective.}
\label{ED_fig1}
\end{figure*}

\begin{figure*}[!htb]
\centering
\includegraphics[width=0.8\linewidth]{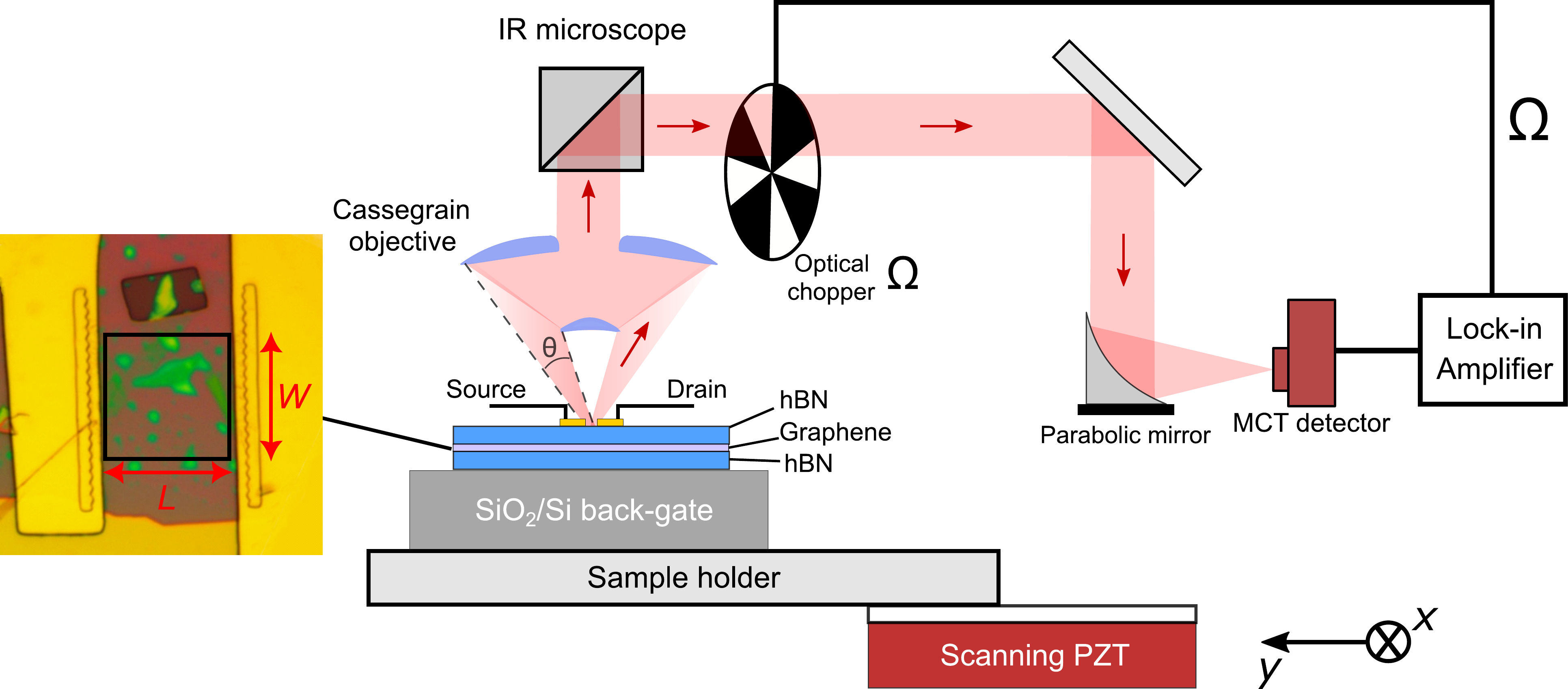}
\caption{\textbf{Schematic illustration of the detection technique used for characterising the mid-IR signal} ($P_{\mathrm{MIR}}$, integrated over the full spectral range of the detector) of an hBN-encapsulated graphene device (left, TR2 device with $L\times W = 35\times 35\, \mu \text{m}^{2}$) due to the radiative emission of the SiO$_{2}$ back-gate. The setup is utilised to perform the pyrometric characterisation of Fig.~\ref{fig3} of the main text.}
\label{ED_fig2}
\end{figure*}

\begin{table*}[!htb]
\centering
\hspace*{-0.7cm}
    \begin{tabular}{| p{4.0cm} |  c | c| c|c| c| c|}
    \hline
    \textbf{Device} & \textbf{L}  & \textbf{W} & \textbf{$t_{\text{hBN,top}}$}& \textbf{$t_{\text{hBN,bottom}}$}& \textbf{$\mu$} & \textbf{$R_c$}  \\
       & ($\mu$m) & ($\mu$m) & (nm) & (nm) & $(\mathrm{m^2.V^{-1}.s^{-1}})$ & $(\Omega)$\\ \hline\hline
    \textbf{TR1 (HPHT-hBN)}& 9.2 & 15.5 & 52 & 30 & 6 & 250 \\ \hline       
    \textbf{TR2 (HPHT-hBN)}& 35 & 35 & 82 & 65 & 10 & 150 \\ \hline         
    \textbf{TR3 (HPHT-hBN)}& 9 & 8 & 54 & 30 & 9 & 190 \\ \hline         
    \textbf{TR4 (HPHT-hBN)}& 8.1 & 12.6 & 91 & 41 & 5.6 & 220 \\ \hline         
    \textbf{F161 (HPHT-hBN)}& 6 & 10 & 54 & 152 & 16 & 45 \\ \hline
    \textbf{Lyon1 (PDC-hBN)}& 9 & 9 & 49 & 142 & 4 & 230 \\ \hline
    \textbf{KSU-hBN}& 10 & 16 & 191 & 151 & 1.4 & 800    \\ \hline  
    \end{tabular}
\caption{\textbf{Geometrical and electronic properties of the various graphene transistors investigated in this study.} $L$ and $W$ are the length and width of the graphene channel, $t_{\text{hBN,top}}$ and $t_{\text{hBN,bottom}}$ are the thicknesses of hBN encapsulates, $\mu$ is the zero-bias electronic mobility and $R_c$ is the contact resistance of the sample.}
\label{liste_devices}
\end{table*}

\begin{figure*}[!htb]
\centering
\includegraphics[width=0.8\linewidth]{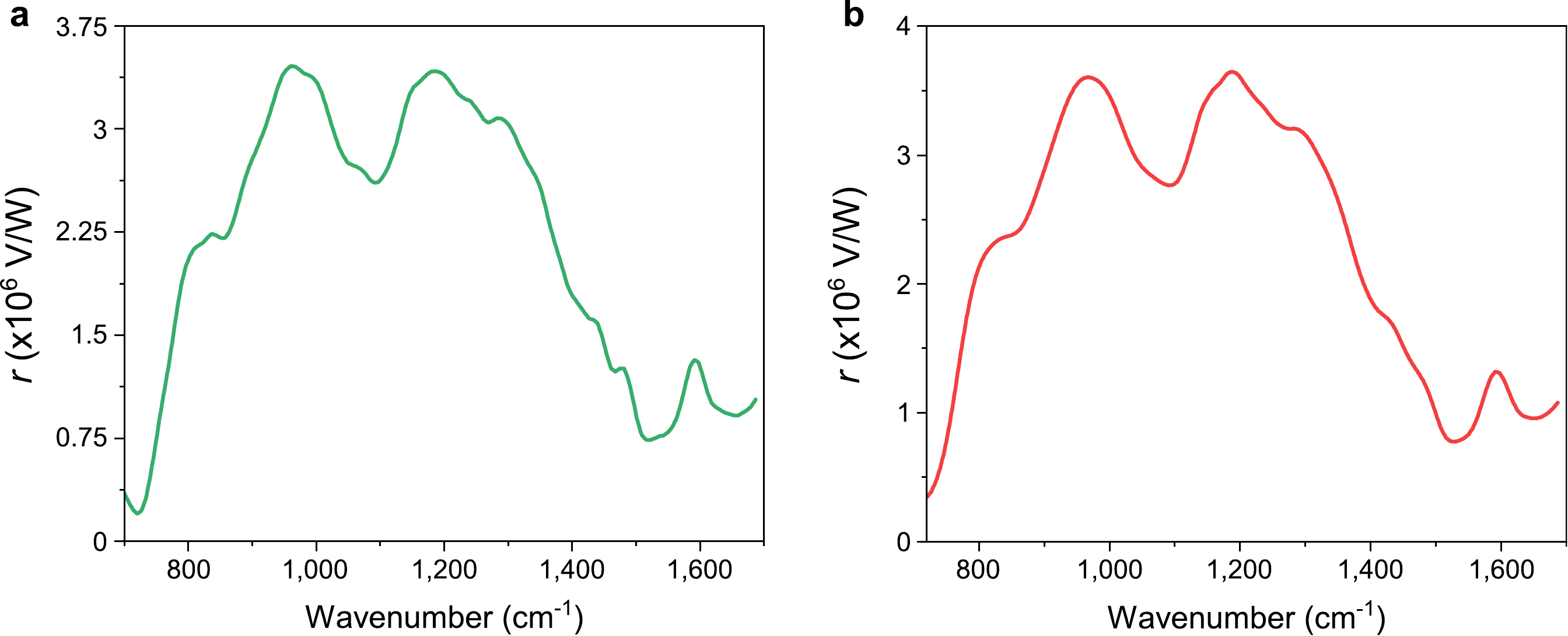}
\caption{\textbf{Spectral response function ($r$) of the mid-IR setup for the two detection techniques utilised in this study.}
\textbf{a}, Spectral response function for modulation with an optical chopper.
\textbf{b}, Spectral response function of the infrared spatial modulation spectroscopy (IR-SMS) technique. }
\label{ED_fig3}
\end{figure*}

\begin{table*}[!htb]
    \centering
    \begin{tabular}{cc}
    \hline
    \hline
       $V_{ds} (V)$ & $T_{e}$ (K) \\
        \hline
      4.3 & 480  \\
      5.2 & 510 \\
      6.2 & 550 \\
      7.5 & 600 \\
      8.2 & 625 \\
       \hline
       \hline
       \end{tabular}
    \caption{\textbf{Measured electron temperature ($T_{e}$) as a function of drain-source voltage ($V_{ds}$) in the TR1 device.}}
    \label{table:Te}
\end{table*}

\begin{figure*}[!htb]
\centering
\includegraphics[width=\linewidth]{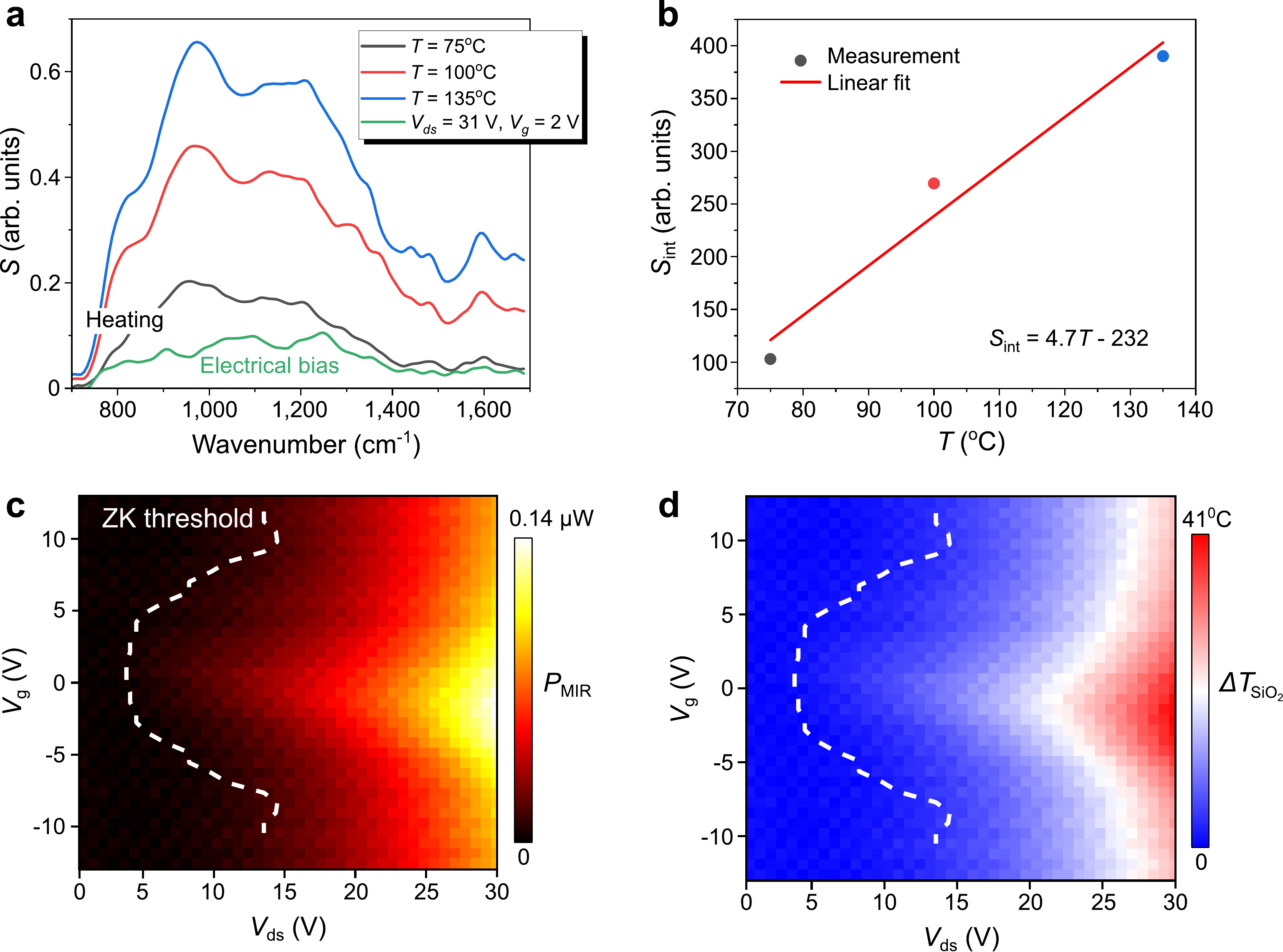}
\caption{\textbf{Calibration of the pyrometric technique.}
\textbf{a},~Comparison between the mid-IR spectra of the hBN-encapsulated device TR2 with an SiO$_{2}$ back-gate under electrical bias and by heating the device via a hot plate at various temperatures $T$. The mid-IR signal is measured using an optical chopper (see Extended Data Fig.~\ref{ED_fig2}). The plotted spectra are corrected by a reference measurement, which removes the contribution of the chopper to the detected signal, and then normalized by an instrumental response function (see supplementary information of Ref.~\cite{schmitt2023electroluminescence}).
\textbf{b},~Linear fit of the spectrally-integrated signal $S_{\mathrm{int}}$ obtained with the hot plate as a function of temperature. $S_{\mathrm{int}}$ is found by integrating the spectra $S$ from panel~\textbf{a} over the detector's spectral bandwidth. From the comparison of panel \textbf{a} we find that the temperature of the SiO$_{2}$ back-gate varies between $\Delta T_{\mathrm{SiO}_{2}} \simeq 0-40$~$^{\circ}$C when the device is electrically biased.
\textbf{c},~Scan of the mid-IR spectrally-integrated power $P_{\mathrm{MIR}}$ emitted by the TR2 device as a function of bias.
\textbf{d},~Increase in the temperature of the device's SiO$_{2}$ back-gate ($\Delta T_{\mathrm{SiO}_{2}}$) as a function of electrical bias, obtained from the mid-IR signal to sample temperature calibration.}
\label{ED_fig4}
\end{figure*}

\begin{figure*}[!htb]
\centering
\includegraphics[width=0.5\linewidth]{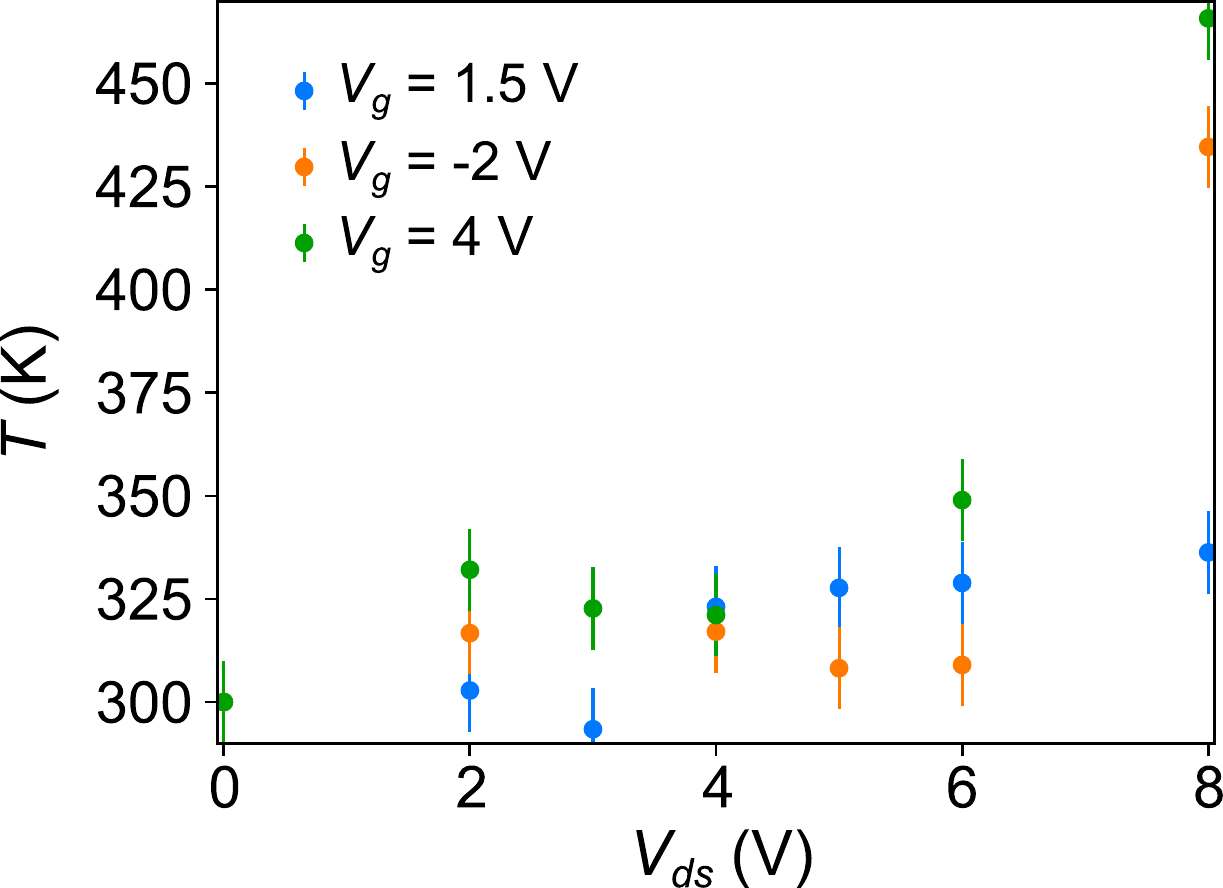}
\caption{\textbf{Raman Stokes-anti-Stokes (SAS) thermometry of the TR1 device.} The temperature of hBN optical phonons measured by SAS thermometry (see Methods) is plotted as a function of bias for three different dopings ($V_g=1.5~\mathrm{V}$ corresponding to charge neutrality). At charge neutrality, the temperature of hBN's optical phonons is found to be $\lesssim 75 ^\circ \mathrm{C}$.}
\label{ED_fig5}
\end{figure*}

\begin{figure*}[!htb]
\centering
\includegraphics[width=0.6\linewidth]{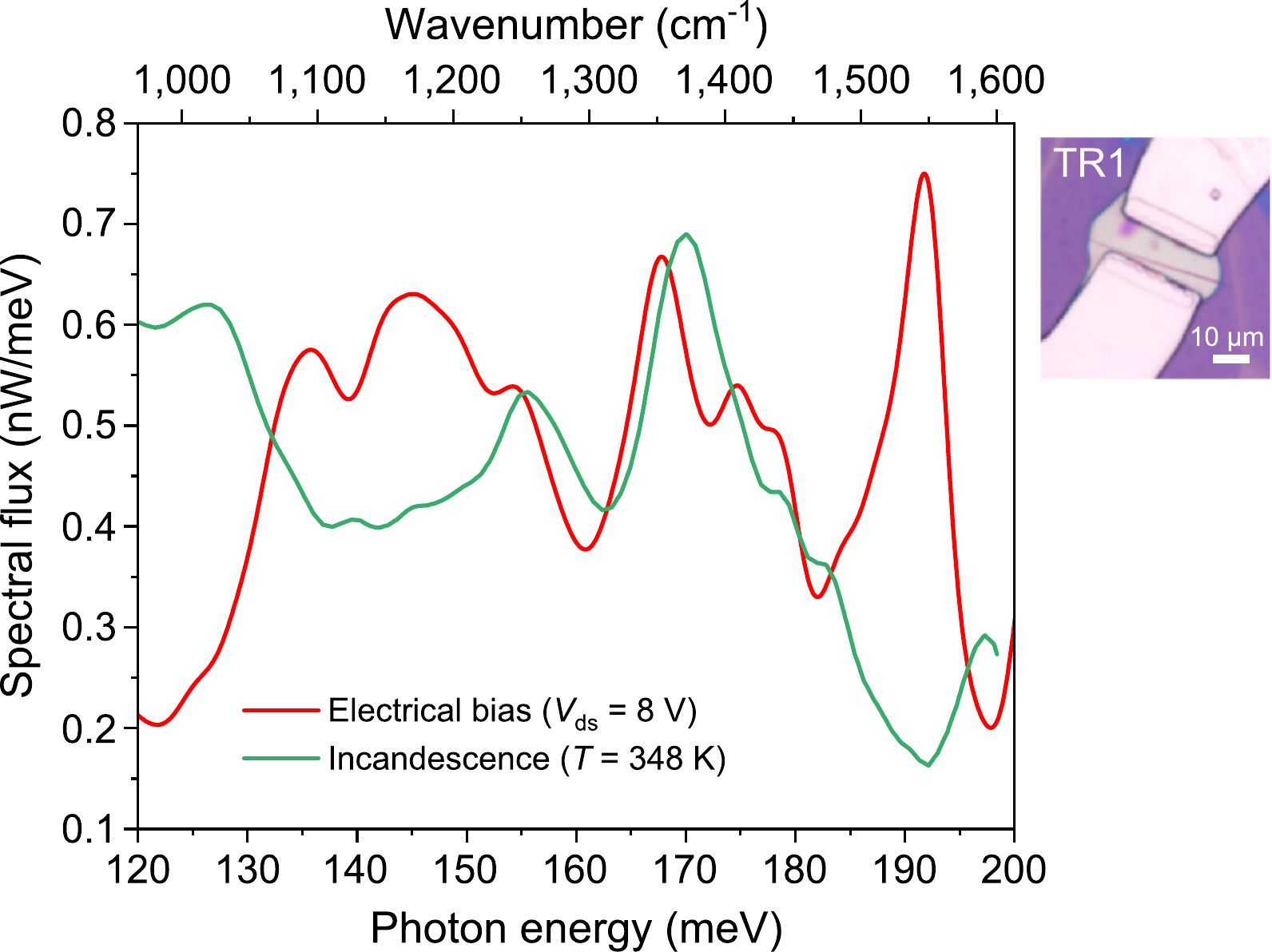}
\caption{\textbf{Electroluminescence vs hBN incandescence.} Comparison between the IR-SMS spectral flux of TR1 under electrical bias (red curve) and its incandescence when heated via a hot plate to the temperature of hBN's optical phonons ($\sim 75^{\circ}$C, green curve). This comparison reveals that the main polariton peak at $\sim 190$~meV, which is involved in the electroluminescence process, is notably absent from the thermal emission spectrum (green curve). Thus, incandescence from hBN's optical phonons can be dismissed as the origin of the mid-IR signal.}
\label{ED_fig6}
\end{figure*}

\begin{figure*}[!htb]
\centering
\includegraphics[width=\linewidth]{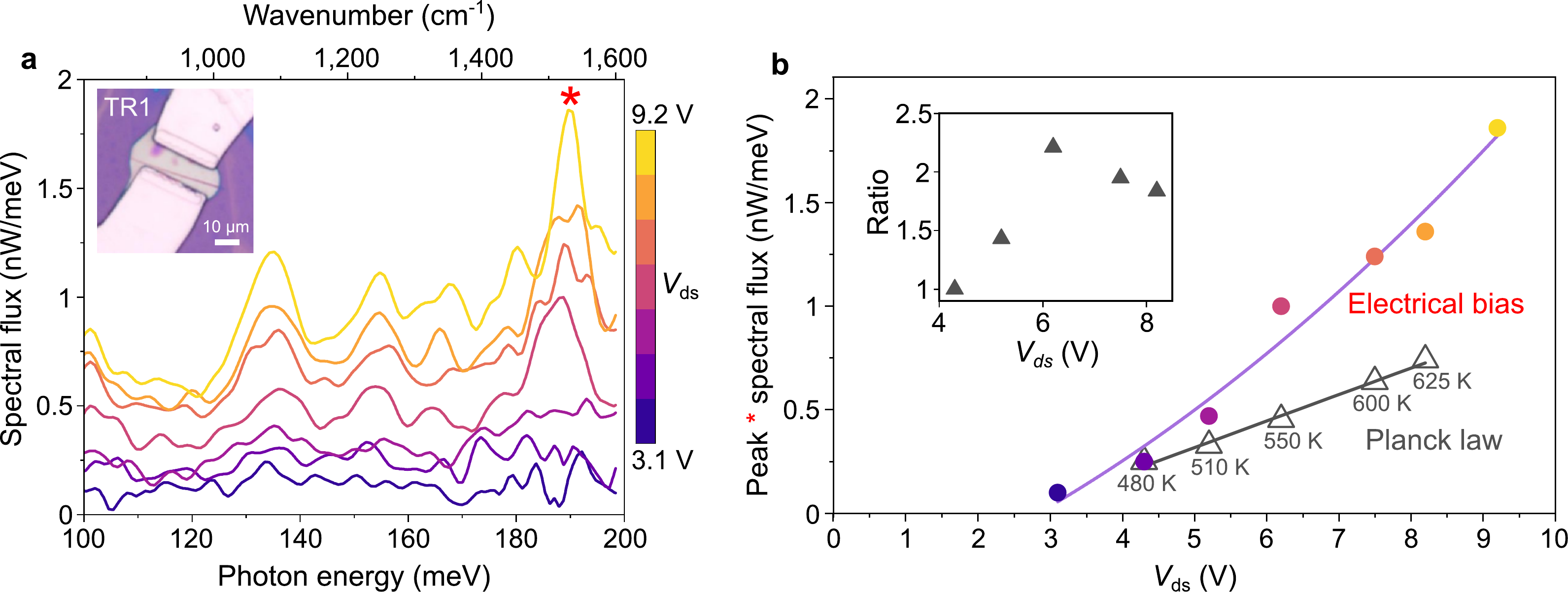}
\caption{\textbf{Electroluminescence vs graphene's hot electron incandescence.}
\textbf{a},~Sweep of the mid-infrared spectral emission of the TR1 device (inset) under electrical bias, measured via IR-SMS. 
\textbf{b},~Spectral flux at the HPhP resonance, marked by an asterisk in panel \textbf{a}, as a function of applied bias. \eb{}{Both experiments were performed at $V_g=0~\mathrm{V}.$}
The hollow grey triangles are the Planck law fit given by $A E^{3} / [ \exp{ (E/k_{B} T_{e}})  -1  ]$, in which $A$ is a fitting constant, $E = 192$~meV is the polariton energy, and $T_{e}$ is the electron temperature given in Extended Data Table~\ref{table:Te}.
$A$ is determined by equating the spectral flux at $V_{ds} = 4.3$~V to the value of the measured spectral flux at the same point. The magenta and grey curves are a second-order polynomial fit and a linear fit, respectively, corresponding to the data points. The inset shows the ratio of the measured spectral flux to that obtained from the Planck law fit.
The measured spectral flux with increasing bias is about two times larger than that expected from incandescence in the bias range $6-8$~V, thus precluding graphene's incandescence as a possible source of the measured signal.
Note that this fitting procedure represents a worst-case scenario, in which it is assumed that the measured mid-IR flux at a low value of bias (4.3~V) is equal to that of the incandescence flux and the evolution of the two fluxes is compared as a function of increasing bias.}
\label{ED_fig7}
\end{figure*}

\begin{figure*}[!htb]
\centering
\includegraphics[width=\linewidth]{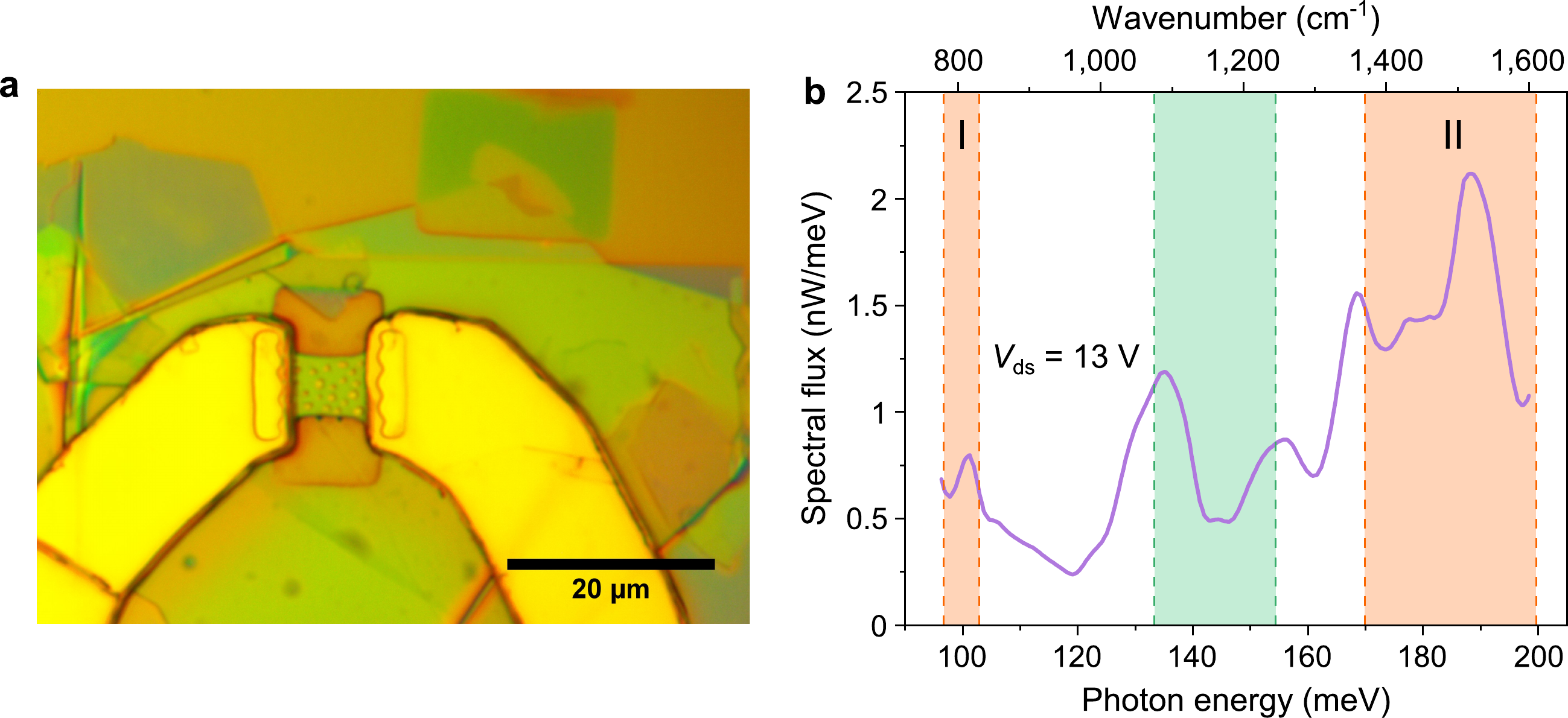}
\caption{\textbf{Electroluminescence of device TR3 with scatterers exhibiting a peak around $800\  \mathrm{cm^{-1}}$ in hBN's first \emph{Reststrahlen} band.}
\textbf{a}, Optical microscope image of the electroluminescent HPHT-hBN device TR3 resolved with a $\times 100$ objective. The device includes scatterers in the form of subwavelength gold disks and etches in the hBN layer that are randomly distributed across the device's channel. \textbf{b}, Mid-IR spectrum of TR3 under bias ($V_{ds} = 13$~V, $V_{g} = 0$~V) measured via IR-SMS.
}
\label{ED_fig8}
\end{figure*}

\begin{figure*}[!htb]
\centering
\includegraphics[width=\linewidth]{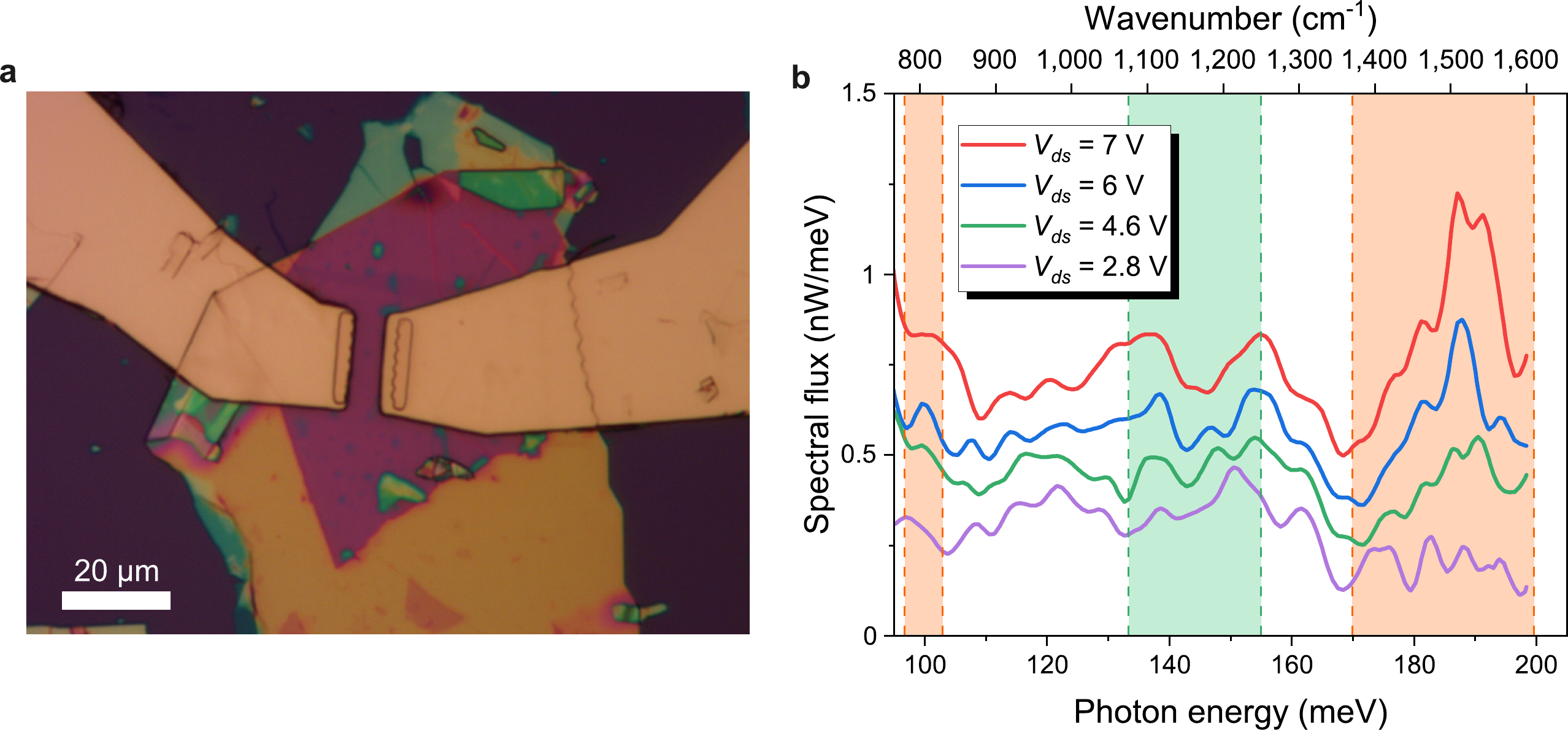}
\caption{\textbf{Electroluminescence of device TR4 without scatterers exhibiting a peak around $190\  \mathrm{meV}$ in hBN's second \emph{Reststrahlen} band.}
\textbf{a}, Optical microscope image of electroluminescent HPHT-hBN device TR4 resolved with a $\times 50$ objective. Notably, the device does not include scatterers in the form of subwavelength gold disks. Thus, only the metallic contacts can scatter HPhPs and the short channel length further enhances this effect. \textbf{b}, Mid-IR spectrum of TR4 at zero gate voltage and variable drain-souce bias ($V_{ds}$) measured via IR-SMS.
}
\label{ED_fig9}
\end{figure*}

\begin{figure*}[!htb]
\centering
\includegraphics[width=\linewidth]{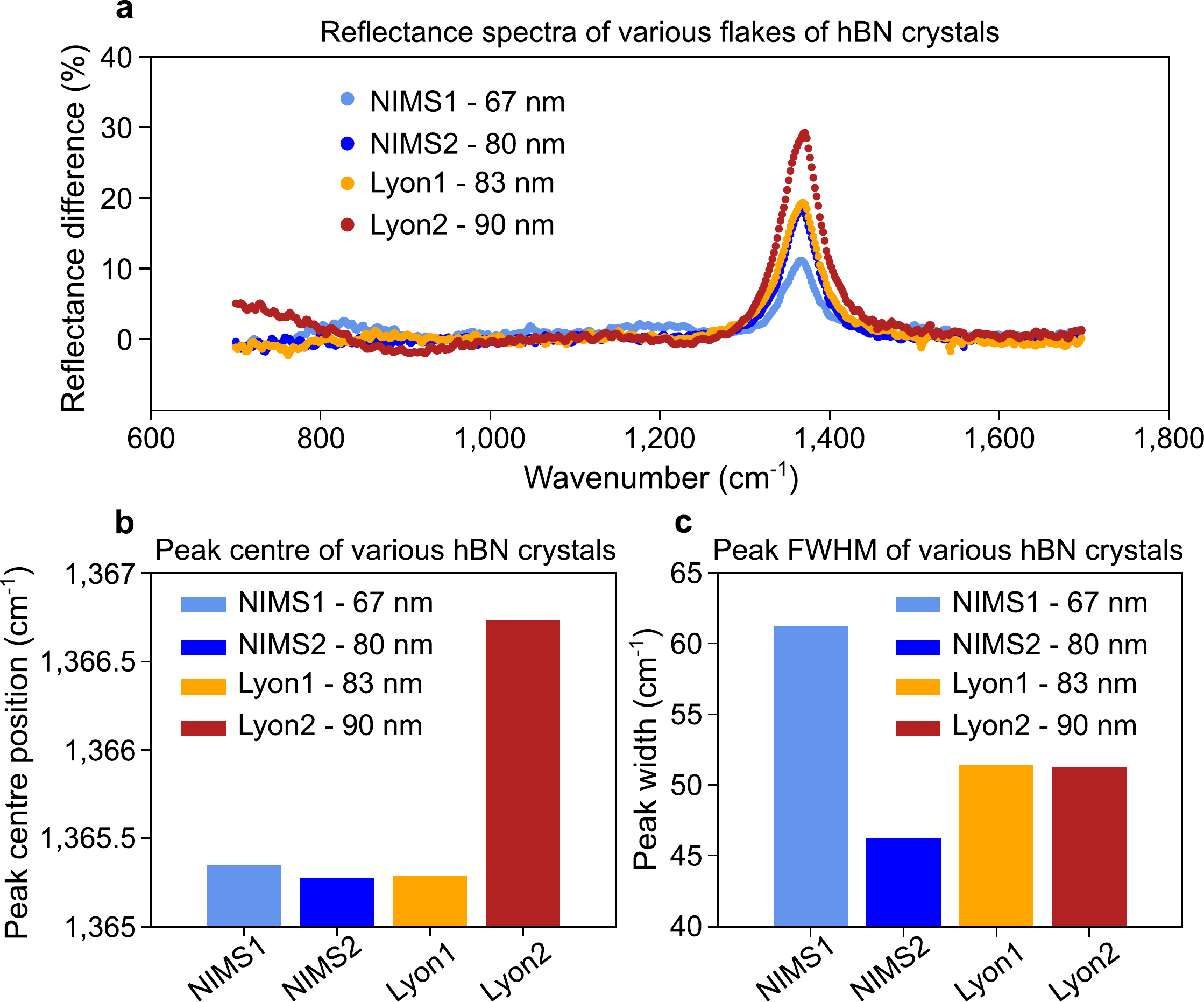}
\caption{\textbf{Comparison of the differential reflectance peaks of hBN flakes of various crystalline quality.} \textbf{a}, Reflectance difference in the mid-IR between the reflectance spectra of hBN flakes on a diamond substrate and the reflectance spectra of a diamond reference for two different HPHT-hBN flakes labelled NIMS and two different PDC-hBN flakes labeled Lyon. The signal is proportional to the absorptance with an amplitude related to the flake thickness. \textbf{b and c}, Bar plots comparing the absorption peak centre energy (panel \textbf{b}) and linewidth (panel \textbf{c}), fitted by a Lorentzian function from the differential reflectance of the 4 samples studied here. Measurements were performed with a Cassegrain objective with magnification $\times 15$, a numerical aperture of 0.4, and a \emph{Brüker Hyperion} spectrometer. The spectral resolution is 2~$\mathrm{cm}^{-1}$.
}
\label{ED_fig10}
\end{figure*}

\begin{figure*}[!htb]
\centering
\includegraphics[width=\linewidth]{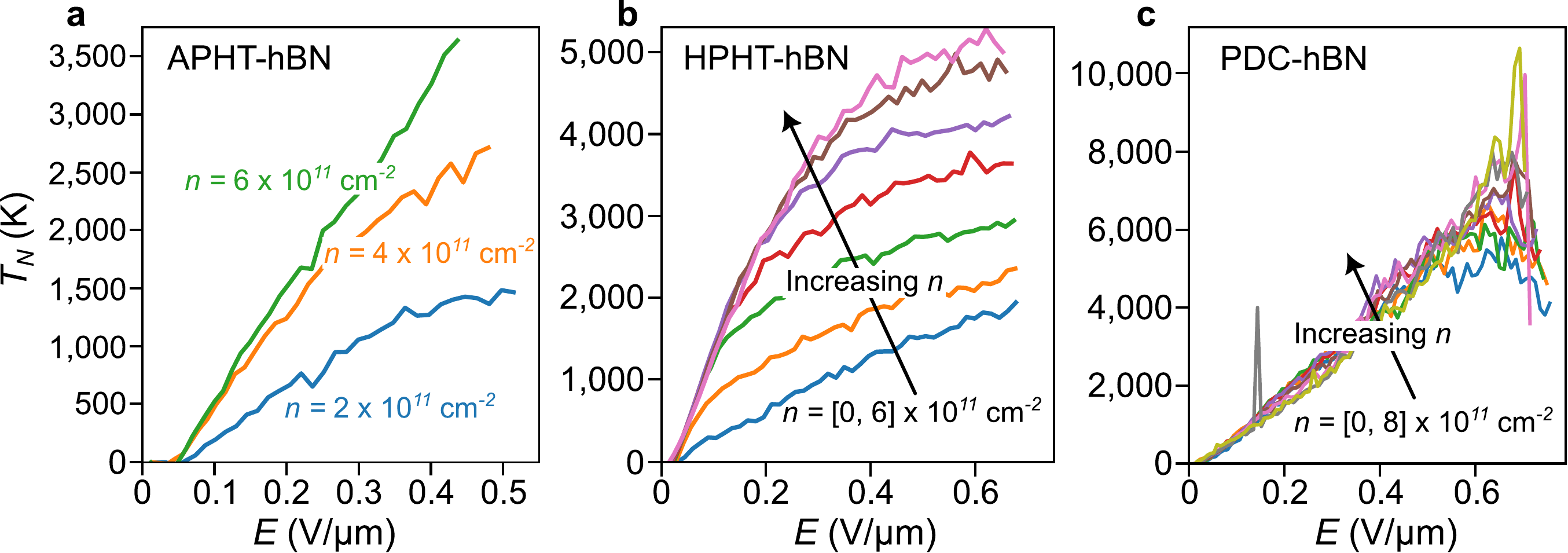}
\caption{\textbf{Electronic noise temperature ($T_{N}$) as a function of bias for 3 graphene transistors each made with a different type of hBN crystal:}  HPHT-hBN from NIMS, PDC-hBN from Lyon and APHT mono-isotopic h$^{11}$BN from KSU.
\textbf{a}, Electronic noise temperature of a device made with mono-isotopic boron nitride from KSU, USA ($>99$\% $^{11}$B, fabricated at high temperature and atmospheric pressure).
\textbf{b}, Electronic noise temperature of a device made with HPHT-hBN from NIMS, Japan.
\textbf{c}, Electronic noise temperature of a device made with PDC-hBN from Lyon, France. 
While the Zener tunneling regime leads to reduced electronic heating for both APHT- and HPHT-hBN at large bias, there is no such signature in PDC-hBN, confirming that radiative energy transfer is significantly reduced in this material.
}
\label{ED_fig11}
\end{figure*}

\renewcommand{\figurename}{Fig.~SI-}

\renewcommand{\tablename}{Table~SI-}

\newpage
\clearpage
\clearpage
\clearpage
\clearpage
\clearpage
\clearpage
\clearpage

\section*{Supplementary information}

\renewcommand{\figurename}{\textbf{Fig. SI-}}
\renewcommand{\tablename}{\textbf{Table SI-}}
\setcounter{figure}{0} 
\setcounter{table}{0}

\section{Modeling the out-of-plane energy transfer in a graphene/hBN heterostructure} \label{model_therm}

\noindent
In this section, we detail various thermal models that describe energy transfer in a graphene/hBN device with an SiO$_{2}$/Si back-gate, starting from the simplest case (in which only the thermal resistance of the substrate is considered) then making our way through a more refined approach, in which the whole structure is modeled. The latter model is utilised to obtain the results presented in the main text.  

\begin{figure}[t!]
\centering
\includegraphics[width=\linewidth]{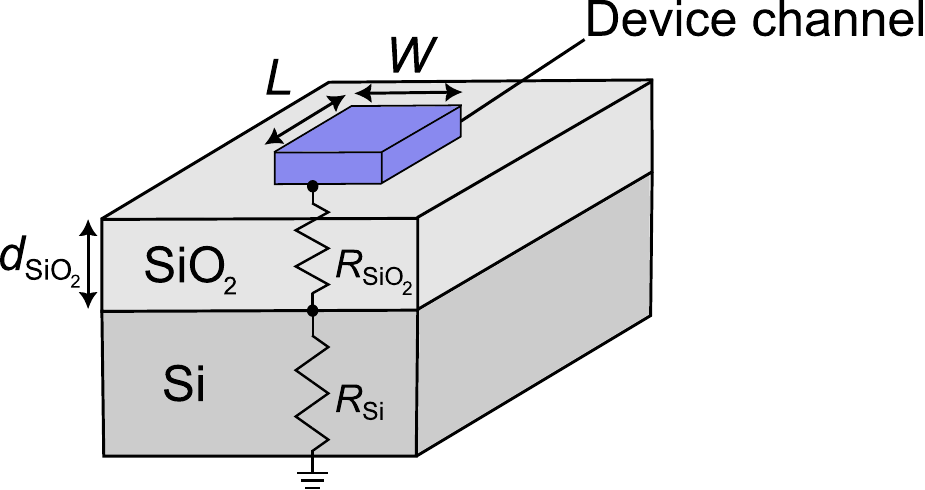}
\caption{Sketch of a simple thermal model used to compute the total thermal resistance $R_{\mathrm{th}} = R_{\mathrm{SiO}_{2}} + R_{\mathrm{Si}}$ ruling the out-of-plane energy transfer in an hBN-encapsulated graphene device with an SiO$_{2}$/Si back-gate.}
\label{therm model}
\end{figure}

The out-of-plane radiative power can be calculated via a simple thermal model (introduced in Ref. \cite{pop2010energy}) which only takes into account the thermal resistances of the SiO$_{2}$ and Si layers of the device's back-gate (see Fig.~SI-~\ref{therm model}). To do so we define the altitude $z$ in reference to the wafer substrate. The temperature profile within the SiO$_{2}$ layer of thickness $d_{\mathrm{SiO}_{2}}$ for a given out-of-plane power exchange $ P_{\mathrm{out\mbox{-}of\mbox{-}plane}}$ is obtained from the thermal ohm's law:

\begin{equation}
    T_{\mathrm{SiO}_{2}}(z)=\left[ R_{\mathrm{SiO}_{2}}(z) + R_{\mathrm{Si}} \right] P_{\mathrm{out\mbox{-}of\mbox{-}plane}},
\end{equation}
where $R_{\mathrm{SiO}_{2}}(z)=\frac{z+d_{\mathrm{SiO}_{2}}}{d_{\mathrm{SiO}_{2}}}R_{\mathrm{SiO}_{2}},$ and $R_{\mathrm{SiO}_{2}}=R_{\mathrm{SiO}_{2}}(0)$ is the total thermal resistance of the SiO$_{2}$ layer. 
For the small temperature increase considered in the pyrometry experiment conducted here, the incandescent signal is proportional to the average temperature increase of SiO$_{2}$ with respect to the substrate temperature $T_0$, which is given by $\Delta T_{\text{SiO}_{2}} = \frac{1}{d_{\text{SiO}_{2}}} \int_{-d_{\text{SiO}_{2}}}^{0} T_{\text{SiO}_{2}}(z) \text{d}z  - T_{0}$. In this framework, the out-of-plane power is given by

\begin{align}
    P_{\mathrm{out\mbox{-}of\mbox{-}plane}} &= \frac{\Delta T_{\mathrm{SiO}_{2}}}{R_{\mathrm{th}}} \nonumber \\
   & = \frac{\Delta T_{\mathrm{SiO}_{2}}}{\frac{1}{2}R_{\mathrm{SiO}_{2}} +   R_{\mathrm{Si}} },
   \label{Rtot}
\end{align}

\noindent
 where $R_{\mathrm{th}}$ is the total effective thermal resistance of the layers and $R_{\mathrm{SiO}_{2}}$ and $R_{\mathrm{Si}}$ are the effective thermal resistances of the SiO$_{2}$ and Si layer, respectively (see Fig.~SI-~\ref{therm model}), which are given by

 \begin{equation}
  R_{\mathrm{SiO}_{2}} = \frac{d_{\mathrm{SiO}_{2}}}{\kappa_{\mathrm{SiO}_{2}}  LW},
  \label{RSiO}
 \end{equation}

 \noindent
 and

 \begin{equation}
  R_{\mathrm{Si}} = \frac{1}{2 \kappa_{\mathrm{Si}}\sqrt{LW}}.
  \label{RSi}
 \end{equation}

\noindent
 In the above equations $d_{\mathrm{SiO}_{2}} = 285$~nm is the thickness of the SiO$_{2}$ layer, $\kappa_{\mathrm{SiO}_{2}}  = 1.4\ \rm{W.m^{-1}.K^{-1}}$ and $\kappa_{\mathrm{Si}}  = 150\ \rm{W.m^{-1}.K^{-1}}$ are the thermal conductivities of SiO$_{2}$ and Si, and $L = 35$~$\mu$m and $W= 35$~$\mu$m give the dimensions of the device's graphene channel (see Fig. SI-\ref{therm model}). From Eqs.~(\ref{RSiO}) and (\ref{RSi}), we obtain a total thermal resistance $R_{\mathrm{th}}\simeq 178\ \rm{K.W^{-1}}$. This results in an out-of-plane power $P_{\mathrm{out\mbox{-}of\mbox{-}plane}}$ that reaches 68~\% of the total electrical power injected into the device. Note that this simple model is the same as that reported in Ref.~\cite{pop2010energy}, however, in our case, there is an additional factor of 1/2 in the denominator of Eq.~(\ref{Rtot}) to account for the fact that the thermal radiation emitted by the device results from the average temperature of the SiO$_{2}$ layer.

\begin{figure*}[t!]
\centering
\includegraphics[width=\linewidth]{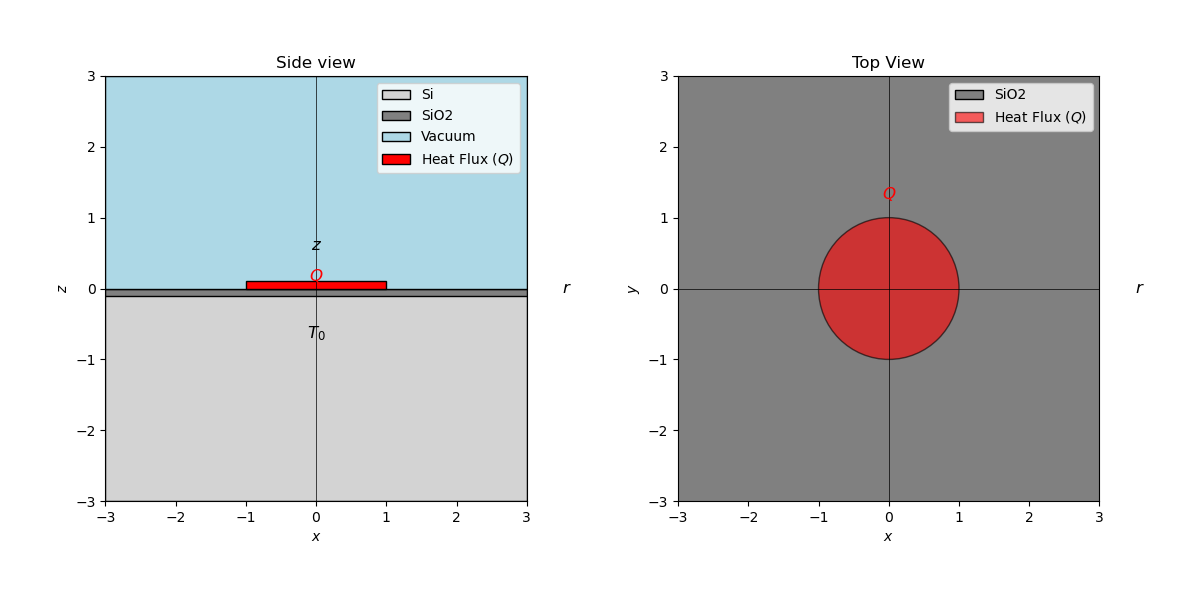}
\caption{Side view (left panel) and top view (right panel) of the considered simple geometry, where a disk-shaped heat flux is applied on top of a $285~\mathrm{nm}$-thick SiO$_2$ layer and a semi-infinite Si substrate. Scales are in units of the disk radius $R$.}
\label{modele_Manu}
\end{figure*}

A more detailed approach for calculating $P_{\mathrm{out\mbox{-}of\mbox{-}plane}}$ involves solving the general heat
equation for a more complicated geometry. In its three-dimensional Cartesian form, the transient heat conduction equation
is expressed as 

\begin{equation}
    \rho\, C_p \frac{\partial T}{\partial t} = \nabla \cdot (\kappa \nabla T) +Q,
\end{equation}

\noindent where $\rho$ represents the material density, $C_p$ the specific heat capacity, $T$ the temperature, $t$ the time, $\kappa$ the thermal conductivity, $\nabla \cdot$ the divergence and $Q$ the heat generation term. 

We consider in what follows a steady-state scenario with a cylindrical geometry (Fig.~SI-~\ref{modele_Manu}), where the graphene transistor under large bias generates a disk-shaped heat flux (represented in red) on top of the SiO$_2$/Si substrate.

\noindent In this case the heat equation is given by 

\begin{equation}
    \frac{1}{r} \frac{\partial}{\partial r} \left[ r \kappa_{rr}(r,z) \frac{\partial T}{\partial r}\right] + \frac{\partial}{\partial z} \left[ \kappa_{zz} (r,z) \frac{\partial T}{\partial z} \right] = -Q,
\end{equation}

\noindent where $\kappa_{rr}(r,z)$ and $\kappa_{zz}(r,z)$ are the spatially dependent thermal conductivities (taking into account uniaxial materials such as hBN).

\noindent Table SI-\ref{tab:conduc} summarizes the thermal conductivities of the different materials considered in the heat transfer problem.

\begin{table}[h]
\centering
\hspace*{-0.7cm}
    \begin{tabular}{| p{4.0cm} |  c | }
    \hline
    \textbf{Material} & \textbf{Thermal conductivity}   \\
       & $(\mathrm{W.m^{-1}.K^{-1}})$\\ \hline\hline
    \textbf{SiO$_{2}$}& 1.4 \\ \hline
    \textbf{Si}& 150 \\ \hline
    \textbf{hBN (in-plane)\cite{Mercado2020}}& 400 \\ \hline
    \textbf{hBN (out-of-plane)\cite{Jaffe2023}}& 8 \\ \hline
    \textbf{Au}& 320 \\ \hline
    \end{tabular}
\caption{Thermal conductivities of the materials making up the studied devices.}
\label{tab:conduc}
\end{table}

\begin{figure}[ht!]
\centering
\includegraphics[width=\linewidth]{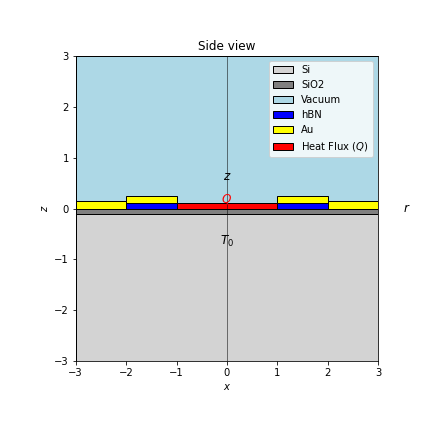}
\caption{Side view of a more realistic geometry, where gold metallizations and a hollow hBN disk are present. Scales are in units of the disk radius $R$.}
\label{modele_Manu2}
\end{figure}

 We considered two different geometries:

\begin{enumerate}[label=(\roman*)]
    \item  The simple geometry considered in the model of Ref. \cite{pop2010energy}, where a silicon dioxide layer extends from $z=0$ to $z=-285~\mathrm{nm}$ on top of a semi-infinite (actual thickness: 525~$\mu$m) Si substrate. A small disk-shaped heat flux is applied at the position of the graphene channel under bias (disk between $r=0$ and $r=R$, with the radius of the disk $R=19.7$~$\mu$m chosen to mimic the total surface of the graphene transistor TR2 under study). Both layers extend from $r=0$ to $r=1\,\mathrm{mm}$ and have distinct thermal properties. Such geometry is displayed in Fig.~SI-~\ref{modele_Manu}. \\
 \item  A more realistic geometry shown in Fig.~SI-~\ref{modele_Manu2}, where we add an additional layer of hBN situated on top of the SiO$_2$ layer and located radially from $r=R$ to $r=2R$ (hBN hollow disk) and vertically from $z=0$ to $z=80\,\mathrm{nm}$. This layer models the fraction of hBN outside of the graphene region that has been etched to create edge contacts. We also add gold metallizations, with an Au disk located atop the hBN hollow disk (thickness: 150~nm) and an outer Au disk located everywhere outside the hBN region (from $r=2R$ to $r=1\,\mathrm{mm}$). These two disks represent the gold access deposited for the electrodes. 
\end{enumerate}

\vspace{1mm}

 The aforementioned steady-state heat equation is solved numerically using a finite-element method implemented in Mathematica. The solution domain is discretized into finite elements and the equation is solved over these elements.
The boundary conditions define the fixed heat flux on top and the fixed temperature at the bottom of the structure. Finally, the spatial variation in the thermal conductivity is encoded by defining a piece-wise function for $\kappa_{rr}$
and $\kappa_{zz}$ to account for these variations across different regions of the domain.

 Table SI-\ref{tab:val} compares the thermal resistances obtained for the different scenarios discussed above and the simple model introduced at the beginning of this section, where we consider the average temperature for the SiO$_{2}$ layer. The results underline the fact that the simple model is in excellent agreement with a more refined treatment of the whole structure and confirms that the out-of-plane radiative power reaches almost $70 \%$ of the total dissipated electrical power. Note that this result relies on the following four assumptions

\begin{enumerate}[label=(\roman*)]
\item  The SiO$_2$ layer is thin enough so that we consider it almost transparent, making the emission power additive with the SiO$_2$ layer depth. \\
\item Interference effects are negligible in the contribution from each layer to the far-field. \\
\item The elevation in temperature is small compared to room temperature so that we can use the linearized version of the emitted power with respect to temperature elevation. We also assume an affine temperature profile within SiO$_2$ (which is confirmed by simulations).\\
\item Interfacial thermal resistances are neglected.
\end{enumerate}

\begin{figure*}[ht]
\centering
\includegraphics[width=\linewidth]{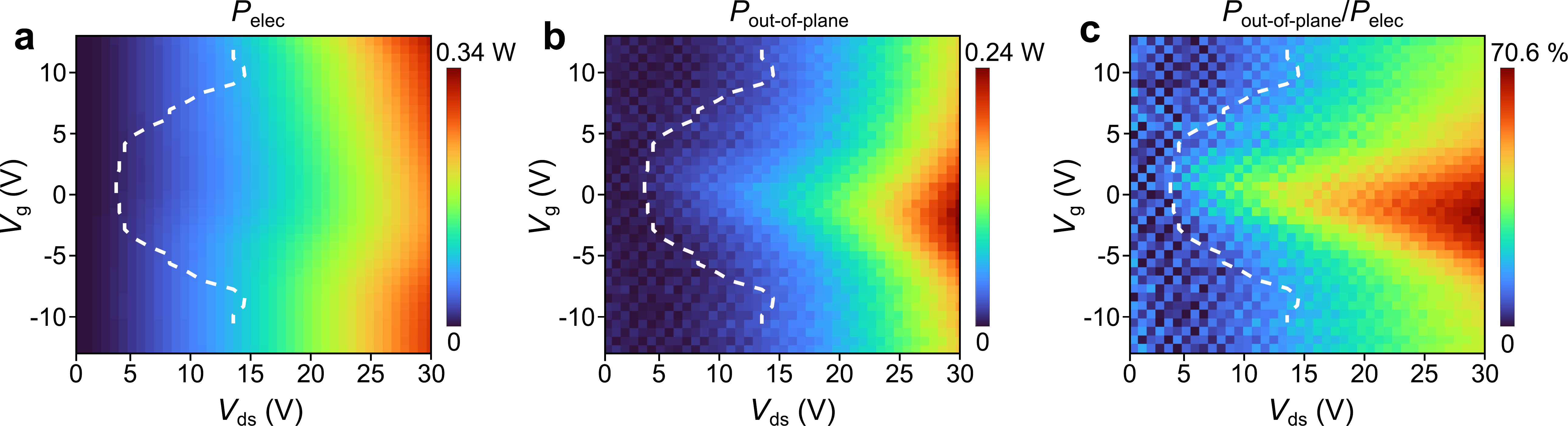}
\caption{\textbf{a}, Total electrical power ($P_{\mathrm{elec}}$) injected into the TR2 device as a function of bias. \textbf{b}, Out-of-plane dissipated power ($P_{\mathrm{out\mbox{-}of\mbox{-}plane}}$) in the TR2 device as a function of electrical bias, computed from $P_{\mathrm{out}} =  \Delta T_{\mathrm{SiO}_{2}} / R_{\mathrm{th}}$, where $ \Delta T_{\mathrm{SiO}_{2}} \simeq 0-40$~$^{\circ}$C and $R_{\mathrm{th}} = 171.4$~K.W$^{-1}$ (see Table SI-\ref{tab:val}). \textbf{c}, Percentage giving the contribution of the out-of-plane radiative energy transfer mechanism to the total power budget obtained from the ratio $\frac{P_{\mathrm{out\mbox{-}of\mbox{-}plane}}}{ P_{\mathrm{elec}}} = \frac{P_{\mathrm{out\mbox{-}of\mbox{-}plane}}}{ P_{\mathrm{out\mbox{-}of\mbox{-}plane}} + P_{\mathrm{in-plane}}}$, where $P_{\mathrm{in-plane}}$ is the in-plane dissipated power. }
\label{all powers}
\end{figure*}

\begin{table}[t!]
\centering
\hspace*{-0.7cm}
    \begin{tabular}{| p{6.0cm} |  c | }
    \hline
    \textbf{Model} & \textbf{Thermal resistance}   \\
       & $(\mathrm{K.W^{-1}})$\\ \hline\hline
    Basic model\cite{pop2010energy}& 178.33 \\ \hline
    Heat equation w/ SiO$_2$ and Si (Fig. SI- \ref{modele_Manu})& 172.72 \\ \hline
    Addition of hBN hollow disk& 171.9 \\ \hline
    Introduction of Au metallization (Fig. SI-~\ref{modele_Manu2})& 171.44 \\ \hline

    \end{tabular}
\caption{Thermal resistances obtained for the different models (for $R=19.7~\mathrm{\mu m}$ corresponding to that of the TR2 device)}
\label{tab:val}
\end{table}

The out-of-plane power resulting from the solution of the full thermal problem is plotted in Fig.~SI-~\ref{all powers}\textbf{b} along with the ratio between $P_{\mathrm{out\mbox{-}of\mbox{-}plane}}$ and the total electrical power $P_{\mathrm{elec}}$ injected into the device (Fig.~SI-~\ref{all powers}\textbf{c}), which gives the contribution of out-of-plane radiative near-field energy transfer to the device's total power budget (Fig.~SI-~\ref{all powers}\textbf{a}).

Crucially, we note that, while the near-field electroluminescent energy transfer occurs between graphene and the first neighbouring layers of the top and bottom hBN layers, the pyrometric measurement of $P_{\mathrm{out\mbox{-}of\mbox{-}plane}}$ given by Eq.~(\ref{Rtot}) is insensitive to the thermal resistance of the bottom hBN layer $R_{\mathrm{hBN}}$. This is simply a consequence of the thermal Ohm's law.

\section{Determination of the far-field imaging resolution of the mid-IR setup} \label{opt res sec}

\begin{figure*}[t!]
\centering
\includegraphics[width=\linewidth]{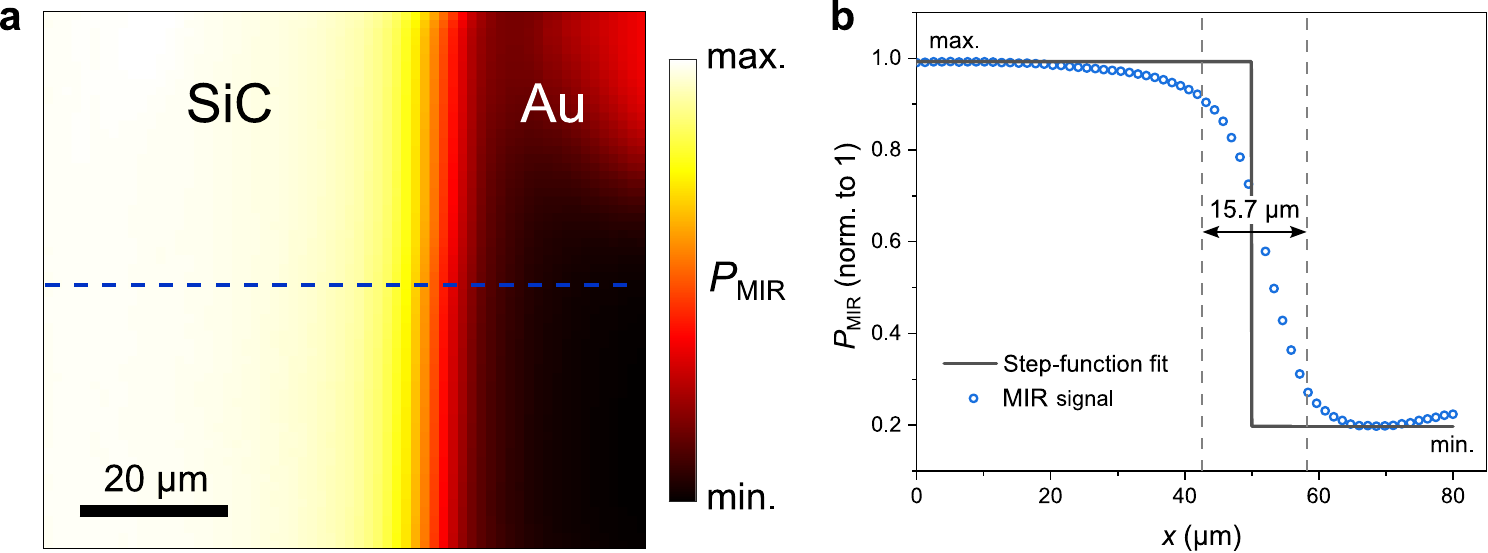}
\caption{\textbf{a}, Spatial scan of the spectrally-integrated mid-infrared signal ($P_{\textrm{MIR}}$) measured at the interface between a SiC substrate and a large Au pattern. \textbf{b}, Average horizontal line scan of panel \textbf{a}. The two vertical lines correspond to the positions at which the signal reaches 10$\,\%$ and 90$\, \%$ of its maximal value. The imaging resolution is estimated to be 15.7~$\mu$m.}
\label{spatial_resolution}
\end{figure*}

The optical resolution of the mid-IR experimental setup is determined through the knife-edge method described in Fig. SI-\ref{spatial_resolution}. A spatial scan of the mid-IR signal from a straight silicon carbide/gold interface is measured (Fig. SI-\ref{spatial_resolution}\textbf{a}) and an average horizontal line scan is extracted, whose profile resembles a step function. The resolution is defined by the distance between two vertical lines on the profile, one when the signal reaches 90\% and one when it reaches 10\% (see Fig. SI-\ref{spatial_resolution}\textbf{b}). The optical resolution is thus measured to be 15.7~$\mu$m.

\section{Hyperbolic plasmon-phonon-polariton dispersion curves in hBN-encapsulated graphene}
\label{polariton dispersion}

\begin{figure}[ht!]
\centering
\includegraphics[width=\linewidth]{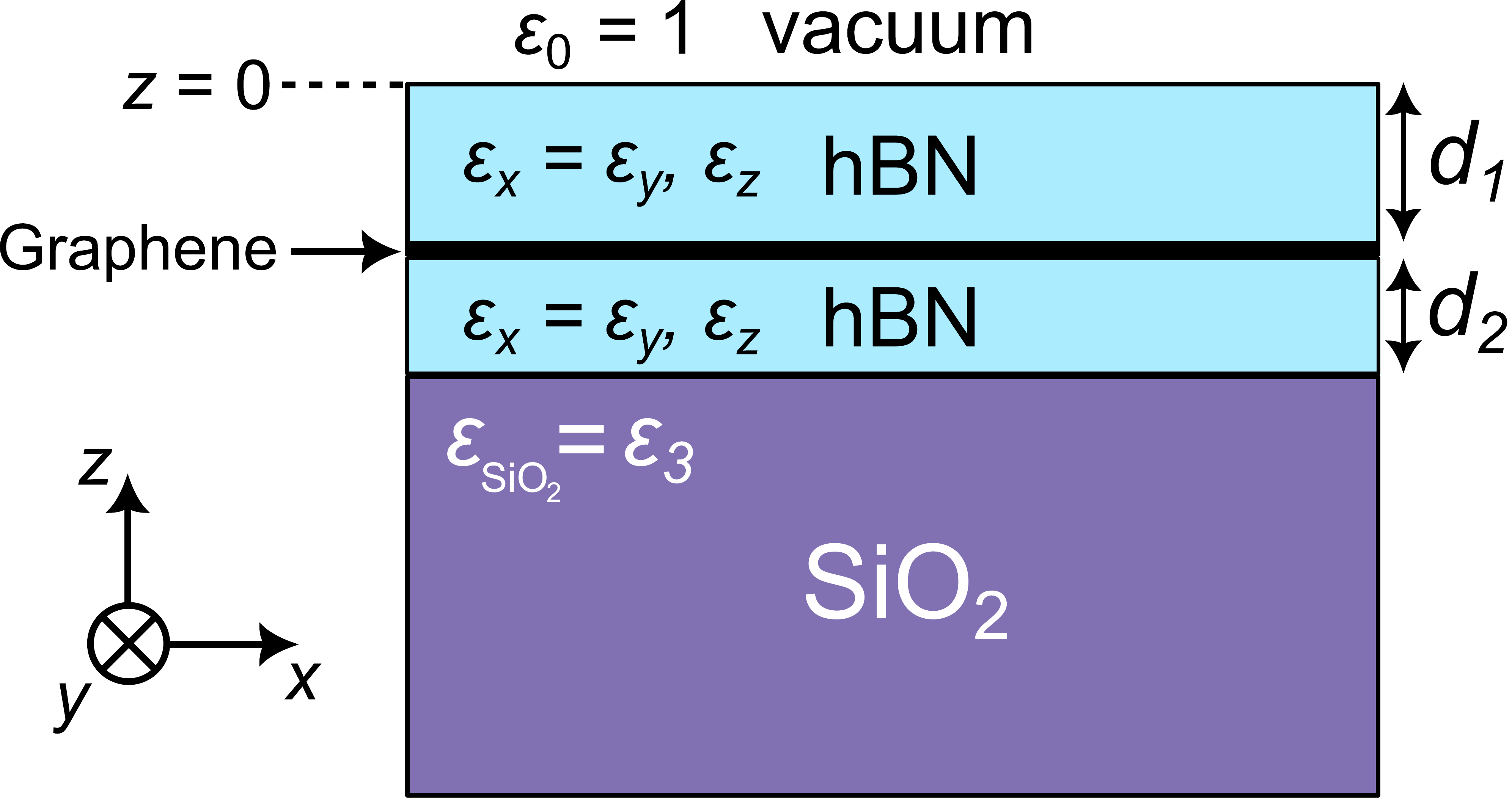}
\caption{Sketch of a graphene layer encapsulated between two hBN thin films of thicknesses $d_{1}$ and $d_{2}$ on an SiO$_{2}$ substrate. }
\label{hBN_graph}
\end{figure}

\noindent
The polaritonic dispersion curves of our samples were obtained by evaluating the poles of the reflection coefficient $r_{p}$ of the sample's multi-layer heterostructure for $p$-polarized light. A similar approach was utilised in Refs.~\cite{dai2014tunable, dai2015graphene}.

\begin{figure*}[t!]
\centering
\includegraphics[width=\linewidth]{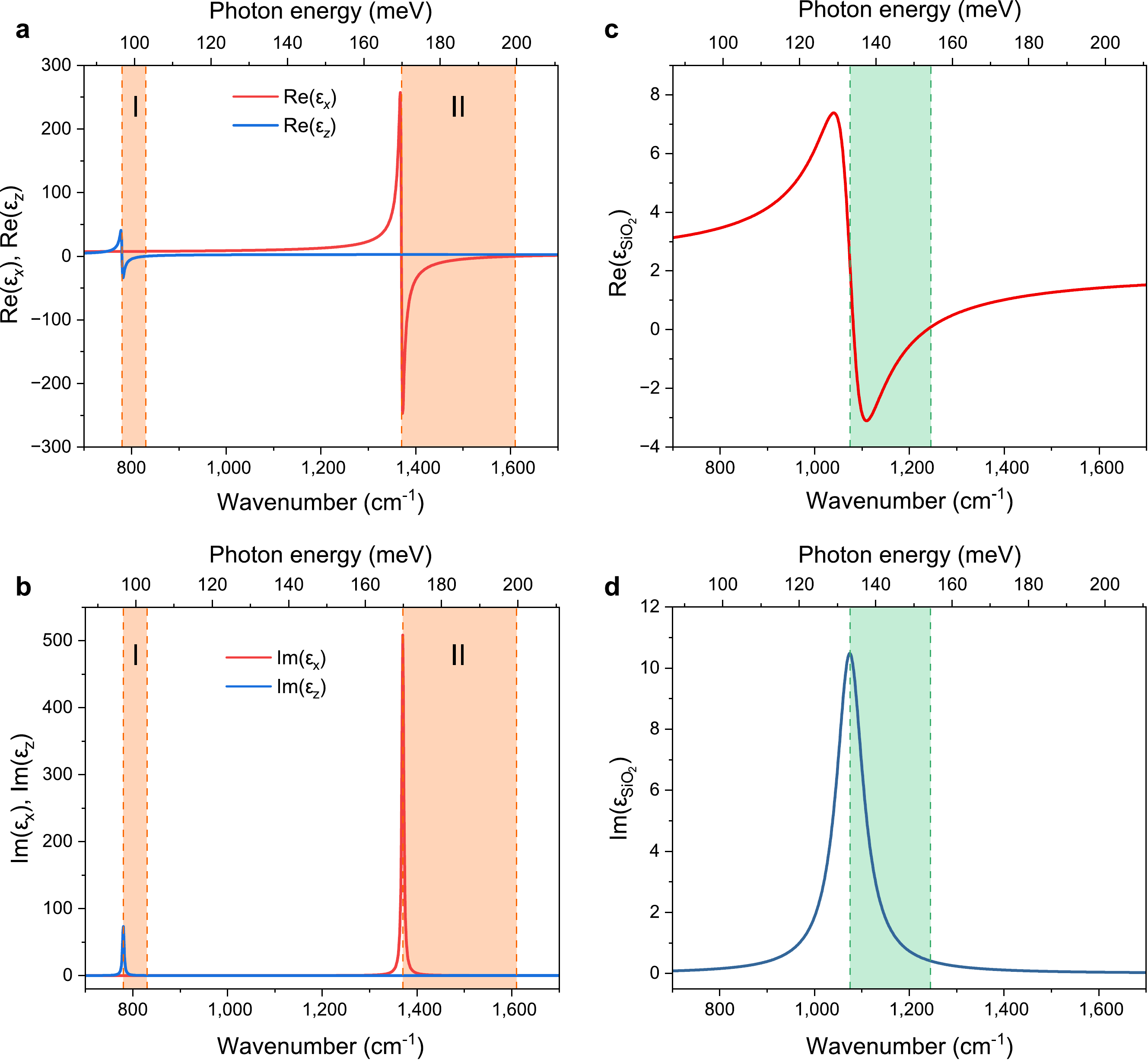}
\caption{
\textbf{a} \& \textbf{b}, Real and imaginary parts of the dielectric function of hBN, obtained by applying Eq.~(\ref{eps DL}) for the in-plane ($x$~\&~$y$) and out-of-plane ($z
$) directions. The orange-shaded regions indicate the two \emph{Reststrahlen} bands of hBN. The optical constants appropriate for hBN are\cite{kumar2015tunable} $\varepsilon_{\infty}^{z}=2.95$, $\varepsilon_{0}^{z}=3.34$,  $\omega_{\mathrm{TO}}^{z}=780$~cm$^{-1}$, $\omega_{\mathrm{LO}}^{z}=830$~cm$^{-1}$, and $\gamma^{z} = 4$~cm$^{-1}$, for the out-of-plane direction (type~I \emph{Reststrahlen} band),  and $\varepsilon_{\infty}^{x}=4.87$, $\varepsilon_{0}^{x}=6.72$, $\omega_{\mathrm{TO}}^{x}=1,370$~cm$^{-1}$, $\omega_{\mathrm{LO}}^{x}=1,610$~cm$^{-1}$, and $\gamma^{x} = 5$~cm$^{-1}$ for the in-plane direction (type~II \emph{Reststrahlen} band).
\textbf{c} \& \textbf{d}, Real and imaginary parts of the dielectric function of SiO$_{2}$. The green-shaded region indicates the \emph{Reststrahlen} band of SiO$_{2}$. Optical constants:\cite{palik1998handbook} $\varepsilon_{\infty}^{\mathrm{SiO}_{2}}=1.97$, and $\varepsilon_{0}^{\mathrm{SiO}_{2}}=2.64$, $\omega_{\mathrm{TO}}^{\mathrm{SiO}_{2}}=1,075$~cm$^{-1}$, $\omega_{\mathrm{LO}}^{\mathrm{SiO}_{2}}=1,245$~cm$^{-1}$, and $\gamma^{\mathrm{SiO}_{2}}=69$~cm$^{-1}$.}
\label{eps}
\end{figure*}

\begin{figure*}[t!]
\centering
\includegraphics[width=\linewidth]{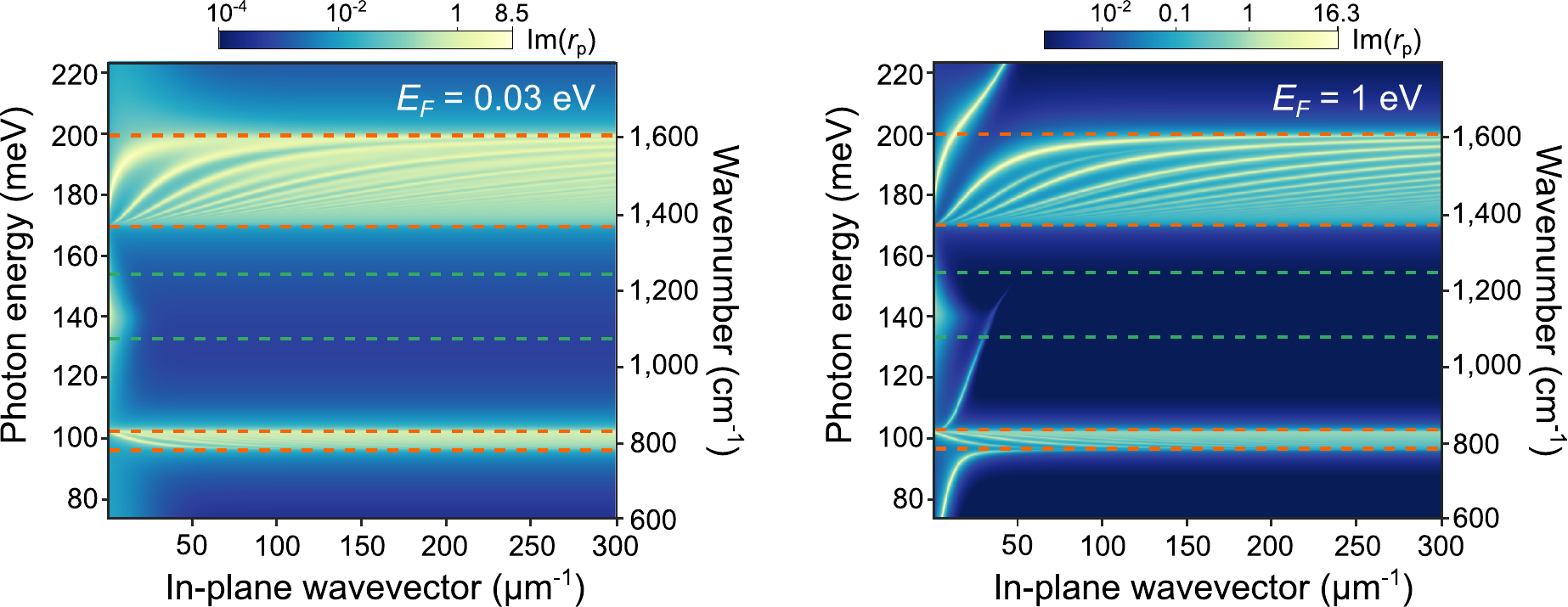}
\caption{HPhP-plasmon dispersion curves visualized through the maxima of the imaginary part of the reflection coefficient $r_{p}$ (plotted in a logarithmic scale), for two different values of the Fermi energy ($E_{F} = 30$~meV, left; $E_{F} = 1$~eV, right), electron mobility $\mu_{e} = 6.6\ \rm{m^2.V^{-1}.s^{-1}}$, and electron temperature $T_{e} = 625$~K. The thicknesses of the hBN layers are $d_{1} = 52$~nm and $d_{2} = 30$~nm. The dashed orange and green lines enclose the \emph{Reststrahlen} bands of hBN and SiO$_{2}$, respectively.}
\label{HPhP disp}
\end{figure*}

Our samples consist of a mono-layer graphene encapsulated by two thin hBN flakes and placed on an SiO$_{2}$/Si substrate with a 285~nm-thick SiO$_{2}$ layer and an Si layer of bulk thickness. If we consider the SiO$_{2}$ thin film on the Si substrate only, then the surface phonon-polaritons supported by this layered system should exhibit two polariton branches $\omega_{+}$ and $\omega_{-}$ with large wavevector asymptotic values of their dispersion curves reached when $\textrm{Re}(\varepsilon_{\textrm{SiO}_{{2}}})= -1$ for the upper polariton $\omega_{+}$, and $\textrm{Re}(\varepsilon_{\textrm{SiO}_{{2}}})=-\varepsilon_{\textrm{Si}}=-11.7$ for the lower polariton branch $\omega_{-}$ \cite{kliewer1974theory, agranovich2012surface}, where $\varepsilon_{\textrm{SiO}_{{2}}}$ and $\varepsilon_{\textrm{Si}}$ are the dielectric functions of SiO$_{2}$ and Si, respectively. Nevertheless, the large damping inherent to amorphous SiO$_{2}$ precludes the excitation of the lower polariton branch, as the minimal value of the real part of $\varepsilon_{\textrm{SiO}_{{2}}}$ is $\textrm{min}\big\{  \textrm{Re}(\varepsilon_{\textrm{SiO}_{2}})  \big\} \approx -3.1$ (see Fig.~SI-~\ref{eps}\textbf{c}). The polaritonic response, in this case, is independent of the SiO$_{2}$ layer thickness, \emph{i.e.}, the response is the same as that of bulk SiO$_{2}$. Thus, to fully describe the polaritonic response of our samples, it is sufficient to consider a four-layered hBN/graphene/hBN/SiO$_{2}$ heterostructure, as sketched in Fig.~SI-~\ref{hBN_graph}.

We use a \emph{Drude-Lorentz} model with a single oscillator to describe the in-plane ($\varepsilon_{x} = \varepsilon_{y}$) and out-of-plane ($\varepsilon_{z}$) dielectric functions of hBN, as well as the dielectric function of SiO$_{2}$ ($\varepsilon_{\mathrm{SiO}_{2}}$), that is

\begin{align}
    \varepsilon_{\mu} (\omega) &= \varepsilon_{\infty}^{\mu} +  \frac{\varepsilon_{0}^{\mu}-\varepsilon_{\infty}^{\mu}}{(\omega_{\textrm{TO}}^{\mu})^{2}-\omega^{2}- i \gamma^{\mu} \omega}(\omega_{\textrm{TO}}^{\mu})^{2}\label{eps DL} \\
    &\mathrm{with} \quad   \mu = x, z, \mathrm{or} \,\mathrm{SiO}_{2}. \nonumber   
\end{align}

\noindent
Here, $\omega_{\textrm{TO}}^{\mu}$ is the transverse optical phonon frequency, $\gamma^{\mu}$ is a damping constant, and $\varepsilon_{\infty}^{\mu}$ and $\varepsilon_{0}^{\mu}$ are
the values of $\varepsilon_{\mu}(\omega)$ for $\omega \gg \omega_{\textrm{TO}}^{\mu}$, and $\omega \ll \omega_{\textrm{TO}}^{\mu}$, respectively. The longitudinal optical phonon frequency $\omega_{\textrm{LO}}^{\mu}$ is the frequency for which $\varepsilon_{\mu}(\omega) = 0$ and is given in terms of $\omega_{\textrm{TO}}^{\mu}$ by the Lyddane-Sachs-Teller relation \cite{lyddane1941polar}: $\omega_{\textrm{LO}}^{\mu} = \omega_{\textrm{TO}}^{\mu}(\varepsilon_{0}^{\mu}/\varepsilon_{\infty}^{\mu})^{1/2}$. The real and imaginary parts of the dielectric functions of hBN and SiO$_{2}$, calculated from Eq.~(\ref{eps DL}), are shown in Fig.~SI-~\ref{eps}.

The influence of the graphene mono-layer is entailed through the inclusion of its sheet conductivity $\sigma$. This is obtained by summing the intraband ($\sigma_{\mathrm{intra}}$) and the interband ($\sigma_{\mathrm{inter}}$) contributions to the sheet conductivity, \emph{i.e.}, \cite{falkovsky2008optical} 

\begin{align}
    \sigma(\omega) &= \sigma_{\mathrm{intra}}(\omega) + \sigma_{\mathrm{inter}}(\omega), \\
    \sigma_{\mathrm{intra}}(\omega) &=   \frac{i}{\omega + i \Gamma} \frac{2e^{2}k_{B}T_{e}}{\pi \hbar^{2}}   \mathrm{Ln} \bigg( 2 \cosh{\frac{E_{F}}{2k_{B}T_{e}}} \bigg), \\
     \sigma_{\mathrm{inter}}(\omega) &= \frac{e^{2}}{4\hbar} \Bigg[G\bigg(\frac{\hbar \omega}{2}\bigg)+ i\frac{4\hbar \omega}{\pi}  \int_{0}^{\infty}  \frac{G(\xi)-G\big(\frac{\hbar \omega}{2}\big)}{(\hbar \omega)^{2} 
  - 4\xi^{2} }d\xi 
    \Bigg],
\end{align}

\noindent
where
\begin{equation}
G(\xi)=\frac{1-\exp{\left(-\frac{2\xi}{k_B T_e}\right)}}{\exp{\left(\frac{E_F-\xi}{k_B T_e}\right)}+\exp{\left(-\frac{E_F+\xi}{k_B T_e}\right)}+1+\exp{\left(-\frac{2\xi}{k_B T_e}\right)}}.
\end{equation}

\noindent
In the above equations, $E_{F}$ is the Fermi energy, $T_{e}\sim 600$~K is the electron temperature, and $\Gamma =  e v_{F}^{2}/\mu_{e}  E_{F}$, represents the phenomenological scattering rate or damping constant of electrons due to impurity scattering, in which $\mu_{e}$ is the electron mobility, $v_{F} \approx 10^{6}$~m/s is the Fermi velocity, and $e$ is the electron charge.

 Assuming that the thicknesses of the encapsulating hBN layers are $d_{1}$ and $d_{2}$, we may divide the hBN/graphene/hBN/SiO$_{2}$ heterostructure of Fig.~SI-~\ref{hBN_graph} into the following four regions: $z>0$ ($j=0$, vacuum), $-d_{1}<z<0$ ($j=1$, hBN), $-d_{1}-d_{2}<z<-d_{1}$ ($j=2$, hBN), and $z<-d_{1}-d_{2}$ ($j=3$, SiO$_{2}$). The total reflection coefficient of the structure can then be obtained by adding the two hBN thin films successively as follows. The reflection coefficient for the first film is written as

\begin{equation}
    r_{p1}  =     \frac{r_{12}  +r_{23}(1-r_{12}-  r_{21})e^{2ik_{z}d_{2}}}{1 - r_{21}r_{23}e^{2ik_{z}d_{2}}}.
\end{equation}

\noindent
The addition of the second film yields the total reflection coefficient of the stack:

\begin{equation}
r_{p2} = r_{p}  =     \frac{r_{01}  +r_{p1}(1-r_{01}-  r_{10})e^{2ik_{z}d_{1}}}{1 - r_{10}r_{p1}e^{2ik_{z}d_{1}}}. \label{rp encaps graph}
\end{equation}
\noindent
In the above two equations, $r_{12} =(\Lambda_{2} - \Lambda_{1} + S)/(\Lambda_{1} + \Lambda_{2} + S)$ and $r_{21} =(\Lambda_{1} - \Lambda_{2} + S)/(\Lambda_{1} + \Lambda_{2} + S)$ are the reflection coefficients at the graphene/hBN boundaries, and $r_{mn}   = - r_{nm} =(\Lambda_{n} - \Lambda_{m} )/(\Lambda_{n} + \Lambda_{m} ) $ ($mn \neq 12,21$) are the reflection coefficients at the boundary between regions $m$ and $n$, in which the following definitions were used 

\begin{align*}
      &k_{z} = \sqrt{  \varepsilon_{x} \bigg( \frac{\omega^{2}}{c^{2}}  -  \frac{k_{x}^{2}}{\varepsilon_{z}}  \bigg)    },  \quad  k_{z0,3} = \sqrt{  \varepsilon_{0,3}  \frac{\omega^{2}}{c^{2}}  -  k_{x}^{2}   }, \\
     & \textrm{Im}( k_{z}, k_{z0,3})>0,  \quad \Lambda_{1,2} = \frac{\varepsilon_{x}}{k_{z}},   \quad  
 \Lambda_{0,3} = \frac{\varepsilon_{0,3}}{k_{z0,3}},\\
      &  \varepsilon_{0} = 1, \quad   \varepsilon_{3} = \varepsilon_{\mathrm{SiO}_{2}}, \quad \textrm{and}  \quad S = \frac{1}{\epsilon_{\circ}}\frac{\sigma(k_{x},\omega)}{\omega}.
\label{Qs}
\end{align*}

\noindent
Note that $\epsilon_{\circ} = 8.854 \times 10^{-12}\ \rm{F.m^{-1}}$ is the permittivity of free space, and $k_{z}$ and $k_{z0,3}$ give the out-of-plane polariton wavevectors for $j=1,2$ and $j=0,3$, respectively.

The polariton branches, visualized from the maxima of the imaginary part of the reflection coefficient $r_{p}$ [Eq.~(\ref{rp encaps graph})], are shown in Fig.~SI-~\ref{HPhP disp} for two different values of the Fermi energy (0.03~eV and 1~eV). As presented in Fig.~SI-~\ref{HPhP disp}, the influence of graphene's plasmons on the HPhP branches of hBN becomes significant only for high Fermi energies ($E_{F} =1$~eV) and is almost completely negligible when the graphene is close to neutrality (see Fig.~SI-~\ref{HPhP disp}, $E_{F}= 0.03$~eV). It should be noted here that the devices investigated in this study were measured near charge neutrality. 

\begin{figure}[t!]
\centering
\includegraphics[width=\linewidth]{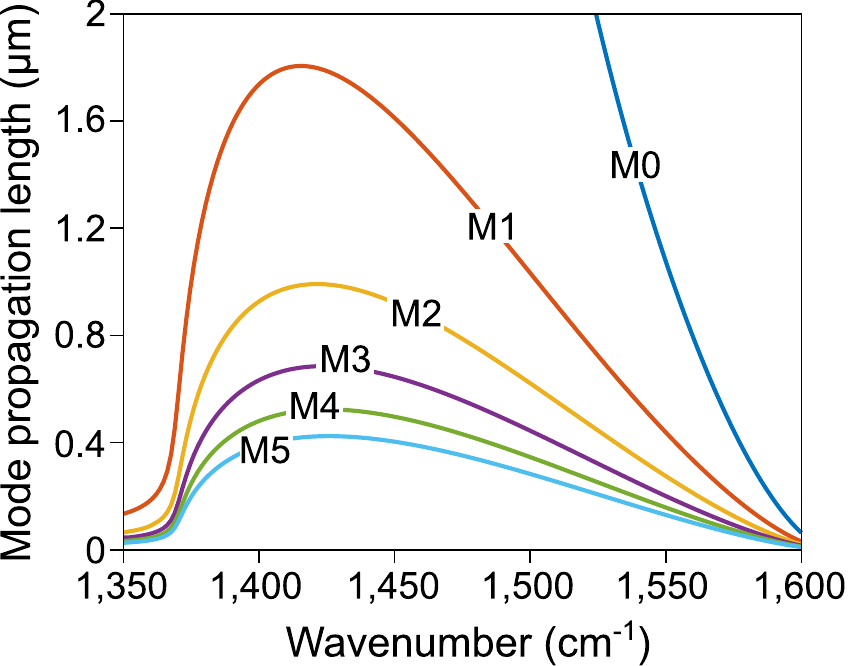}
\caption{Calculated lateral propagation length of the first six HPhP thin film modes.}
\label{6modespropagation}
\end{figure}

Using Eq.~(\ref{rp encaps graph}) of the complex reflection coefficient, one can solve for the complex poles $k_{n}$ in $\mathbf{k}$ at each frequency. Writing $k_{n}=k_{n}^{\prime}+ik_{n}^{\prime\prime}$, the electromagnetic fields of mode M$n$ (see Fig. 2a in the main text) then scale as $e^{i k_{n}^{\prime} r -k_{n}^{\prime\prime} r}$, where $r$ is a spatial coordinate. The quantity $L_{n}=1/k_{n}^{\prime\prime}$ thus defines the lateral propagation length for the field of the $n^{th}$ mode, which is represented in Fig. SI-\ref{6modespropagation} for the first 6 modes. Both the M0 and M1 modes propagate at micrometric lengthscales ($>1$~$\mu$m), while higher order modes propagate at shorter distances ($<1$~$\mu$m). Therefore, for a scatterer located less than 600~nm from the emission point, energy is conveyed by HPhP rays, whereas for scatterers further away, only thin film modes M0, M1, and to a lesser extent M2 are involved.



\section{Stokes-anti-Stokes characterisation of PDC-hBN} \label{PDC SAS sec}

\begin{figure}[h!]
\centering
\includegraphics[width=\linewidth]{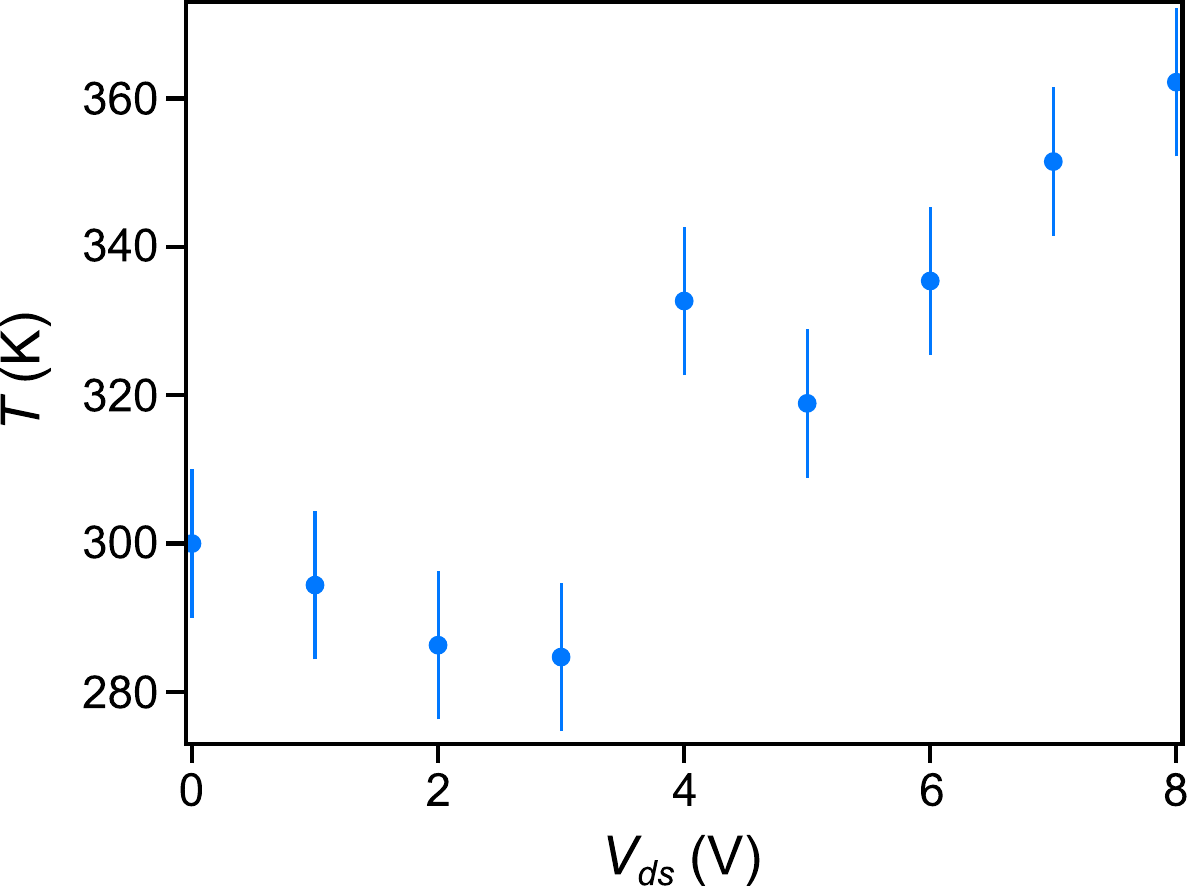}
\caption{\textbf{Stokes-anti-Stokes Raman characterisation of the PDC-hBN device Lyon1.} Plotted is the temperature of hBN's optical phonons from the $E_{2g}$ mode (1,365 $\mathrm{cm}^{-1}$) as a function of bias. The electronic charge density is $n = 6 \times 10^{11}$~cm$^{-2}$.}
\label{SAS_Lyon}
\end{figure}

\begin{figure*}[t!]
\centering
\includegraphics[width=\linewidth]{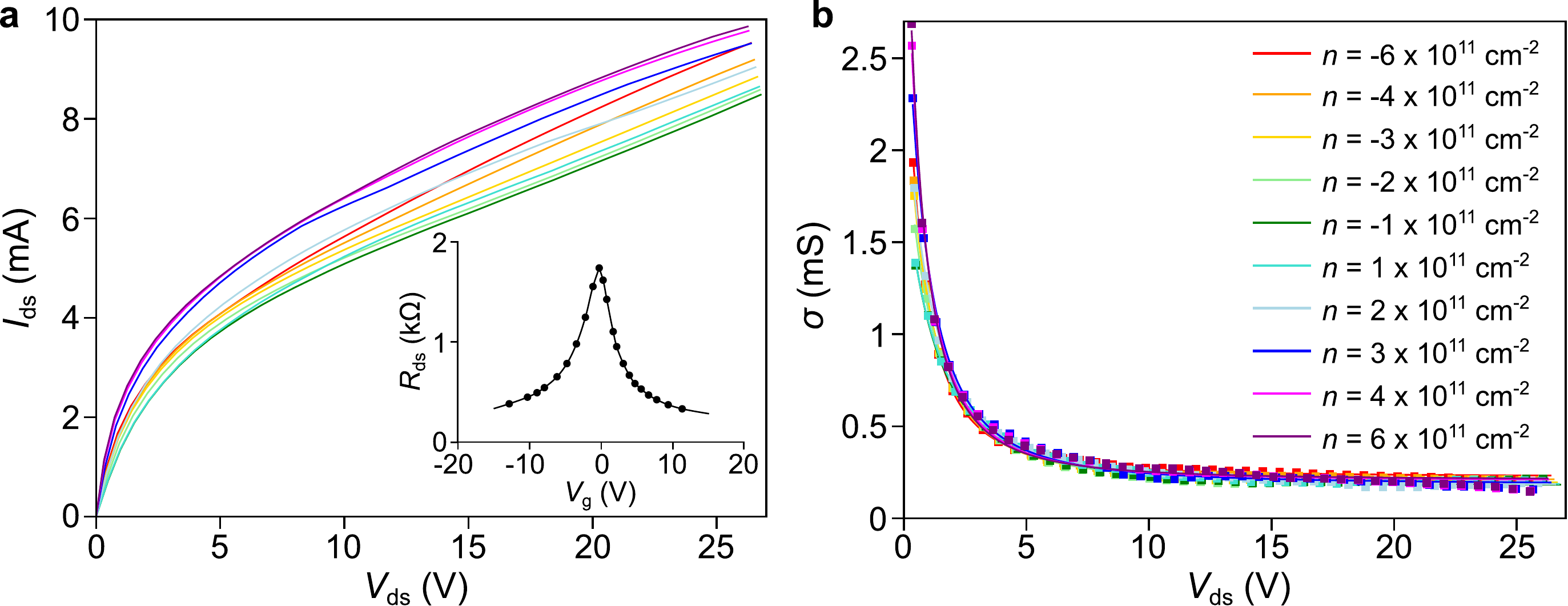}
\caption{\textbf{Electronic transport in the TR2 device}. \textbf{a}, Current-voltage curves for a constant charge carrier density $n=[-6,6]\times 10^{11}~\mathrm{cm^{-2}}$. Note that the curves are measured at constant value of gate voltage ($V_{g}$) and are then corrected from drain-gating effects. The inset shows the transfer curve of the device. \textbf{b}, Differential conductance $\sigma$ plotted as a function of bias voltage for the same doping range as in panel \textbf{a}. Fits using Eq.~(\ref{eq1}) [solid lines] are in good agreement with experimental data [squares].}
\label{transport_TR2}
\end{figure*}

\noindent
We performed a Stokes-anti-Stokes \emph{in situ} determination of the temperature of hBN's optical phonons for devices made with PDC-hBN under large bias (see Fig. SI-~\ref{SAS_Lyon}). The temperatures obtained are comparable to the ones obtained with HPHT-hBN and confirm that the decay of hyperbolic phonon-polaritons into optical phonons is similar for both hBN crystals.

\section{Electronic transport characterisation} \label{elec trans sec}

\noindent
In this section, we illustrate the transport properties of the investigated devices. The TR2 device is chosen as a representative device here since all the studied devices exhibit identical transport behaviour. The current-voltage curves are first corrected from drain-gating effects and are plotted in Fig. SI-\ref{transport_TR2}\textbf{a}. At low bias, the devices show linear ohmic behavior, with a strong increase of the intraband current with doping. Due to the high mobility, the current rapidly reaches significant values and saturates for $V_{ds} > V_{\text{sat}}$, where $V_{\text{sat}}$ is the saturation voltage. This saturation of intraband current can be consistently attributed to the interaction with hyperbolic phonon-polaritons of the lower \emph{Reststrahlen} band of hBN ($\hbar \omega_{I}= 90-100$~meV) \cite{yang2018graphene}\cite{Baudin2020adfm}.
This velocity saturation regime is followed at large bias by the onset of Zener interband conduction, characterised by a constant bias- and doping-independent differential conductivity $\sigma_{Z}\lesssim1\;\mathrm{mS}$, and the onset of near-field radiative energy transfer (see main text). 
The differential conductivity is fitted using the standard dependence \cite{yang2018graphene}\cite{Meric2008}:

\begin{equation}
\label{eq1}
    \sigma(E)= \sigma_{\text{intra}} + \sigma_{\text{inter}} = \frac{ne\mu}{(1+E/E_{\text{sat}})^{2}}+ \sigma_{Z}
\end{equation}

\noindent
where $\mu$ is the electronic mobility at zero bias, $n$ is the charge carrier density, and $E=V/L$ is the local electric field in which $V$ is the bias voltage and $L$ is the device channel length. Fits of the differential conductivity are presented in Fig. SI-~\ref{transport_TR2}\textbf{b}, showing good agreement with experimental data. The existence of a doping-dependent voltage threshold for interband Zener conduction related to Pauli blocking is unaccounted for in Eq.~(\ref{eq1}), which is a valid approximation as the intraband conductivity is usually much larger than $\sigma_Z$ at low bias. In order to extract an estimate of the voltage threshold ($V_{\text{threshold}}$) for the onset of the Zener-Klein tunneling regime from these fits, we use the empirical criterion:

\begin{equation}
\sigma_{\text{intra}}=\sigma_Z ~~ \mathrm{when}~~ V=V_{\text{threshold}}
\end{equation}
which proves to be quite accurate. Values of the voltage threshold are represented by dashed lines in the maps in Figure 3 of the main text.

\section{An atomic-force microscopy map of the hBN fold on TR1}
\label{AFM sec}

\begin{figure*}[th!]
\centering
\includegraphics[width=\linewidth]{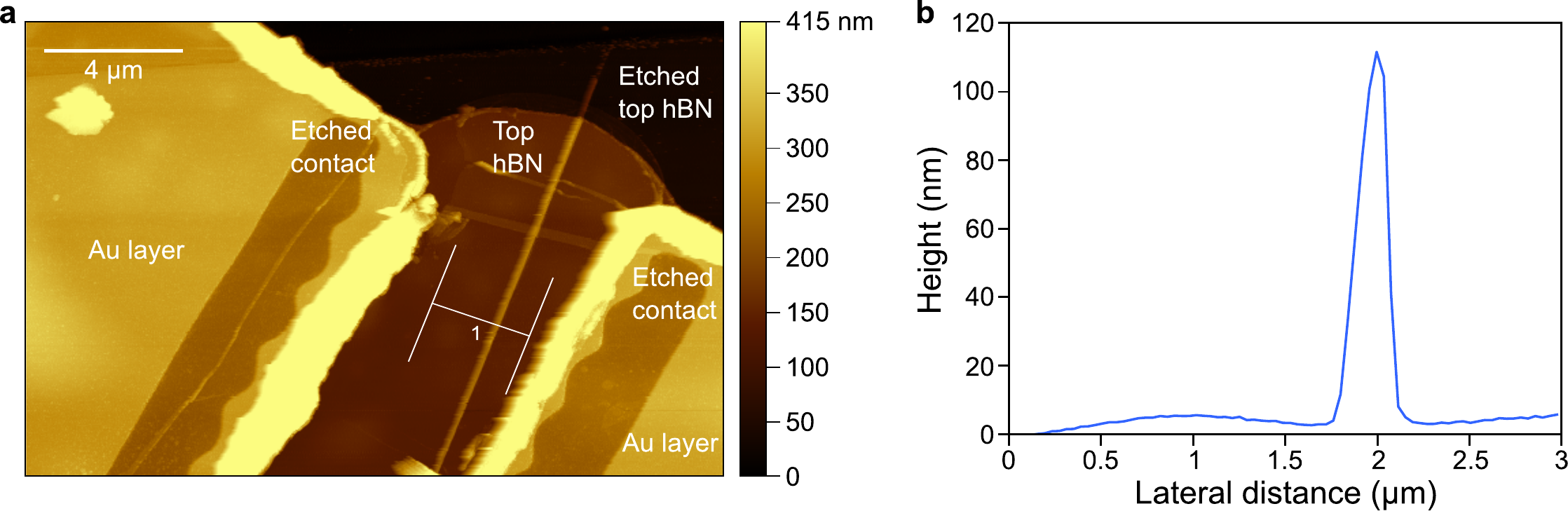}
\caption{\textbf{Atomic-force microscopy (AFM) measurement of transistor TR1 with a fold in its top hBN layer.} \textbf{a}, AFM map of the channel with the fold and gold contacts. The white line indicated by the number 1 is where a line profile was extracted.
\textbf{b}, Profile extracted from panel \textbf{a}, averaged over 100 pixels and shifted down by 140~nm so that the baseline is near zero for ease of reading. The measured fold height and width are 108~nm and 370~nm, respectively.}
\label{TR1_AFM}
\end{figure*}

We investigated the topology of the channel of the TR1 device via atomic-force microscopy (AFM) to obtain the geometric parameters of the fold in the top hBN layer. The AFM map of TR1 is shown in Fig. SI-~\ref{TR1_AFM}\textbf{a}. A line profile of the map, giving a cross-section of the fold, was extracted and averaged over 100 pixels to reduce noise (see Fig. SI-\ref{TR1_AFM}\textbf{b}). The fold was found to be 108~nm high and 370~nm thick. Similar values were used to compute the layer-by-layer emissivity of the heterostructure with this type of scatterer (see Fig.~SI-~\ref{layer_by_layer_emittance_simu}d and Fig. SI-\ref{scatterer_types}\textbf{b}).

\section{The origin of the observed peaks in the emission spectrum under electrical bias}
\label{spec features}

To understand the origin of the spectral features observed in emission in Fig.~2c of the main text, we turn to the layer-by-layer simulations of the emissivity of a heterostructure with the geometry of the TR1 device. We successively introduce models with increasing complexity to reveal the role of each element in the final emission spectrum.

\begin{figure*}[t!]
\centering
\includegraphics[width=\linewidth]{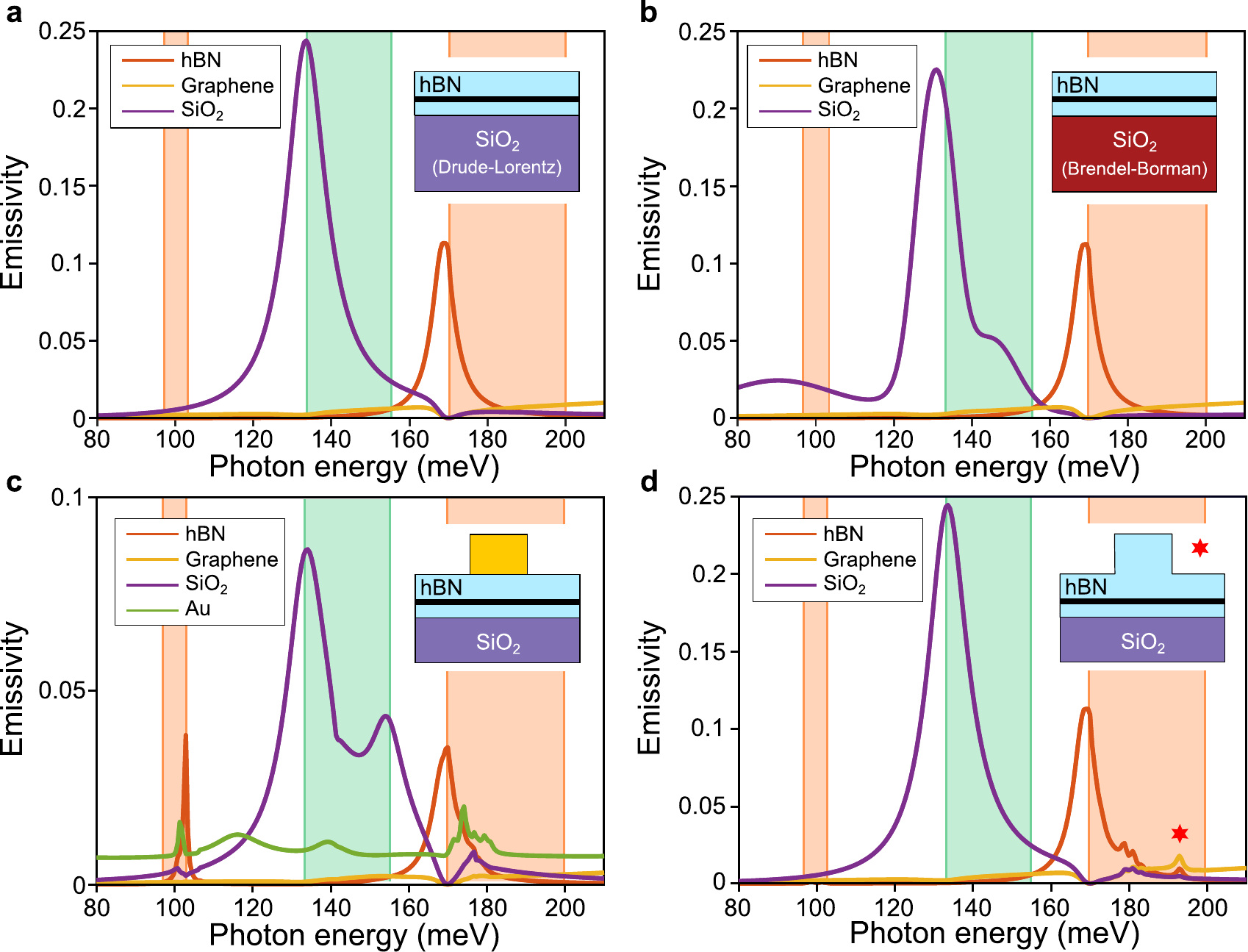}
\caption{Simulation of the layer-by-layer emissivity of the TR1 heterostructure, considering the individual roles of different elements. The layer thicknesses are as follows: 285 nm of $\mathrm{SiO_2}$, and 52 nm and 30 nm for the top and bottom hBN flakes, respectively.\textbf{a}, The pristine heterostructure with the electromagnetic response of $\mathrm{SiO_2}$ modeled using a Drude-Lorentz model. \textbf{b}, The same simulation as in \textbf{a}, but with the electromagnetic response of $\mathrm{SiO_2}$ modeled using the more complex Bormann-Brendel approach described in the text. \textbf{c}, Effect of Au contacts: A gold stripe (height, 100 nm; width, 24 $\mu$m; periodic boundary condition, 30 $\mu$m) is considered on the surface. The simulation uses the Drude-Lorentz model for $\mathrm{SiO_2}$. \textbf{d}, Effect of a fold on the top hBN: A rectangular-shaped fold (height, \SI{100}{\nano\meter}; width, \SI{250}{\nano\meter}) is on the surface. The simulation uses the Drude-Lorentz model for $\mathrm{SiO_2}$.}
\label{layer_by_layer_emittance_simu}
\end{figure*}

Figure SI-~\ref{layer_by_layer_emittance_simu}a represents the pristine heterostructure's emissivity using the Drude-Lorentz model described in section~\ref{polariton dispersion} of this supplemental document for the electromagnetic response of SiO$_{2}$ [see Eq.~(\ref{eps DL})]. The chosen spectral range corresponds to the mid-IR spectral window of our study and is centered on SiO$_{2}$’s \emph{Reststrahlen} band (green-shaded region in Fig. SI-~\ref{layer_by_layer_emittance_simu}) and covers hBN’s first and second \emph{Reststrahlen} bands as well (orange-shaded regions). The simulation demonstrates that the two broad peaks observed experimentally at 135~meV and 167~meV originate from the SiO$_{2}$ substrate and hBN layers' absorption, respectively. These peaks correspond to the transverse optical phonon energies of SiO$_{2}$ ($\omega_{\text{TO}}^{\text{SiO}_{2}}$) and in-plane phonons of hBN ($\omega_{\text{TO}}^{x}$). 

The emission spectra of devices in SiO$_{2}$’s \emph{Reststrahlen} band present two additional peaks at larger energies. We first start by revisiting the idealized Drude-Lorentz model of SiO$_{2}$. This model cannot capture in detail the full electromagnetic response of the SiO$_{2}$ glass on a wafer, which is in a non-crystalline form and therefore includes modifications of its vibrational modes due to disorder-induced coupling. This effect has been thoroughly explained in Ref.~\onlinecite{kirk1988prb} and can be captured by a Brendel-Bormann dielectric permittivity model\cite{brendel1992prb}. We use the following Brendel-Bormann model 

$$\varepsilon = \varepsilon_\infty + \sum_j 
\frac{\nu_{p,j}^2}{\sqrt{8 \pi}\sigma_{j} a_{j}}
\Big[ I_0(z_{+,j}) + I_0(z_{-,j}) \Big],$$
where $\varepsilon_\infty=2.15137$, $z_{\pm,j}=\frac{a_j\pm \nu_{0,j}}{\sqrt{2 \sigma_j}}$, $a_j = \sqrt{\nu^2 + i \Gamma_j \nu}$, and $I_0(z)=\int_{-\infty}^{\infty} \frac{e^{-x^2}}{x-z} \mathrm{d}x=i\pi e^{-z^2}\mathrm{erfc}(-iz).$

\begin{table}[h!]
\centering
\begin{tabular}{ccccc}
\toprule
$j$ & $\nu_0$ (\si{\per\centi\meter}) & $\Gamma$ (\si{\per\centi\meter}) & $\nu_p$ (\si{\per\centi\meter}) & FWHM (\si{\per\centi\meter}) \\
\midrule
1 & 724.653 & 2.5189 & 336.91 & 275.850 \\
2* & 1,053.249 & 15.1579 & 735.42 & 69.049 \\
3 & 1,872.704 & 1,603.9714 & 237.80 & 132.663 \\
4 & 433.254 & 21.4118 & 429.95 & 62.608 \\
5* & 1,172.866 & 0.0669 & 266.93 & 110.664 \\
6 & 585.871 & 105.4864 & 2.53 & 102.901 \\
\bottomrule
\end{tabular}
\caption{Parameters of the Brendel-Bormann model of $\mathrm{SiO_2}$'s dielectric permittivity used for the layer-by-layer emissivity simulations of the TR1 heterostructure. The Gaussian broadening parameters are $\sigma=\mathrm{FWHM}/\sqrt{8 \ln(2)}$. The * symbol indicates the modes participating in the SiO$_{2}$ emission in the spectral range of interest. }
\label{tab:simulation_parameters}
\end{table}

\noindent
The simulation result using the above model is shown in Fig. SI-~\ref{layer_by_layer_emittance_simu}b. Spectral weight from the main SiO$_{2}$ phonon resonance at 135~meV  is transferred to an additional peak at 145~meV. The amplitude of this new peak, which is associated with disorder-induced coupled bright and dark vibrational modes AS$_1$-AS$_2$ of the O-Si-O groups, is strongly dependent on the SiO$_{2}$ disorder and is therefore sample-dependent. As such, the amplitude of this peak varies experimentally from sample to sample and can approach that of the main SiO$_{2}$ emissivity peak at  135~meV in some cases.

We now return to the simple Drude-Lorentz SiO$_{2}$ model and consider the effect of gold contacts which are made of large (micrometric) stripes of gold. Below the gold layer, an epsilon-near-zero SiO$_{2}$ surface phonon-polariton mode develops. \cite{vassant2012berreman} The simulation in Fig. SI-~\ref{layer_by_layer_emittance_simu}c reveals the presence of this mode due to scattering to the far-field, manifesting as an emissivity peak at 154~meV, near the longitudinal optical phonon frequency of the SiO$_{2}$ ($\omega_{\text{LO}}^{\text{SiO}_{2}}$). This peak can be clearly distinguished in experimental spectra. 

Finally, we consider the fold geometry of the TR1 device of the main text. In Fig.  SI-~\ref{layer_by_layer_emittance_simu}d, we observe the emergence of an emissivity peak at 192 meV associated with graphene and hBN. 

The four peaks illustrated above are the main contributors to the emission spectral features observed under electrical bias in Fig. 2 of the main text. 

Let us now turn towards additional contributions to the emission spectra due to scatterers or defects. First, we note that the emissivity peak due to the scattering of type I hyperbolic phonon-polaritons (HPhPs) in hBN's first \emph{Reststrahlen} band (see Fig.~SI-~\ref{layer_by_layer_emittance_simu}c) is only observed when gold is deposited on top of the heterostructure. With this configuration, it is possible to observe the electroluminescence resulting from Cherenkov radiation of graphene’s accelerated electrons. A small peak at $\sim100$~meV is indeed observed in Fig. 2c of the main text in the measured spectra and in other devices as well (see Extended Data Fig. 8 and 9).

\begin{figure*}[t!]
\centering
\includegraphics[width=\linewidth]{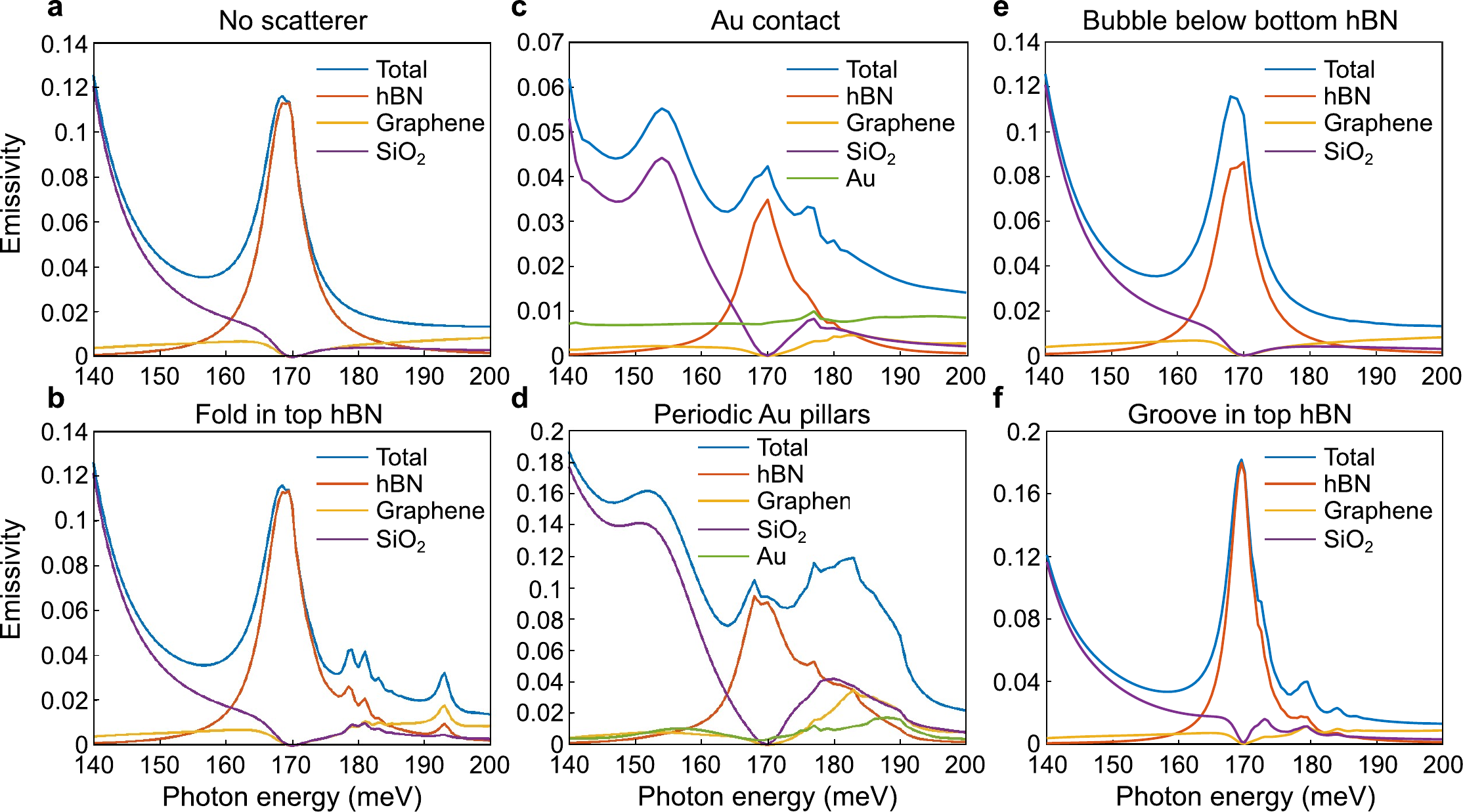}
\caption{\textbf{Computed emissivity of various scatterers for a device with the geometry of TR1.}
\textbf{a}, Emissivity of the heterostructure without any scatterer.
\textbf{b}, Emissivity of the heterostructure with a single fold in the top hBN layer (rectangular shaped, height, 100~nm; width, 250~nm).
\textbf{c}, Emissivity of the heterostructure with gold contacts (height, 100~nm; width, 24~$\mu$m; periodic boundary condition, 30~$\mu$m).
\textbf{d}, Emissivity of the heterostructure with periodic gold pillars (height, 100~nm; diameter, 3.5~$\mu$m; periodic boundary condition, 6.5~$\mu$m).
\textbf{e}, Emissivity of the heterostructure with an air bubble below the bottom hBN layer (20~nm vertical translation of the heterostructure over a disk of 3.5~$\mu$m in diameter, periodic boundary condition period 6.5~$\mu$m).
\textbf{f}, Emissivity of the heterostructure with a groove in the top hBN layer down to graphene (3.5~$\mu$m wide; periodic boundary conditions, 6.5~$\mu$m).}
\label{scatterer_types}
\end{figure*}

In Fig. SI-~\ref{scatterer_types}, we simulate various scatterer configurations to outline their effect on the emissivity in hBN's second \emph{Reststrahlen} band compared to the pristine heterostructure (Fig. SI-~\ref{scatterer_types}a). We find that the hBN fold (Fig. SI-~\ref{scatterer_types}b) produces an isolated high-energy peak at 192~meV. In contrast, gold discontinuities on the structure act as efficient scatterers but produce signals in the low-energy end (167 to 180~meV) of hBN's second \emph{Reststrahlen} band (see Fig. SI-~\ref{scatterer_types}c and d). In practice, this emission is concomitant with hBN's bulk incandescent emission, which precludes the precise identification of the radiation's origin. We must emphasize here that the spectra are rather sensitive to scatterer size. In particular, for scatterers smaller than 300 nm, well-defined confined modes can be identified.
Another unintentional defect that appears experimentally is the presence of bubbles within the heterostructure. Bubbles commonly result from the nanofabrication process and consist of a mixture of water and hydrocarbons \cite{haigh2012cross}. Such bubbles distinctly appear as green spots on the TR2 device (see the bottom micrograph to the right of Fig. 2c in the main text). They have no effect on electrical transport, but in the mid-IR, they may result in the emergence of surface phonon-polaritons from the SiO$_2$/bubble interface which can, in principle, be scattered to the far-field within SiO$_2$'s \emph{Reststrahlen} band. However, this mode was found to be optically-inactive in the numerical simulations presented in Fig. SI-\ref{scatterer_types}e. Finally, a groove in the top hBN layer was found to be an inefficient scatterer (see Fig. SI-~\ref{scatterer_types}f).

\section{Emission spectrum under uniform heating}\label{uni heat}

The quantification of the out-of-plane electroluminescent energy transfer from graphene to the device's substrate is achieved by monitoring the device's substrate mid-IR signal (emission occurring in $\mathrm{SiO_2}$'s \textit{Reststrahlen} band) under bias, which provides access to the temperature increase of $\mathrm{SiO_2}$, $\Delta T_{\mathrm{SiO_2}}$. The temperature increase was calibrated by comparing the device's substrate mid-IR signal under bias to the signal measured at null bias for a known substrate temperature. This pyrometry technique has two limitations: First, the $\mathrm{SiO_2}$ emission under bias could \emph{a priori} originate from electroluminescence of $\mathrm{SiO_2}$ polaritons rather than from its temperature increase. We consider that the $\mathrm{SiO_2}$ signal is always thermal in our experimental conditions, based on the following experimental observations:

\begin{itemize}
    \item The broad $\mathrm{SiO_2}$ peak at 134~meV and its satellite peak at 145~meV in Fig.~\ref{fig2}c correspond to its bulk absorption, as confirmed by the numerical simulation of Fig.~\ref{fig2}d (green curve) of the main text (see also section~\ref{spec features} of this supplement). This simulation considers the translationally invariant case in which phonon-polaritons are optically inactive. Consequently, the two peaks at 134~meV and 145~meV are due to incandescent emission from $\mathrm{SiO_2}$;
    \item The $\mathrm{SiO_2}$ emission peak at 154~meV in Fig.~\ref{fig2}c (main text) is due to the scattering of phonon-polaritons by defects or gold contacts. If this peak were due to electroluminescence, its relative spectral weight with respect to the $\mathrm{SiO_2}$ bulk absorption peaks would be much larger when the device is biased compared to when it is heated. However, we have observed relatively similar spectral weights in both cases (Fig.~\ref{fig2}c under bias, and Extended Data Fig.~\ref{ED_fig6} heated at 75°C and under bias), indicating at least that the $\mathrm{SiO_2}$ signal is mostly thermal in origin.
\end{itemize}

Based on the above observations, we conclude that the $\mathrm{SiO_2}$ apparent radiation temperature under electrical bias is the average $\mathrm{SiO_2}$ temperature. Moreover, as mentioned in the main text, the electroluminescence from HPhPs, at the edge of the spectral window of the HgCdTe detector used, is limited to $11\%$ of the integrated signal.

\begin{figure}[t!]
\centering
\includegraphics[width=\linewidth]{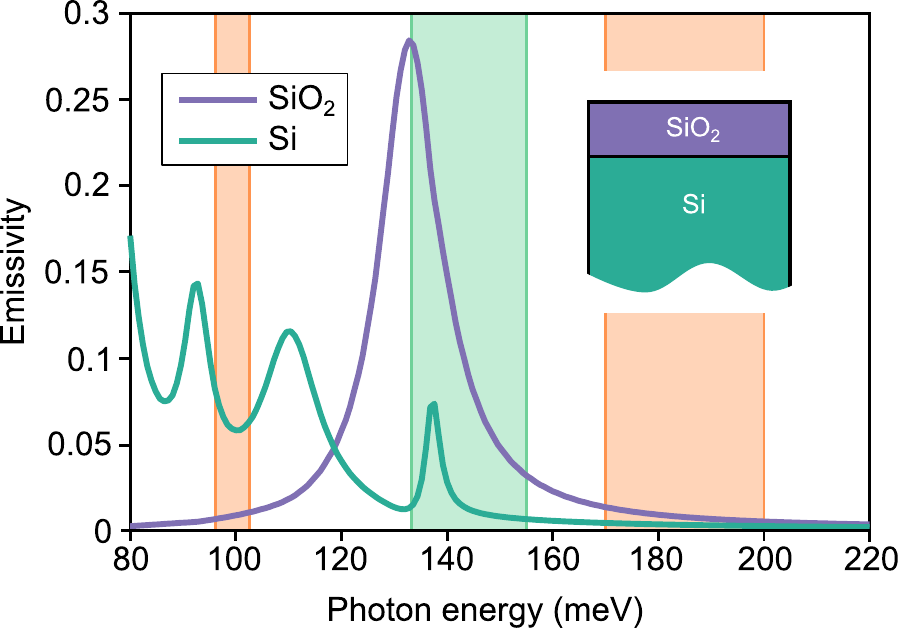}
\caption{\textbf{Simulation of the layer-by-layer emissivity of the $\mathrm{SiO_2/Si}$ wafers used in this work.}}
\label{Effect_of_Si}
\end{figure}

The second limitation is due to the reference measurements at known substrate temperatures, which are plotted in Extended Data Fig.~4a. We observe that, while their spectral profiles are similar, the low-energy end of the incandescent spectra is more pronounced than the one obtained under electrical bias. This effect is simply due to the extra contribution of the silicon wafer to the emission: Silicon has an extremely small absorptivity in the mid-IR, which is compensated for by its large thickness (0.6~mm) in our samples compared to the $\mathrm{SiO_2}$ thickness. Under electrical bias, the silicon incandescence signal is negligible because the local heating produced by the transistor operation under bias is limited to a depth on the order of its lateral width, \textit{i.e.}, $\lesssim  35~\mathrm{\mu m}$. However, under uniform warming of the sample, the silicon bulk brings an extra contribution around a spectral energy of 110~meV (887~$\text{cm}^{-1}$, see Fig.~SI-~\ref{Effect_of_Si}). Note that our detector's detectivity is maximal at $\sim 1,000$~cm$^{-1}$, making it rather sensitive to the sample's absorptivity in the vicinity of this frequency. 

In our pyrometry analysis, by equating the apparent temperature radiation and $\mathrm{SiO_2}$ temperature, we do not consider this effect. If this effect is taken into account, the out-of-plane energy transfer estimate increases by $\sim 20\%$.

\bibliography{references} 

\begin{thebibliography}{10}

\bibitem{akasaki2015nobel}
Isamu Akasaki.
\newblock Nobel lecture: Fascinated journeys into blue light.
\newblock {\em Rev. Mod. Phys.}, 87(4):1119, 2015.

\bibitem{rosencher2002optoelectronics}
Emmanuel Rosencher and Borge Vinter.
\newblock {\em Optoelectronics}.
\newblock Cambridge University Press, 2002.

\bibitem{kalt2019semiconductor}
H~Kalt and C~Klingshirn.
\newblock {\em Semiconductor Optics}.
\newblock Springer, 2019.

\bibitem{low2017polaritons}
T.~Low, A~Chaves, J~D Caldwell, A~Kumar, N~X Fang, Ph~Avouris, T~F Heinz, F~Guinea, L~Martin-Moreno, and F~Koppens.
\newblock Polaritons in layered two-dimensional materials.
\newblock {\em Nat. Mater.}, 16(2):182--194, 2017.

\bibitem{Dai2014science}
S~Dai, Z~Fei, Q~Ma, AS~Rodin, M~Wagner, AS~McLeod, MK~Liu, W~Gannett, W~Regan, K~Watanabe, T~Taniguchi, M~Thiemens, G~Dominguez, AH~Castro Neto, A~Zettl, F~Keilmann, P~Jarillo-Herrero, MM~Fogler, and DN~Basov.
\newblock Tunable phonon polaritons in atomically thin van der {W}aals crystals of {B}oron {N}itride.
\newblock {\em Science}, 343:1125--1129, 2014.

\bibitem{Dai2015nnano}
S~Dai, Q~Ma, MK~Liu, T~Andersen, Z~Fei, MD~Goldflam, M~Wagner, K~Watanabe, T~Taniguchi, M~Thiemens, et~al.
\newblock Graphene on hexagonal boron nitride as a tunable hyperbolic metamaterial.
\newblock {\em Nat. Nanotechnol.}, 10(8):682--686, 2015.

\bibitem{Wei2018Nnano}
W~Yang, S~Berthou, X~Lu, Q~Wilmart, A~Denis, M~Rosticher, T~Taniguchi, K~Watanabe, G~Fève, JM~Berroir, G~Zhang, C~Voisin, E~Baudin, and B~Plaçais.
\newblock A graphene {Z}ener–{K}lein transistor cooled by a hyperbolic substrate.
\newblock {\em Nat. Nanotechnol.}, 13:47--52, 2018.

\bibitem{Baudin2020adfm}
E~Baudin, C~Voisin, and B~Plaçais.
\newblock Hyperbolic phonon polariton electroluminescence as an electronic cooling pathway.
\newblock {\em Adv. Funct. Mater.}, 30:1904783, 2020.

\bibitem{Li2020jmcc}
J~Li, C~Elias, G~Ye, D~Evans, S~Liu, R~He, G~Cassabois, B~Gil, P~Valvin, B~Liu, and JH~Edgar.
\newblock Single crystal growth of monoisotopic hexagonal boron nitride from a {Fe–Cr} flux.
\newblock {\em J. Mater. Chem. C}, 8:9931--9935, 2020.

\bibitem{Messina2016prb}
R~Messina, P~Ben-Abdallah, B~Guizal, M~Antezza, and SA~Biehs.
\newblock Hyperbolic waveguide for long-distance transport of near-field heat flux.
\newblock {\em Phys. Rev. B}, 94:104301, 2016.

\bibitem{Shi2015nlett}
J~Shi, B~Liu, P~Li, LY~Ng, and S~Shen.
\newblock Near-field energy extraction with hyperbolic metamaterials.
\newblock {\em Nano Lett.}, 15:1217--1221, 2015.

\bibitem{shen2009surface}
S~Shen, A~Narayanaswamy, and G~Chen.
\newblock Surface phonon polaritons mediated energy transfer between nanoscale gaps.
\newblock {\em Nano Lett.}, 9(8):2909--2913, 2009.

\bibitem{Tielrooij2018Nnano}
KJ~Tielrooij, N~CH Hesp, A~Principi, MB~Lundeberg, EA~A Pogna, L~Banszerus, Z~Mics, M~Massicotte, P~Schmidt, D~Davydovskaya, D~G Purdie, I~Goykhman, G~Soavi, A~Lombardo, K~Watanabe, T~Taniguchi, M~Bonn, D~Turchinovich, C~Stampfer, A~C Ferrari, G~Cerullo, M~Polini, and F~HL Koppens.
\newblock Out-of-plane heat transfer in van der {W}aals stacks through electron–hyperbolic phonon coupling.
\newblock {\em Nat. Nanotechnol.}, 13:41--46, 2018.

\bibitem{Principi2017prl}
A~Principi, MB~Lundeberg, NCH Hesp, KJ~Tielrooij, F~HL. Koppens, and M~Polini.
\newblock Super-{P}lanckian electron cooling in a van der {W}aals stack.
\newblock {\em Phys. Rev. Lett.}, 118:126804, 2017.

\bibitem{Pogna2021ACS}
EA~A Pogna, X~Jia, A~Principi, A~Block, L~Banszerus, J~Zhang, X~Liu, T~Sohier, S~Forti, K~Soundarapandian, B~Terrés, JD~Mehew, C~Trovatello, C~Coletti, F~HL Koppens, M~Bonn, HI~Wang, N~van Hulst, MJ~Verstraete, H~Peng, Z~Liu, C~Stampfer, G~Cerullo, and KJ~Tielrooij.
\newblock Hot-carrier cooling in high-quality graphene is intrinsically limited by optical phonons.
\newblock {\em ACS Nano}, 15:11285--11295, 2021.

\bibitem{maciel2020probing}
C~Maciel-Escudero, A~Kone{\v{c}}n{\'a}, R~Hillenbrand, and J~Aizpurua.
\newblock Probing and steering bulk and surface phonon polaritons in uniaxial materials using fast electrons: hexagonal boron nitride.
\newblock {\em Phys. Rev. B}, 102(11):115431, 2020.

\bibitem{guo2023hyperbolic}
Q~Guo, I~Esin, Ch~Li, Ch~Chen, S~Liu, J~H Edgar, S~Zhou, E~Demler, G~Refael, and F~Xia.
\newblock Hyperbolic phonon-polariton electroluminescence in graphene-h{BN} van der waals heterostructures.
\newblock {\em arXiv preprint arXiv:2310.03926}, 2023.

\bibitem{Schmitt2023NPhys}
A~Schmitt, P~Vallet, D~Mele, M~Rosticher, T~Taniguchi, K~Watanabe, E~Bocquillon, G~Fève, JM~Berroir, C~Voisin, J~Cayssol, MO~Goerbig, J~Troost, E~Baudin, and B~Plaçais.
\newblock Mesoscopic {K}lein-{S}chwinger effect in graphene.
\newblock {\em Nat. Phys.}, 19:830--835, 2023.

\bibitem{Vandecasteele2010prb}
N~Vandecasteele, A~Barreiro, M~Lazzeri, A~Bachtold, and F~Mauri.
\newblock Current-voltage characteristics of graphene devices: Interplay between {Z}ener-{K}lein tunneling and defects.
\newblock {\em Phys. Rev. B}, 82:045416, 2010.

\bibitem{sadi2020thermophotonic}
Toufik Sadi, Ivan Radevici, and Jani Oksanen.
\newblock Thermophotonic cooling with light-emitting diodes.
\newblock {\em Nat. Photonics}, 14(4):205--214, 2020.

\bibitem{Li2018prl}
C~Li, V~Krachmalnicoff, P~Bouchon, J~Jaeck, N~Bardou, R~Haidar, and Y~De Wilde.
\newblock Near-field and far-field thermal emission of an individual patch nanoantenna.
\newblock {\em Phys. Rev. Lett.}, 121:243901, 2018.

\bibitem{Abou2021ol}
L~Abou-Hamdan, C~Li, R~Haidar, V~Krachmalnicoff, P~Bouchon, and Y~De~Wilde.
\newblock Hybrid modes in a single thermally excited asymmetric dimer antenna.
\newblock {\em Opt. Lett.}, 46(5):981--984, 2021.

\bibitem{Abou2022acs}
L~Abou-Hamdan, L~Coudrat, S~Bidault, V~Krachmalnicoff, R~Ha{\"\i}dar, P~Bouchon, and Y~De~Wilde.
\newblock Transition from phononic to geometrical mie modes measured in single subwavelength polar dielectric spheres.
\newblock {\em ACS Photonics}, 9(7):2295--2303, 2022.

\bibitem{gilburd2017}
L~Gilburd, K~S Kim, K~Ho, D~Trajanoski, A~Maiti, D~Halverson, S~de~Beer, and G~C Walker.
\newblock Hexagonal boron nitride self-launches hyperbolic phonon polaritons.
\newblock {\em J. Phys. Chem. Lett.}, 8(10):2158--2162, 2017.

\bibitem{Greffet2018prx}
JJ~Greffet, P~Bouchon, G~Brucoli, and F~Marquier.
\newblock Light emission by nonequilibrium bodies: Local {K}irchhoff law.
\newblock {\em Phys. Rev. X}, 8:021008, 2018.

\bibitem{footnote1}
Note that all phonons including surface phonon-polaritons are in thermal equilibrium as shown by comparison of the emission spectrum with a control thermal emission spectrum with the same device.

\bibitem{Olsson2016ieee}
A~Olsson, J~Tiira, M~Partanen, T~Hakkarainen, E~Koivusalo, A~Tukiainen, M~Guinea, and J~Oksanen.
\newblock Optical energy transfer and loss mechanisms in coupled intracavity light emitters.
\newblock {\em IEEE Trans. Electron Devices}, 63:3567--3573, 2016.

\bibitem{Radevici2018ssc}
I~Radevici, J~Tiira, T~Sadi, and J~Oksanen.
\newblock Influence of photo-generated carriers on current spreading in double diode structures for electroluminescent cooling.
\newblock {\em Semicond. Sci. Technol.}, 33:05LT01, 2018.

\bibitem{Sadi2020NPhot}
T~Sadi, I~Radevici, and J~Oksanen.
\newblock Thermophotonic cooling with light-emitting diodes.
\newblock {\em Nat. Photonics}, 14:205--214, 2020.

\bibitem{Taniguchi2007jcg}
T~Taniguchi and K~Watanabe.
\newblock Synthesis of high-purity boron nitride single crystals under high pressure by using {Ba–BN} solvent.
\newblock {\em J. Cryst. Growth}, 2:303, 2007.

\bibitem{Maestre20222D}
C~Maestre, Y~Li, V~Garnier, P~Steyer, S~Roux, A~Plaud, A~Loiseau, J~Barjon, L~Ren, C~Robert, B~Han, X~Marie, C~Journet, and B~Toury.
\newblock From the synthesis of h{BN} crystals to their use as nanosheets in van der {W}aals heterostructures.
\newblock {\em 2D Mater.}, 9:035008, 2022.

\bibitem{roux2021}
S~Roux, Ch~Arnold, F~Paleari, L~Sponza, E~Janzen, J~H Edgar, B~Toury, C~Journet, V~Garnier, Ph~Steyer, T~Taniguchi, K~Watanabe, F~Ducastelle, A~Loiseau, and J~Barjon.
\newblock Radiative lifetime of free excitons in hexagonal boron nitride.
\newblock {\em Phys. Rev. B}, 104(16):L161203, 2021.

\bibitem{schmitt2023electroluminescence}
A~Schmitt, L~Abou-Hamdan, M~Tharrault, S~Rossetti, D~Mele, R~Bretel, A~Pierret, M~Rosticher, P~Morfin, T~Taniguchi, et~al.
\newblock Electroluminescence of the graphene 2d semi-metal.
\newblock {\em arXiv preprint arXiv:2306.05351}, 2023.

\bibitem{sapienza2011}
R~Sapienza, P~Bondareff, R~Pierrat, B~Habert, R~Carminati, and NF~Van~Hulst.
\newblock Long-tail statistics of the purcell factor in disordered media driven by near-field interactions.
\newblock {\em Phys. Rev. Lett.}, 106(16):163902, 2011.

\bibitem{Kim2006prl}
W~Kim, J~Zide, A~Gossard, D~Klenov, S~Stemmer, A~Shakouri, and A~Majumdar.
\newblock Thermal conductivity reduction and thermoelectric figure of merit increase by embedding nanoparticles in crystalline semiconductors.
\newblock {\em Phys. Rev. Lett.}, 96:045901, 2006.

\bibitem{poudel2008high}
B~Poudel, Q~Hao, Y~Ma, Y~Lan, A~Minnich, B~Yu, X~Yan, D~Wang, A~Muto, D~Vashaee, et~al.
\newblock High-thermoelectric performance of nanostructured bismuth antimony telluride bulk alloys.
\newblock {\em Science}, 320(5876):634--638, 2008.

\bibitem{pierret2022dielectric}
A~Pierret, D~Mele, H~Graef, J~Palomo, T~Taniguchi, K~Watanabe, Y~Li, B~Toury, C~Journet, P~Steyer, V~Garnier, A~Loiseau, J-M Berroir, E~Bocquillon, G~F{\`e}ve, C~Voisin, E~Baudin, M~Rosticher, and B~Pla\c{c}ais.
\newblock Dielectric permittivity, conductivity and breakdown field of hexagonal boron nitride.
\newblock {\em Mater. Res. Express}, 9(6):065901, 2022.

\bibitem{Schmitt2023prb}
A~Schmitt, D~Mele, M~Rosticher, T~Taniguchi, K~Watanabe, C~Maestre, C~Journet, V~Garnier, G~F`eve, JM~Berroir, C~Voisin, B~Pla\c cais, and E~Baudin.
\newblock High-field 1/f noise in h{BN}-encapsulated graphene transistors.
\newblock {\em Phys. Rev. B}, 107:L161104, 2023.

\bibitem{hugonin2021reticolo}
JP~Hugonin and P~Lalanne.
\newblock Reticolo software for grating analysis.
\newblock {\em arXiv preprint arXiv:2101.00901}, 2021.

\bibitem{pop2010energy}
E~Pop.
\newblock Energy dissipation and transport in nanoscale devices.
\newblock {\em Nano Res.}, 3(3):147--169, 2010.

\bibitem{Mercado2020}
E~Mercado, C~Yuan, Y~Zhou, J~Li, JH~Edgar, and M~Kuball.
\newblock Isotopically enhanced thermal conductivity in few-layer hexagonal {B}oron {N}itride: Implications for thermal management.
\newblock {\em ACS Appl. Nano Mater.}, 3(12):12148--12156, 2020.

\bibitem{Jaffe2023}
GR~Jaffe, KJ~Smith, K~Watanabe, T~Taniguchi, MG~Lagally, MA~Eriksson, and VW~Brar.
\newblock Thickness-dependent cross-plane thermal conductivity measurements of exfoliated hexagonal {B}oron {N}itride.
\newblock {\em ACS Appl. Mater. Interfaces}, 15(9):12545--12550, 2023.

\bibitem{dai2014tunable}
S~Dai, Z~Fei, Q~Ma, AS~Rodin, M~Wagner, AS~McLeod, MK~Liu, W~Gannett, W~Regan, K~Watanabe, et~al.
\newblock Tunable phonon polaritons in atomically thin van der waals crystals of boron nitride.
\newblock {\em Science}, 343(6175):1125--1129, 2014.

\bibitem{dai2015graphene}
S~Dai, Q~Ma, MK~Liu, T~Andersen, Z~Fei, MD~Goldflam, M~Wagner, K~Watanabe, T~Taniguchi, M~Thiemens, et~al.
\newblock Graphene on hexagonal boron nitride as a tunable hyperbolic metamaterial.
\newblock {\em Nat. Nanotechnol.}, 10(8):682--686, 2015.

\bibitem{kumar2015tunable}
A~Kumar, T~Low, KH~Fung, P~Avouris, and NX~Fang.
\newblock Tunable light--matter interaction and the role of hyperbolicity in graphene--hbn system.
\newblock {\em Nano Lett.}, 15(5):3172--3180, 2015.

\bibitem{palik1998handbook}
ED~Palik.
\newblock {\em Handbook of optical constants of solids}, volume~3.
\newblock Academic press, 1998.

\bibitem{kliewer1974theory}
KL~Kliewer and R~Fuchs.
\newblock {\em Theory of dynamical properties of dielectric surfaces}, volume~27.
\newblock Wiley, UK, 1974.

\bibitem{agranovich2012surface}
VM~Agranovich.
\newblock {\em Surface polaritons}.
\newblock Elsevier, 2012.

\bibitem{lyddane1941polar}
RH~Lyddane, RG~Sachs, and E~Teller.
\newblock On the polar vibrations of alkali halides.
\newblock {\em Phys. Rev.}, 59(8):673, 1941.

\bibitem{falkovsky2008optical}
Leonid~A Falkovsky.
\newblock Optical properties of graphene.
\newblock In {\em J. Phys. Conf. Ser.}, volume 129, page 012004. IOP Publishing, 2008.

\bibitem{yang2018graphene}
W~Yang, S~Berthou, X~Lu, Q~Wilmart, A~Denis, M~Rosticher, T~Taniguchi, K~Watanabe, G~F{\`e}ve, JM~Berroir, et~al.
\newblock A graphene zener--klein transistor cooled by a hyperbolic substrate.
\newblock {\em Nat. Nanotechnol.}, 13(1):47--52, 2018.

\bibitem{Meric2008}
I.~Meric, M.~Y. Han, A.~F. Young, B.~Ozyilmaz, P.~Kim, and K.~L. Shepard.
\newblock Current saturation in zero-bandgap, top-gated graphene field-effect transistors.
\newblock {\em Nat. Nanotechnol.}, 3:654--659, 2008.

\bibitem{kirk1988prb}
CT~Kirk.
\newblock Quantitative analysis of the effect of disorder-induced mode coupling on infrared absorption in silica.
\newblock {\em Phys. Rev. B}, 38(2):1255, 1988.

\bibitem{brendel1992prb}
Rolf Brendel and D~Bormann.
\newblock An infrared dielectric function model for amorphous solids.
\newblock {\em J. Appl. Phys.}, 71(1):1--6, 1992.

\bibitem{vassant2012berreman}
Simon Vassant, Jean-Paul Hugonin, Francois Marquier, and Jean-Jacques Greffet.
\newblock Berreman mode and epsilon near zero mode.
\newblock {\em Opt. Express}, 20(21):23971--23977, 2012.

\bibitem{haigh2012cross}
Sarah~J Haigh, Ali Gholinia, Rashid Jalil, Simon Romani, Liam Britnell, Daniel~C Elias, Konstantin~S Novoselov, Leonid~A Ponomarenko, Andre~K Geim, and Roman Gorbachev.
\newblock Cross-sectional imaging of individual layers and buried interfaces of graphene-based heterostructures and superlattices.
\newblock {\em Nat. Mater.}, 11(9):764--767, 2012.

\end{thebibliography}
\bibliographystyle{unsrt}

\end{document}